\documentclass[11pt,english,article, table]{article}
\usepackage{lmodern}

\usepackage[T1]{fontenc}
\usepackage[latin9]{inputenc}
\usepackage{xcolor}
\usepackage{array}
\usepackage{float}
\usepackage{amsmath}
\usepackage{amsthm}
\usepackage{amssymb}
\usepackage{graphicx}
\usepackage{geometry}
\usepackage{setspace}
\usepackage[authoryear]{natbib}
\makeatletter

\providecommand{\tabularnewline}{\\}

\theoremstyle{definition}
\newtheorem{example}{\protect\examplename}

\usepackage{booktabs}
\usepackage{epstopdf}
\setcitestyle{round}
\usepackage{lscape}
\usepackage{longtable}

\date{}
\usepackage{array,tabularx,multirow,graphicx}
\usepackage{float}
\usepackage{pdflscape}
\usepackage{xcolor}

\usepackage{pifont}
\newcommand{\cmark}{\ding{51}}%
\newcommand{\xmark}{\ding{55}}%

\newlength\mylen
\usepackage{tabularx}
\newcolumntype{P}[1]{>{\centering\arraybackslash}p{#1}}
\newcolumntype{Y}{>{\raggedleft\arraybackslash}X}
\newcolumntype{C}{>{\centering\arraybackslash}X}

\newcommand{\nhphantom}[1]{\sbox0{#1}\hspace{-\the\wd0}} % command to add negative space, like off-set a minussign

\ifdefined\showcaptionsetup
 \PassOptionsToPackage{caption=false}{subfig}
\fi
\usepackage{subfig}
\AtBeginDocument{

}

\makeatother

\usepackage{babel}
\providecommand{\examplename}{Example}

\begin{document}
\title{{\huge\textbf{A Multivariate Realized GARCH Model}}\thanks{We thank participants at seminars and conferences, including the Bank
of England in June, 2019, Data Science in Finance, Sao Paulo in Oct.
2019, CFE in Dec. 2019, and Singapore Management University in March,
2020, for their valuable feedback. We are also grateful to the editor,
the guest associate editor, and three anonymous referees for their
valuable comments. Ilya Archakov acknowledges financial support from
the Austrian Science Fund (FWF), research project P33484-G.}{\normalsize\emph{\medskip{}
}}}
\author{\textbf{Ilya Archakov} $^{a}$\textbf{ }$\qquad$\textbf{Peter Reinhard
Hansen} $^{b}$$\qquad$\textbf{Asger Lunde} $^{b,c}$\bigskip{}
\\
{\normalsize$^{a}$}{\normalsize\emph{York University\smallskip{}
}}\\
{\normalsize$^{b}$}{\normalsize\emph{University of North Carolina\smallskip{}
}}\\
{\normalsize$^{c}$}{\normalsize\emph{Copenhagen Economics\smallskip{}
}}}
\date{\today}
\maketitle
\begin{abstract}
We propose a novel class of multivariate GARCH models that incorporate
realized measures of volatility and correlations. The key innovation
is an unconstrained vector parametrization of the conditional correlation
matrix, which enables the use of factor models for correlations. This
approach elegantly addresses the main challenge faced by multivariate
GARCH models in high-dimensional settings. As an illustration, we
explore block correlation matrices that naturally simplify to linear
factor models for the conditional correlations. The model is applied
to the returns of nine assets, and its in-sample and out-of-sample
performance compares favorably against several popular benchmarks.

\bigskip{}
\end{abstract}
{\small\textit{Keywords:}}{\small{} Multivariate GARCH, Realized Measures,
Correlation Factor Model, Block Correlation Matrix}{\small\par}

\noindent{\small\textit{JEL Classification:}}{\small{} G11, G17, C32,
C58 \newpage}{\small\par}

\section{Introduction}

Univariate GARCH models have enjoyed considerable empirical success
since the ARCH model was introduced by \citet{Engle:1982}. Subsequently,
many univariate GARCH-type models have been proposed in the literature,
whereas the research on multivariate GARCH models is less extensive.
Generalizing a univariate GARCH model to higher dimensions involves
several choices, and it is not always clear how to do this in the
most natural way. One challenge to modeling the conditional covariance
matrix, $H_{t}=\mathrm{var}(r_{t}|\mathcal{F}_{t-1})$, is the need
for $H_{t}$ to be positive (semi) definite. This requirement amounts
to nonlinear restrictions across all the elements of $H_{t}$. Another
obstacle is that the number of covariance terms increases with $n^{2}$,
where $n$ is the dimension of the system. This becomes computationally
challenging unless $n$ is relatively small. Multivariate GARCH models
have been introduced to address these issues, see \citet{BauwensLaurentRombouts:2006},
\citet{SilvennoinenTerasvirta:2009}, and \citet{FrancqZakoian2019}
for reviews of this literature. In this paper, we adopt a novel approach
that guarantees a positive definite $H_{t}$, while the complexity
of the model can be contained with a simple factor model. We model
the conditional variances and the conditional correlations separately,
similar to the Dynamic Conditional Correlation (DCC) model by \citet{Engle2002}.
See also \citet{EngleSheppard:2001}, \citet{PakelShephardSheppardEngle:2021},
\citet{Aielli:2013}, and \citet{EngleLedoitWolf:2019}, and see \citet{EngleKelly:2012}
for the DCC variants known as DECO (Dynamic Equicorrelation Correlation)
and Block-DECO.

Our main contributions are the following. First, we develop a new
class of multivariate GARCH models that facilitate flexible modeling
of the correlation structure, while positive definiteness is assured
as an innate property. We refer to these as Multivariate Realized
GARCH (MRG) models. The main methodological contribution is the dynamic
model for the correlation matrix, which can accommodate a simple factor
structure and utilize realized measures of correlations in the modeling.
Conveniently, the factor approach can greatly reduce the number of
latent variables and parameters to be estimated. Second, we show that
this factor structure arises naturally with block correlation matrices,
including equicorrelation matrices. A block correlation structure
may be motivated by the sector classification for companies, or some
other partitioning of assets into clusters. Not only is the block
correlation specification equivalent to a simple linear factor model,
the log-likelihood function is also readily available and easy to
evaluate, even for high dimensions. Third, we demonstrate the usefulness
of the framework in an empirical application with nine assets. We
find that the MRG model improves empirical fit, both in-sample and
out-of-sample, relative to the DCC, DECO, and Block-DECO models. The
predicted covariance matrices can be used for portfolio construction,
such as variance minimization. In an out-of-sample comparison, we
find that the empirical portfolio variance is reduced by a factor
of two relative to the equal-weighted portfolio. Fourth, we make an
interesting auxiliary empirical observation. We find the vector representation
of the realized correlation matrix is approximately Gaussian distributed.
This result is analogous to existing results for the logarithmically
transformed realized variances, see \citet{ABDE:2001} and \citet{ABDL:2001}.
An auxiliary result of our analysis is a framework that makes it possible
to estimate the Block-DECO model by maximum likelihood for any number
of blocks.

Early multivariate GARCH models relied solely on daily returns to
update the conditional covariance matrix. The MRG model, introduced
in this paper, incorporates realized measures of volatilities and
correlations computed from high frequency data. Realized measures
are beneficial because they provide accurate signals for dynamic modeling
of conditional variances and correlations. These measures gained prominence
following their empirical applications in \citet{AndersenBollerslev:1998a}
and subsequent theoretical results by \citet{ABDL:2001}, \citet{BNS:2002},
\citet{ABDL:2003}, \citet{BNS:2004}, see also \citet{HansenLundeVolForecastingHandbook}
and references therein. Realized measures were initially used to evaluate
the performance of GARCH models, see \citet{AndersenBollerslev:1998a}.
A very natural progression was to incorporate realized measures into
GARCH models. This was explored in \citet{Engle2002b} who found that
adding the realized variance as an exogenous variable, leads to significant
improvements in the empirical fit. This development was followed by
more comprehensive models that specified dynamic processes for the
realized measures themselves, including the MEM by \citet{engle-gallo:06},
the HEAVY model by \citet{ShephardSheppard:2010}, and the Realized
GARCH model by \citet{HansenHuangShek:2012}. Multivariate extensions
of these models were proposed in \citet{NoureldinShephardSheppard:2012},
\citet{HansenLundeVoev:2014}, and \citet{GorgiHansenJanusKoopman:2019}.
Another way to incorporate realized measures in multivariate GARCH
models is explored in \citet{BauwensStortiViolante:2012}, who build
on the Conditional Autoregressive Wishart model of \citet{Golosnoy_Gribisch_Liesenfeld_2012}.

Our approach to modeling correlations could, with some modifications,
be implemented using only daily returns. However, incorporating realized
measures into the modeling offers significant advantages. For example,
including realized measures makes a GARCH model more responsive to
sudden shifts in volatility, leading to substantial improvements in
empirical fit and predictive performance, see \citet{HansenHuang:2016}.
The Realized GARCH framework facilitates the incorporation of realized
measures of volatility into modeling. The model proposed in this paper
represents the first multivariate generalization of the Realized GARCH
framework, facilitating the incorporation of realized measures of
correlations without imposing additional restrictions on the covariance
structure..

The new class of multivariate GARCH models is based on the vector
parametrization, $\gamma_{t}=\gamma(C_{t})$, where the mapping $\gamma=\gamma(C)$
is defined by stacking the $d=n(n-1)/2$ below-diagonal elements of
$\log C$ (the matrix logarithm of $C$) into the vector $\gamma$,
see \citet{ArchakovHansen:Correlation}. The mapping $C\mapsto\gamma(C)$
is one-to-one between the set of non-singular correlation matrices
and $\mathbb{R}^{d}$. So, any vector $\gamma\in\mathbb{R}^{d}$ will
map to a unique positive definite correlation matrix, $C(\gamma)$,
without the need for additional restrictions. However, it is easy
to impose additional structure on $\gamma$ (i.e. structure on $C(\gamma)$)
as we will demonstrate with a factor model. This structure makes it
possible to estimate the model with a large number of assets. 

We are not first to use the matrix logarithm in this context. For
instance, \citet{ChiuLeonardTsui:1996}, \citet{Kawakatsu:2006},
and \citet{AsaiSo:2015} applied the matrix logarithm to covariance
matrices. The transformation has also been used in stochastic volatility
models, see \citet{IshiharaOmoriAsai:2016}, and in reduced-form models
of realized covariance matrices, see e.g. \citet{BauerVorkink:2011}
and \citet{Weigand:2014}.\footnote{Additional related literature includes the work by \citet{Liu_2009},
\citet{ChiriacVoev:2011}, \citet{Golosnoy_Gribisch_Liesenfeld_2012},
and \citet{BauwensStortiViolante:2012}.} The logarithmic transformation of conditional covariance matrices
is also related to the dynamic eigenvalue model by \citet{HetlandPedersenRahbek:2023}.
Our approach, which applies the matrix logarithm to the correlation
matrix, allows us to model each of the conditional variances with
univariate GARCH models, which has additional benefits. For instance,
it enables us to explicitly model the empirically important leverage
effect. \citet{HafnerWang:2023} have recently proposed a dynamic
model of correlations that also uses the parametrization by \citet{ArchakovHansen:Correlation}.
Their model uses the score-driven framework by \citet{CrealKoopmanLucas:2013},
whereas we build on the Realized GARCH framework and develop a parsimonious
factor model for the correlation structure.

We proceed as follows. In Section \ref{sec:MultivariateModel}, we
introduce notation, the modeling framework, and discuss how the factor
structure can be imposed on the correlation matrix. Section \ref{sec:Estimation}
details the estimation of the model and how the model can be used
for forecasting. An extensive empirical analysis with nine asset returns
series from three economic sectors is presented in Section \ref{sec:Empirical-Analysis}.
We compare the new model with existing models using out-of-sample
criteria in Section \ref{sec:Out-of-sample}, we conclude in Section
\ref{sec:Conclusion-and-Outline}. We derive analytical expressions
for the derivatives of the log-likelihood estimation in Appendix A.
These greatly speed up the estimation of the model. Appendix B has
step-by-step directions for maximum likelihood estimation of the model.

\section{The Multivariate Realized GARCH Model\label{sec:MultivariateModel}}

In this section, we present the details of the model. The MRG is based
on the vector parametrization of the correlation matrix,
\[
\gamma=\gamma(C)=\mathrm{vecl}(\log C),
\]
where $\log C$ represents the logarithmically transformed correlation
matrix\footnote{For a nonsingular correlation matrix, we have $\log C=Q\log\Lambda Q^{\prime}$,
where $C=Q\Lambda Q^{\prime}$ is the spectral decomposition of $C$,
so that $\Lambda$ is a diagonal matrix with the eigenvalues of $C$.} and $\mathrm{vecl}(\cdot)$ extracts and vectorizes the elements
below the diagonal. To illustrate this parametrization, consider the
following example,\begin{footnotesize}
\[
g\left(\left[\begin{array}{ccc}
1.0 & \bullet & \bullet\\
0.8 & 1.0 & \bullet\\
0.0 & 0.2 & 1.0
\end{array}\right]\right)=\mathrm{vecl}\log\left[\begin{array}{ccc}
1.0 & \bullet & \bullet\\
0.8 & 1.0 & \bullet\\
0.0 & 0.2 & 1.0
\end{array}\right]=\mathrm{vecl}\left[\begin{array}{ccc}
-0.53 & \bullet & \bullet\\
1.14 & -0.57 & \bullet\\
-0.13 & 0.28 & -0.03
\end{array}\right]=\left(\begin{array}{c}
1.14\\
-0.13\\
0.28
\end{array}\right).
\]
\end{footnotesize}

In the bivariate case, $n=2$, we have $\gamma(C)=\frac{1}{2}\log\frac{1-\rho}{1+\rho}$,
which is the Fisher transformed correlation. This parametrization
was used in \citet{HansenLundeVoev:2014} and the model we propose
here can therefore be viewed as a natural generalization of the bivariate
structure in \citet{HansenLundeVoev:2014}.\footnote{\citet{HansenLundeVoev:2014} accommodated the case $n>2$ by fusing
bivariate models to a larger system, which induces a restricted structure
on $C_{t}$. }

Our theoretical results also have useful applications for existing
models. For instance, they make it straightforward to estimate the
Block-DECO model by \citet{EngleKelly:2012} by maximum likelihood.
In \citet{EngleKelly:2012} the correlations within each block were
obtained by averaging over estimated correlations, which were based
on an auxiliary DCC model for the full dimension. They derived an
expression for the log-likelihood for the case with $K=2$, but not
for $K>2$. Instead they proposed to use composite likelihood methods
when $K>2$. The canonical representation of block matrices by \citet{ArchakovHansen:CanonicalBlockMatrix}
makes it straightforward to evaluate the log-likelihood function for
any $K$. The problem is effectively simplified to involve a single,
low-dimensional $K\times K$ matrix. Another advantage of the parametrization,
$\gamma(C)$, is that it is very simple to specify a low-dimensional
factor model that is equivalent to a block structure in $C$. This
avoids having to take averages of elements from an auxiliary and high-dimensional
DCC model. 

\subsection{Notation and Preliminaries}

We let $r_{t}=(r_{1,t},\ldots,r_{n,t})^{\prime}$ denote a $n$-dimensional
vector of returns in period $t$, where $t$ represents a generic
unit of time, such as a trading day. The conditional mean is denoted
by $\mu_{t}=\mathbb{E}(r_{t}|\mathcal{F}_{t-1})$ and the conditional
variance by $H_{t}=\mathrm{var}(r_{t}|\mathcal{F}_{t-1})$, where
$\{\mathcal{F}_{t}\}$ is the natural filtration for $(r_{t},\mathrm{RM}_{t})$.
Here $\mathrm{RM}_{t}$ denotes an ex-post empirical measure of $H_{t}$,
such as the realized covariance matrix, see \citet{BNS:2004}, or
the multivariate realized kernel by \citet{BNHLS-MRK:2011}.

We decompose the conditional covariance matrix into variances and
correlations,
\begin{equation}
H_{t}=\Lambda_{h_{t}}^{1/2}C_{t}\Lambda_{h_{t}}^{1/2},\label{eq:DCCstructure}
\end{equation}
where $\Lambda_{h_{t}}=\mathrm{diag}(h_{1,t},\ldots,h_{n,t})$ with
$h_{i,t}=[H_{t}]_{ii}$, $i=1,\ldots,n$. So, $h_{i,t}$ is the conditional
variance of $r_{i,t}$, $i=1,\ldots,n$, and $C_{t}=\mathrm{corr}(r_{t}|\mathcal{F}_{t-1})$
is the conditional correlation matrix of $r_{t}$. The DCC structure
in (\ref{eq:DCCstructure}) enables us to disentangle the dynamic
properties of the conditional variances from that of the conditional
correlations. We will deviate from the DCC framework in the way we
parametrize $C_{t}$ and by incorporating realized measures of variances
and correlations into the model.

The central component of a GARCH model is the equation that specifies
the dynamic properties of $H_{t}$ and how these are influenced by
lagged returns. This equation can be enhanced to include realized
measures of volatility. The Realized GARCH model is characterized
by measurement equations that specify how the realized measures are
related to the contemporaneous conditional moments, i.e. the elements
of $H_{t}$. 

Here we will model the conditional variances and the conditional correlations
separately. This leads to two sets of GARCH and measurement equations,
that utilize the appropriate realized measures. From $\mathrm{RM}_{t}\in\mathbb{R}^{n\times n}$,
a positive definite realized measure of the covariance matrix in period
$t$, we extract the diagonal elements $x_{t}=\mathrm{diag}(\mathrm{RM}_{t})$
and the corresponding correlation matrix denoted by
\[
Y_{t}=\Lambda_{x_{t}}^{-1/2}\mathrm{RM}_{t}\Lambda_{x_{t}}^{-1/2}.
\]
Here $\Lambda_{x_{t}}=\mathrm{diag}(x_{1,t},\ldots,x_{n,t})$ denotes
the diagonal matrix with the elements of $x_{t}$ on the diagonal,
and it follows that $Y_{t}$ inherits the positive definiteness from
$\mathrm{RM}_{t}$, such that $y_{t}=\gamma(Y_{t})\in\mathbb{R}^{d}$
is well-defined. In summary, $x_{t}$ and $y_{t}$ are the observed
empirical measures of the latent variables, $h_{t}$ and $\gamma_{t}$,
respectively, such that time variation in $x_{t}$ and $y_{t}$ contains
information about time variation in the corresponding latent variables.
The realized measure, $\mathrm{RM}_{t}$, is typically a consistent
estimator of the quadratic variation. But the (ex-post) quadratic
variation is not identical to the (ex-ante) conditional variance,
$H_{t}$. We should therefore expect a non-trivial measurement errors
in $x_{t}$ and $y_{t}$, because neither are perfect measurements
of $h_{t}$ and $\gamma_{t}=\gamma(C_{t})$, respectively.

We let $I_{n}$ denote the $n\times n$ identity matrix and $1_{\{\cdot\}}$
the indicator function, which equals one if the expression within
the curly brackets is true and zero otherwise. With the required notation
in place, we are now ready to introduce the multivariate realized
GARCH (MRG) model.

\subsection{Models of Marginal Distributions\label{sec:Multivariate-Realized-GARCH}}

We specify univariate Realized GARCH models for each return series
and the corresponding realized measures. 

The return equation for each of the returns series takes the form
\begin{equation}
r_{i,t}=\mu_{i}+h_{i,t}^{1/2}z_{i,t},\qquad i=1,\ldots,n,\quad t=1,\ldots,T.\label{eq:return}
\end{equation}
We have here assumed that $\mathbb{E}(r_{i,t}|\mathcal{F}_{t-1})=\mu_{i}$
is constant, as is often done in GARCH models, and it follows that
the standardized return, $z_{i,t}=h_{i,t}^{-1/2}(r_{i,t}-\mu_{i})$,
is such that $\mathbb{E}(z_{i,t}|\mathcal{F}_{t-1})=0$ and $\mathrm{var}(z_{i,t}|\mathcal{F}_{t-1})=1$.
Note that the standardized returns, $z_{i,t}$ and $z_{j,t}$ for
$i\neq j$, are not assumed to be uncorrelated (nor are they likely
to be).

The corresponding GARCH and measurement equations are given by
\begin{eqnarray}
\log h_{i,t} & = & \omega_{i}+\beta_{i}\log h_{i,t-1}+\tau_{i}(z_{i,t-1})+\alpha_{i}\log x_{i,t-1},\label{eq:GARCH}\\
\log x_{i,t} & = & \xi_{i}+\varphi_{i}\log h_{i,t}+\delta_{i}(z_{t})+v_{i,t},\label{eq:Measurement}
\end{eqnarray}
for $i=1,\ldots,n$ and $t=1,\ldots,T$, where $\omega_{i},\beta_{i},\alpha,\xi_{i},\varphi_{i}\in\mathbb{R}$.
The measurement errors, $v_{t}=(v_{1,t},\ldots,v_{n,t})^{\prime}$,
may be dependent and correlated with the corresponding measurement
errors in transformed realized correlations (defined below). The two
functions, $\tau_{i}(z)=\tau_{i,1}z+\tau_{i,2}(z^{2}-1)$ and $\delta_{i}(z)=\delta_{i,1}z+\delta_{i,2}(z^{2}-1)$,
are leverage functions that capture dependencies between returns and
volatility innovations. This dependency is known to be empirically
important, and it is typically labelled the \emph{leverage effect},
see \citet{Black:1976}, \citet{Christie:1982}, \citet{Engle_Ng_1993}.\footnote{The leverage effect is sometimes used to refer to a linear dependence,
i.e. the (usually negative) correlation between returns and changes
in return volatility.} This quadratic form is motivated by results in \citet{HansenHuangShek:2012}
and \citet{HansenHuang:2016} who found that a second-order Hermite
polynomial suffices for capturing the asymmetry dependence between
return shocks and volatility shocks. The three equations, (\ref{eq:return})-(\ref{eq:Measurement}),
define a univariate Realized GARCH model for each asset, $i=1,\ldots,n$,
where the characteristic feature of a Realized GARCH model is the
measurement equation that relates each realized measure to the corresponding
conditional moment in the model.

\subsection{Full Model for Conditional Correlations\label{subsec:Full-Model-for}}

The key novel innovation of the Multivariate Realized GARCH model
is the way we model the dynamic conditional correlation matrix,
\[
C_{t}=\mathrm{var}(z_{t}|\mathcal{F}_{t-1}).
\]
We simply model the elements of $\gamma_{t}=\gamma(C_{t})\in\mathbb{R}^{d}$,
or linear combinations thereof, in the same way we model $h_{i,t}$,
$i=1,\ldots,n$. In its most general specification (Full) we specify
a GARCH equation and a measurement equation for each element, $j=1,\ldots,d$,
\begin{eqnarray}
\gamma_{j,t} & = & \tilde{\omega}_{j}+\tilde{\beta}_{j}\gamma_{j,t-1}+\tilde{\alpha}_{j}y_{j,t-1},\label{eq:GARCHfull}\\
y_{j,t} & = & \tilde{\xi}_{j}+\tilde{\varphi}_{j}\gamma_{j,t}\quad\qquad+\tilde{v}_{j,t}.\label{eq:MeasurementFull}
\end{eqnarray}
Here $y_{t}=\gamma(Y_{t})$ is the transformed realized correlation
matrix, such that $y_{j,t}$ is the appropriate empirical measurement
of $\gamma_{j,t}$, $j=1,\ldots,d$. The measurement error in the
transformed realized correlations, $\tilde{v}_{t}=(\tilde{v}_{1,t},\ldots,\tilde{v}_{d,t})^{\prime}$,
may have dependent elements and may be correlated with the measurement
errors in, $v_{t}$, in (\ref{eq:Measurement}).

\subsection{A Factor Model for the Correlation Structure\label{subsec:DimReduc}}

A drawback of modeling all elements of $\gamma_{t}$ is that the number
of latent variables in $C_{t}$, $d=n(n-1)/2$, becomes unmanageable
unless $n$ is small. While it is possible to estimate the model with
$n=9$ ($d=36$), but (at the time of writing this) it is difficult
to estimate the full model with dimensions much larger than that.
This necessitates additional structure on the model when $n$ is large.
Fortunately, it is simple to impose a factor structure, $\gamma_{t}=\varrho(\zeta_{t})$,
where $\zeta_{t}\in\mathbb{R}^{r}$ is a lower dimensional vector
of factors. The underlying assumption is that the variation in $C_{t}$
is driven by $r<d$ factors. 

A natural starting point is the linear factor model,
\[
\gamma_{t}=A\zeta_{t},
\]
where $A$ is a $d\times r$ matrix. This enables us to reduce the
number of GARCH equations and measurement equations from $d$ to $r$,
by replacing (\ref{eq:GARCHfull}) and (\ref{eq:MeasurementFull})
with 
\begin{eqnarray}
\zeta_{j,t} & = & \check{\omega}_{j}+\check{\beta}_{j}\zeta_{,jt-1}+\check{\alpha}_{j}\check{y}_{j,t-1},\qquad j=1,\ldots,r\label{eq:GARCHfactor}\\
\check{y}_{j,t} & = & \check{\xi}_{j}+\check{\varphi}_{j}\zeta_{j,t}\quad+\check{v}_{j,t},\qquad\qquad j=1,\ldots,r\label{eq:MeasurementFactor}
\end{eqnarray}
respectively, where $\check{y}_{t}=(A^{\prime}A)^{-1}A^{\prime}y_{t}\in\mathbb{R}^{r}$.
From well-known projection arguments, it follows that $\check{y}_{t}$
is the realized quantity that corresponds to $\zeta_{t}$.\footnote{If $A$ has full column rank, $r$, then there exists an $d\times(d-r)$
matrix, $A_{\bot}$ so that $(A,A_{\bot})$ is a full rank matrix,
and $A^{\prime}A_{\bot}=0$. Thus, $\gamma_{t}=A\zeta_{t}$ implies
$A_{\bot}^{\prime}\gamma_{t}=0$ and the identity $I_{d}=A(A^{\prime}A)^{-1}A^{\prime}+A_{\bot}(A_{\bot}^{\prime}A_{\bot})^{-1}A_{\bot}^{\prime}$
shows that $\gamma_{t}=A\zeta_{t}\Rightarrow\gamma_{t}=A(A^{\prime}A)^{-1}A^{\prime}\gamma_{t}$.
The vector of transformed realized correlations, $y_{t}$, is our
empirical ``signal'' about $\gamma_{t}$, and the identity, 
\[
y_{t}=[A(A^{\prime}A)^{-1}A^{\prime}+A_{\bot}(A_{\bot}^{\prime}A_{\bot})^{-1}A_{\bot}^{\prime}]y_{t}=A\check{y}_{t}+A_{\bot}(A_{\bot}^{\prime}A_{\bot})^{-1}A_{\bot}^{\prime}y_{t},
\]
 shows that $\check{y}_{t}$ is the appropriate signal about $\zeta_{t}$,
whenever $\gamma_{t}=A\zeta_{t}$.} The measurement errors, $\check{v}_{t}=(\check{v}_{1,t},\ldots,\check{v}_{r,t})^{\prime}$,
in (\ref{eq:MeasurementFactor}) may have dependent elements and be
correlated with the measurement errors, $v_{t}$, in (\ref{eq:Measurement}).
In the empirical analysis, we define $u_{t}=(v_{t}^{\prime},\tilde{v}_{t}^{\prime})^{\prime}$
and adopt a Gaussian likelihood function with $u_{t}\sim iidN(0,\Sigma)$,
where $\Sigma$ is an arbitrary covariance matrix.

The matrix, $A$, is needed for this implementation and $A$ may be
known in advance or can be determined empirically. Estimating $A$,
including $r$, is an interesting problem that we leave for future
research.\footnote{We also leave other generalizations, such as non-linear factor structures,
$\gamma_{t}=\varrho(\zeta_{t})$, for future research.} In this paper, we focus on the case where $A$ is known, and we show
that any block correlation structure is equivalent to a linear factor
structure with a known $A$-matrix. 

Note that the linear factor model, $\gamma_{t}=A\zeta_{t}$, is characterized
by the subspace spanned by the columns of $A$, not the particular
choice for $A\in\mathbb{R}^{n\times r}$. This follows from the fact
that $\gamma_{t}=\tilde{A}\tilde{\zeta}_{t}$ and $\gamma_{t}=A\zeta_{t}$
are observationally equivalent whenever $\tilde{A}=A\Phi$ and $\Phi\in\mathbb{R}^{r\times r}$
is invertible. (The relation between the two latent factor variables
will be $\tilde{\zeta}_{t}=\Phi^{-1}\zeta_{t}$).

There are situations where the original measurement equation in (\ref{eq:Measurement})
should be used even when the factor structure is adopted in the GARCH
equation (\ref{eq:GARCHfactor}). For instance, if multiple model
specifications are compared in terms of their total log-likelihoods,
then all models should have measurement equations for the same realized
variables to avoid an apples-to-oranges comparison. Models with different
measurement equations may be compared in terms of their partial log-likelihood
for returns, which is often the primary objective in multivariate
GARCH models. The relevant terms of the log-likelihood for these comparisons
are detailed in the next section. 

\subsection{A Special Case: Dynamic Block Correlation \label{subsec:Dynamic-Block-Correlation}}

Let $n=n_{1}+\cdots+n_{K}$, where $n_{1},n_{2},\ldots,n_{K}\geq1$.
A \emph{block correlation matrix} is characterized by the structure,
\[
C=\left[\begin{array}{cccc}
C_{[1,1]} & C_{[1,2]} & \cdots & C_{[1,K]}\\
C_{[2,1]} & C_{[2,2]} &  & C_{[2,K]}\\
\vdots &  & \ddots & \vdots\\
C_{[K,1]} & C_{[K,2]} & \cdots & C_{[K,K]}
\end{array}\right],
\]
where all elements within each $n_{i}\times n_{j}$ matrix, $C_{[i,j]}$,
are identical and equal to $\rho_{ij}$, except for the the diagonal-block,
$C_{[i,i]}$, that have ones along the diagonal and other elements
equal to $\rho_{ii}$. 

Regardless of the number of blocks and their dimensions, $\log C_{t}$
has the same block structure as $C_{t}$, see \citet[corollary 1]{ArchakovHansen:CanonicalBlockMatrix}.\footnote{This result holds for all block matrices, including non-symmetric
block matrices, see \citet{ArchakovHansen:CanonicalBlockMatrix}.} This is illustrated in the following example: \begin{small}
\begin{equation}
\underset{=C}{\underbrace{\left(\begin{array}{cccccc}
1.0 & {\color{teal}0.4} & {\color{teal}0.4} & {\color{purple}0.2} & {\color{purple}0.2} & {\color{purple}0.2}\\
{\color{teal}0.4} & 1.0 & {\color{teal}0.4} & {\color{purple}0.2} & {\color{purple}0.2} & {\color{purple}0.2}\\
{\color{teal}0.4} & {\color{teal}0.4} & 1.0 & {\color{purple}0.2} & {\color{purple}0.2} & {\color{purple}0.2}\\
{\color{purple}0.2} & {\color{purple}0.2} & {\color{purple}0.2} & 1.0 & {\color{blue}0.6} & {\color{blue}0.6}\\
{\color{purple}0.2} & {\color{purple}0.2} & {\color{purple}0.2} & {\color{blue}0.6} & 1.0 & {\color{blue}0.6}\\
{\color{purple}0.2} & {\color{purple}0.2} & {\color{purple}0.2} & {\color{blue}0.6} & {\color{blue}0.6} & 1.0
\end{array}\right)}}\qquad\qquad\underset{\simeq\log C}{\underbrace{\left(\begin{array}{cccccc}
-.16 & {\color{teal}.349} & {\color{teal}.349} & {\color{purple}.104} & {\color{purple}.104} & {\color{purple}.104}\\
{\color{teal}.349} & -.16 & {\color{teal}.349} & {\color{purple}.104} & {\color{purple}.104} & {\color{purple}.104}\\
{\color{teal}.349} & {\color{teal}.349} & -.16 & {\color{purple}.104} & {\color{purple}.104} & {\color{purple}.104}\\
{\color{purple}.104} & {\color{purple}.104} & {\color{purple}.104} & -.36 & {\color{blue}.553} & {\color{blue}.553}\\
{\color{purple}.104} & {\color{purple}.104} & {\color{purple}.104} & {\color{blue}.553} & -.36 & {\color{blue}.553}\\
{\color{purple}.104} & {\color{purple}.104} & {\color{purple}.104} & {\color{blue}.553} & {\color{blue}.553} & -.36
\end{array}\right)}}.\label{eq:BlockExample}
\end{equation}
\end{small}

A block structure arises naturally in applications where the correlation
between two variables is defined by their group classification. An
$n\times n$ correlation matrix has $n(n-1)/2$ correlations, whereas
a block correlation matrix with $K\times K$ blocks has, at most,
$K(K-1)/2+K$ distinct correlations.\footnote{The exact number of distinct correlation is $r=K(K-1)/2+\tilde{K}$,
where $\tilde{K}$ is the number of blocks that contain two or more
elements. The reason for the distinction between $K$ and $\tilde{K}$
is that an $1\times1$ diagonal block does not have a correlation
coefficient.} The fact that $\log C$ inherits the block structure of $C$ facilitates
a parsimonious modeling of dynamic block correlation matrices. The
elements of the transformed matrix can be modeled in an unrestricted
way, without compromising the structure a correlation matrix must
have. This completely bypasses the non-linear cross restrictions on
$C$'s elements ensure positive definiteness. So, a dynamic model
of the (below diagonal) elements of $\log C$ is a very convenient
implementation of the block structure on $C$. 
\begin{example}
Consider the following example with $n=5$ and $K=2$,\begin{footnotesize}
\[
C_{t}=\left(\begin{array}{ccccc}
1\\
a_{t} & 1\\
b_{t} & b_{t} & 1\\
b_{t} & b_{t} & c_{t} & 1\\
b_{t} & b_{t} & c_{t} & c_{t} & 1
\end{array}\right)\qquad\qquad\text{\ensuremath{\log C_{t}}=\ensuremath{\left(\begin{array}{ccccc}
\ast\\
\tilde{a}_{t} & \ast\\
\tilde{b}_{t} & \tilde{b}_{t} & \ast\\
\tilde{b}_{t} & \tilde{b}_{t} & \tilde{c}_{t} & \ast\\
\tilde{b}_{t} & \tilde{b}_{t} & \tilde{c}_{t} & \tilde{c}_{t} & \ast\ 
\end{array}\right)}},
\]
\end{footnotesize}where $a_{t}$, $b_{t}$, and $c_{t}$ denote the
three distinct correlations in $C_{t}$. Since, $\gamma_{t}=(\tilde{a}_{t},\tilde{b}_{t},\tilde{b}_{t},\tilde{b}_{t},\tilde{b}_{t},\tilde{b}_{t},\tilde{b}_{t},\tilde{c}_{t},\tilde{c}_{t},\tilde{c}_{t})^{\prime}$,
we have the simple linear factor structure
\[
\gamma_{t}=A\zeta_{t},\quad\text{where}\quad A^{\prime}=\left(\begin{array}{cccccccccc}
1 & 0 & 0 & 0 & 0 & 0 & 0 & 0 & 0 & 0\\
0 & 1 & 1 & 1 & 1 & 1 & 1 & 0 & 0 & 0\\
0 & 0 & 0 & 0 & 0 & 0 & 0 & 1 & 1 & 1
\end{array}\right),\quad\text{and}\quad\zeta_{t}=\left(\begin{array}{c}
\tilde{a}_{t}\\
\tilde{b}_{t}\\
\tilde{c}_{t}
\end{array}\right).
\]
In this example the dimension is reduced from $d=10$ to $r=3.$ 
\end{example}
The block structure offers a useful dimension reduction, but in order
to make use of it, one has to specify the clusters that define the
block structure. In our empirical analysis, we consider a case with
9 assets, and a correlation structure with $K=1$, $K=3$, and $K=9$.
The sector-based clusters, $K=3$, reduces the number of latent correlation
variables from $d=36$ to $r=6$. 

A correlation matrix with a block structure leads to a very scalable
model, because increasing $n$ does not increase the complexity of
the second stage estimation. The correlation factors, $\zeta_{t}\in\mathbb{R}^{r}$,
can be used to model an arbitrarily large number of assets, so long
as all assets can be classified within the $K$ clusters. 

The block correlation structure also simplifies many computational
aspects of the model. For instance, the inverse correlation matrix
is readily available for any $K$, see \citet{ArchakovHansen:CanonicalBlockMatrix}.
Previously, a closed-form expressions for $C^{-1}$ was only available
for $K=2$, see \citet[lemma 2.3]{EngleKelly:2012}. For a block correlation
matrix, $C$, with $K\times K$ blocks, we define the (symmetric)
$K\times K$ matrix, $B$, whose elements are given by $b_{ii}=1+(n_{i}-1)\rho_{ii}$
and $b_{ij}=\rho_{ij}\sqrt{n_{i}n_{j}}$, for $i\neq j$. It is simple
to verify that $(i,j)$-th block of $C$ can be expressed as
\[
C_{[i,j]}=b_{ij}P_{[i,j]}+1_{\{i=j\}}(1-\rho_{ii})(I_{n_{i}}-P_{[i,i]}),\qquad\text{for}\quad i,j=1,\ldots,K,
\]
where all elements of $P_{[i,j]}\in\mathbb{R}^{n_{i}\times n_{j}}$
equal $\frac{1}{\sqrt{n_{i}n_{j}}}$. The determinant of $C$ and
the $(i,j)$-th block of the inverse correlation matrix, $C^{-1}$,
can be expressed as
\begin{eqnarray}
\det C & = & (\det B)(1-\rho_{11})^{n_{1}-1}\cdots(1-\rho_{KK})^{n_{K}-1},\label{eq:determinantC}\\
C_{[i,j]}^{-1} & = & b_{ij}^{\#}P_{[i,j]}+1_{\{i=j\}}\tfrac{1}{1-\rho_{ii}}(I_{n_{i}}-P_{[i,i]}),\qquad\text{for}\quad i,j=1,\ldots,K,\label{eq:inverseC}
\end{eqnarray}
respectively, where $b_{ij}^{\#}$ is the $ij$-th element of $B^{-1}$,
see \citet[corollary 2]{ArchakovHansen:CanonicalBlockMatrix}. These
closed-form expressions for $C^{-1}$ and $\det C$ facilitate simple
evaluation of the Gaussian log-likelihood function when $C$ has a
block structure.

\section{Estimation\label{sec:Estimation}}

Estimating the parameters of the MRG model is relatively simple, because
the model is observation-driven, with variation in the latent variables,
$h_{t}$ and $\gamma_{t}$, being driven by observable variables,
$r_{t}$, $x_{t}$, and $y_{t}$.

We can factorize the joint density of $(r_{t},x_{t},y_{t})$, conditional
on past observations, into the marginal density for returns and the
density for realized variables, conditional on contemporaneous returns.
Thus, the joint density is expressed as the product, $f_{t-1}(r_{t},x_{t},y_{t})=f_{t-1}(r_{t})f_{t-1}(x_{t},y_{t}|r_{t})$.
The log-likelihood function can therefore be deduced from
\begin{equation}
\sum_{t=1}^{T}\log f_{t-1}(r_{t},x_{t},y_{t})=\sum_{t=1}^{T}\log f_{t-1}(r_{t})+\sum_{t=1}^{T}\log f_{t-1}(x_{t},y_{t}|r_{t}),\label{eq:likelihood}
\end{equation}
and parameters may be estimated by quasi maximum likelihood estimation,
by specifying Gaussian likelihood functions for $z_{t}$ and $u_{t}$.
More specifically, under the assumption that $C_{t}^{-1/2}z_{t}\sim iidN(0,I_{n})$
and $u_{t}\sim iidN(0,\Sigma)$ are mutually independent. The leverage
functions, $\tau_{i}(\cdot)$ and $\delta_{i}(\cdot)$, serve to eliminate
certain forms of dependence between $v_{i,t}$ and $z_{i,t}$, making
the assumed independence somewhat more realistic. 

Let $\theta=(\theta_{1}^{\prime},\theta_{2}^{\prime})^{\prime}$ represent
all unknown parameters in the model, where $\theta_{1}$ includes
the parameters in the multivariate GARCH-X model for the returns and
$\theta_{2}$ represent the parameters in the measurement equations,
which define the model for the realized measures. The Gaussian specification
and (\ref{eq:likelihood}) imply that the log-likelihood function
is given by
\[
\ell(\theta)=\sum_{t=1}^{T}\ell_{r,t}(\theta_{1})+\sum_{t=1}^{T}\ell_{x,y|r,t}(\theta_{2}),
\]
where
\begin{eqnarray}
-2\ell_{r,t}(\theta_{1}) & = & c_{n}+\sum_{k=1}^{n}\log h_{k,t}+\log\det C_{t}+z_{t}^{\prime}C_{t}^{-1}z_{t},\label{eq:logLr}\\
-2\ell_{x,y|r,t}(\theta_{2}) & = & c_{\tfrac{n(n+1)}{2}}+\log\det\Sigma+u_{t}^{\prime}\Sigma^{-1}u_{t},\nonumber 
\end{eqnarray}
with $c_{n}=n\log2\pi$ and
\begin{eqnarray*}
z_{t} & = & \Lambda_{h_{t}}^{-1}(r_{t}-\mu),\quad\text{and}\\
u_{t} & = & \left(\begin{array}{c}
v_{t}\\
\tilde{v}_{t}
\end{array}\right)=\left(\begin{array}{l}
\log x_{t}-\xi-\Phi\log h_{t}-\delta(z_{t})\\
y_{t}-\tilde{\xi}-\tilde{\Phi}\gamma_{t}
\end{array}\right).
\end{eqnarray*}
Here we have used a condensed notation, where $\xi$ is a vector with
elements, $\xi_{i}$, $i=1,\ldots,n$, and similar for $\delta$ and
$\tilde{\xi}$, and $\Phi$ and $\tilde{\Phi}$ are diagonal matrices
with diagonal elements $(\varphi_{1},\ldots,\varphi_{n})$ and $(\tilde{\varphi}_{1},\ldots,\tilde{\varphi}_{d})$,
respectively.

For a particular value of $\theta$ it is straightforward to evaluate
the log-likelihood function. The latent variables, $\{h_{t},C_{t}\}$,
can be computed recursively. Given $h_{t-1}$ and $C_{t-1}$ and the
observable $(r_{t},x_{t},y_{t})$, we compute $h_{t}$ and $C_{t}$
with the GARCH equations and infer $z_{t}$ from the return equation,
(\ref{eq:return}), and $u_{t}$ from the measurement equations. This
is repeated for period $t+1$, and so forth, and the likelihood function
can be evaluated for the full sample. The starting values for the
latent variables, $h_{1}$ and $C_{1}$, can be treated as unknown
parameters (as part of $\theta$), which we recommend. Alternatively,
$h_{1}$ and $C_{1}$, can be assigned particular values, which may
be based on appropriate empirical quantities.

\subsection{Maximum Likelihood Estimation}

The structure of the log-likelihood function allows the maximization
problem to be simplified. Given the residuals, $\hat{u}_{t}$, $t=1,\ldots,T$,
it can be shown that the maximum likelihood estimator of $\Sigma$
is $\hat{\Sigma}=T^{-1}\sum_{t=1}^{T}\hat{u}_{t}\hat{u}_{t}^{\prime}$,
such that
\[
\sum_{t=1}^{T}\hat{u}_{t}^{\prime}\hat{\Sigma}^{-1}\hat{u}_{t}=\mathrm{tr}\{\hat{\Sigma}^{-1}\sum_{t=1}^{T}\hat{u}_{t}\hat{u}_{t}^{\prime}\}=\mathrm{tr}\{TI_{n+k}\}=T(n+k),
\]
where $I_{n+k}$ is the $(n+k)\times(n+k)$ identity matrix. So, the
objective to be maximized is (apart from a constant) given by
\begin{equation}
-\tfrac{1}{2}\sum_{t=1}^{T}\left\{ \sum_{i=1}^{n}\log h_{i,t}+\log\det C_{t}+z_{t}C_{t}^{-1}z_{t}\right\} -\tfrac{T}{2}\log\det\left(\tfrac{1}{T}\sum_{t=1}^{T}u_{t}u_{t}^{\prime}\right),\label{eq:logL2Maximize}
\end{equation}
where the omitted constant is $-\tfrac{T}{2}(c_{n}+c_{d}+n+k)=-\tfrac{n(n+1)}{4}T\log2\pi-T\tfrac{n+k}{2}$.
As stated earlier, both the determinant, $\det C_{t}$, and the inverse,
$C_{t}^{-1}$, are simple to evaluate when $C_{t}$ has a block structure
using (\ref{eq:determinantC}) and (\ref{eq:inverseC}). More details
about the maximum likelihood estimation can be found in Appendix \ref{sec:app_estimation}.

\subsection{Two-Stage Estimation\label{subsec:Two-Stage-Estimation}}

Joint estimation is possible if $n$ is small (say $n<10$), but it
tends to be slow. So, it will often be convenient to adopt a two-stage
estimation method. In the first stage, we estimate the univariate
Realized GARCH models for each of the $n$ return series. From the
estimated models we obtain $h_{i,t}$ and $z_{i,t}=h_{i,t}^{-1/2}(r_{i,t}-\mu_{i})$,
$i=1,\ldots,n$ and $t=1,\ldots,T$, and in the second stage, we estimate
the parameters that relate to the dynamic conditional correlation
matrix, $C_{t}=\mathrm{var}(z_{t})$. 

So, in the first stage, we maximize 
\[
-\frac{1}{2}\sum_{t=1}^{T}\left\{ \log h_{i,t}+\log\hat{\sigma}_{v_{i}}^{2}+z_{i,t}^{2}\right\} ,
\]
with respect to $\vartheta_{1,i}=(\omega_{i},\beta_{i},\tau_{i,1},\tau_{i,2},\alpha_{i},\xi_{i},\varphi_{i},\delta_{i,1},\delta_{i,2})^{\prime},$
for each $i=1,\ldots,n$, where $\hat{\sigma}_{v_{i}}^{2}=\tfrac{1}{T}\sum_{t=1}^{T}v_{i,t}^{2}$,
and in the second stage, we maximize
\begin{equation}
-\frac{1}{2}\sum_{t=1}^{T}\left\{ \log\det C_{t}+z_{t}C_{t}^{-1}z_{t}\right\} -\tfrac{T}{2}\log\det\left(\tfrac{1}{T}\sum_{t=1}^{T}\tilde{v}_{t}\tilde{v}_{t}^{\prime}\right),\label{eq:logLcorrelationmatrix}
\end{equation}
with respect $\vartheta_{2}=(\vartheta_{2,1}^{\prime},\ldots,\vartheta_{2,r}^{\prime})^{\prime}$,
where $\vartheta_{2,j}=(\check{\omega_{j}},\check{\beta}_{j},\check{\alpha}_{j},\check{\xi}_{j},\check{\varphi}_{j})^{\prime}$,
$j=1,\ldots,r$, are the parameters in (\ref{eq:GARCHfactor}) and
(\ref{eq:MeasurementFactor}), with the vector of transformed realized
correlation variables given by $\check{y}_{t}=(A^{\prime}A)^{-1}A^{\prime}y_{t}$
and $u_{t}=(v_{t}^{\prime},\check{v}_{t}^{\prime})^{\prime}$. Note
that the two-stage estimation involves a different partitioning of
the parameters than the one used in (\ref{eq:logLr}). The parameters
in the first-stage, $(\vartheta_{1,1},\ldots,\vartheta_{1,n})$, include
elements from both $\theta_{1}$ and $\theta_{2}$, and the same is
the case for the second-stage parameters, $\vartheta_{2}$.

In the special case, without a factor structure imposed on $C_{t}$,
(Full), we simply have $r=d$, and the parameters are identical to
those in (\ref{eq:GARCHfull}) and (\ref{eq:MeasurementFull}).

In our empirical analysis, we adopt this two-stage estimation method.
All model-specifications use the same first-stage, such that the model
comparisons concern their ability to capture the dynamic correlation
structure without confounding these with aspects that relate to the
marginal distributions.\footnote{In an earlier version of this paper, we estimated the model by maximum
likelihood (using a shorter sample period). The parameter estimates
were very similar, but the estimation was much slower, especially
for the most flexible specification (Full), which has 36 latent variables
to model $C_{t}$.}

For the estimation problem, we have derived analytic expressions for
the gradient vector and the Fisher information matrix. This are presented
in the Appendix. The analytical expressions greatly reduce the time
it takes to estimate the model, as illustrated in Table \ref{tab:Typical-estimation-times}.
The Table presents second-stage estimation times for the computationally
most demanding specification, with an unrestricted correlation matrix.
The model is estimated with $T=4,744$ and $n$ ranging from $2$
to $9$. The variables are subsets of the assets using in our empirical
analysis).{\footnotesize{}
\begin{table}[H]
\begin{centering}
\vspace{0.2cm}
\begin{small}
\begin{tabularx}{\textwidth}{ >{\hsize=2.6\hsize}X >{\hsize=0.8\hsize}C >{\hsize=0.8\hsize}C >{\hsize=0.8\hsize}C >{\hsize=0.8\hsize}C >{\hsize=0.8\hsize}C >{\hsize=0.8\hsize}C >{\hsize=0.8\hsize}C >{\hsize=0.8\hsize}C }
\toprule
\multicolumn{9}{c}{Estimation time for MRG with dimension $n$} \\
\midrule
\\[-0.2cm]
Dimension:                               & $n=2$    & $n=3$       & $n=4$       & $n=5$       & $n=6$         & $n=7$       & $n=8$       & $n=9$       \\[2mm]
Number of parameters:                   & 5        & 15      & 30      & 50      & 75        & 105     & 140     & 180     \\[2mm]
\multicolumn{9}{l}{Derivatives based on...} \\[1mm]         
...numerical methods:                   & --       & 78s     & --      & --      & >30m      & --      & --      & >4h     \\[2mm]
...analytical expressions:                  & 4.6s     & 16.7s   & 22.1s   & 37.7s   & 55.2s     & 149s    & 267s    & 403s    \\
\\[-0.2cm]
\midrule
\bottomrule
\end{tabularx}
\end{small}
\par\end{centering}
{\footnotesize\caption{\label{tab:Typical-estimation-times} \footnotesize  Estimation time
for MRG with an unrestricted correlation structure (Full model) with
$n=2,\ldots,9$, where s, m, and h refer to \emph{seconds}, \emph{minutes},
and \emph{hours}, respectively. The variables are subsets of the nine
assets in our empirical application for the same sample period (4,744
days). We report the second-stage estimation times for a desktop computer
(Intel Core i9-13900K, 32 MB RAM) with Matlab R2023b, using \textit{fminunc}
to maximize the log-likelihood function (the computation time for
the first-stage estimation is negligible).}
}{\footnotesize\par}
\end{table}
}{\footnotesize\par}

Without analytical derivatives (which was used in an earlier version
of this paper) the estimation of the full model is relatively slow.
Using numerical derivatives, it takes about 78 seconds to estimated
the model with $n=3$, more than 30 minutes to estimate the model
with $n=6$, and more than four hours to estimate the model with $n=9$.
In contrast, it only takes 7 minutes to estimate the model with $n=9$
when the analytical derivatives are used. \footnote{The model was estimated with Matlab's optimization function, \textit{fminunc},
and the computation of numerical derivatives was optimized to use
all (24) cores in parallel. Estimation with analytical expressions
uses the quasi-newton method with the analytical gradient in the first
25 iterations, after which the trust-region method is used with the
analytical gradient and the analytical approximation for the corresponding
Hessian.} 

The analytical derivatives, derived in the appendix will also be useful
for computing standard errors, testing for parameter stability, see
\citet{Nyblom89}, and other statistical applications.

\subsection{Forecasting}

One-step ahead forecasting of the return distributions from the model
is straightforward. All dynamic variables are specified with an observation-driven
structure, such that forecasts are given from known functions of lagged
variables. From the observed variables in period $t$, all the conditional
variances and correlations for period $t+1$ can be computed from
the GARCH equations. The elements of $H_{t+h}$, are not predetermined
beyond horizon $h=1$, because they also depend on future realizations
of $z_{t}$ and $u_{t}$. It is nevertheless straightforward to compute
a distributional forecasts for $H_{t+h}$ using simulation or bootstrap
methods. So, multi-step ahead forecasts can be inferred from the estimated
model for any forecasting horizon. Forecasting schemes for the Realized
GARCH models of this kind are detailed in \citet{LundeOlesenEnergyRealizedGARCH}
and \citet{HansenLundeVoev:2014}. In this context, a bootstrap method
will typically be preferred because it is more robust to distributional
misspecification. 

\section{Empirical Analysis\label{sec:Empirical-Analysis}}

\subsection{Data Description}

Our empirical analysis spans a sample period from January 2, 2002
to December 31, 2020, which has 4,744 trading days after the removal
of holidays and trading days with reduced trading hours. We use daily
close-to-close returns and compute realized variances and correlations
from high-frequency data.

We include nine stocks in our analysis. Three stocks from the energy
sector, CVX, MRO, and OXY, three stocks from the Health Care sector,
JNJ, LLY, and MRK, and three stocks from the Information Technology
sector, AAPL, MU, and ORCL.

We construct close-to-close daily returns for the individual stocks
using closing prices, adjusted for stock splits and dividends, from
the CRSP US Stock Database. Intraday transaction data were obtained
from the TAQ database, and these were cleaned in accordance with the
methodology detailed  in \citet{BNHLS-RKpractice:2009}. From the
high-frequency data, we compute the $9\times9$ multivariate realized
kernel estimates for each trading day, $\mathrm{RM}_{t}\in\mathbb{R}^{9\times9}$,
and these are used to define $x_{t}\in\mathbb{R}^{9}$ and $y_{t}\in\mathbb{R}^{36}$. 

\subsection{Summary Statistics}

We present summary statistics for the nine return series and their
corresponding realized variance measures in Table \ref{tab:SumStats}.
These statistics are consistent with typical estimates for such time
series. We note that Health Care stocks (middle three columns) had
the lowest volatility, whereas IT stocks (the last three columns)
had the largest average volatility in the sample period. This can
be seen from the standard deviations in the second row, and the means
and medians (Q-50\%) of the Realized volatilities in the lower part
of Table \ref{tab:SumStats}.

\begin{table}[H]
\begin{centering}

\scriptsize
\begin{tabularx}{\textwidth}{ >{\hsize=1.18\hsize}Y  >{\hsize=0.98\hsize}Y >{\hsize=0.98\hsize}Y >{\hsize=0.98\hsize}Y >{\hsize=0.98\hsize}Y >{\hsize=0.98\hsize}Y >{\hsize=0.98\hsize}Y >{\hsize=0.98\hsize}Y >{\hsize=0.98\hsize}Y >{\hsize=0.98\hsize}Y }
\\%[-1.0cm]
\toprule
\midrule
  &  \multicolumn{3}{c}{Energy} & \multicolumn{3}{c}{Health Care} & \multicolumn{3}{c}{Information Technology} \\
\cmidrule(l){2-4}
\cmidrule(l){5-7}
\cmidrule(l){8-10}
  & CVX & MRO & OXY & JNJ & LLY & MRK & AAPL & MU & ORCL \\
\midrule
\\[-0.1cm]
  &  \multicolumn{5}{l}{\textit{ Daily returns ($\times$100) }} &  &  &  &   \\
\\[-0.1cm]
 Mean & 0.045 & 0.046 & 0.054 & 0.037 & 0.039 & 0.034 & 0.145 & 0.068 & 0.055 \\
 Std. & 1.765 & 2.788 & 2.527 & 1.166 & 1.569 & 1.650 & 2.203 & 3.347 & 1.993 \\
 Skewness & 0.129 & -0.540 & -0.815 & -0.290 & 0.424 & -0.928 & 0.031 & 0.006 & 0.329 \\
 Kurtosis & 26.237 & 26.892 & 52.357 & 19.431 & 13.159 & 25.732 & 8.206 & 7.296 & 10.724 \\
\\[-0.1cm]
 Min & -22.125 & -46.852 & -50.484 & -15.846 & -12.348 & -26.781 & -17.920 & -23.042 & -14.509 \\
 Q-05\% & -2.477 & -4.074 & -3.406 & -1.644 & -2.279 & -2.265 & -3.244 & -5.084 & -2.919 \\
 Q-25\% & -0.774 & -1.235 & -0.978 & -0.481 & -0.712 & -0.704 & -0.912 & -1.635 & -0.837 \\
 Q-50\% & 0.078 & 0.077 & 0.045 & 0.027 & 0.053 & 0.031 & 0.100 & 0.000 & 0.050 \\
 Q-75\% & 0.866 & 1.383 & 1.112 & 0.580 & 0.777 & 0.824 & 1.232 & 1.808 & 0.970 \\
 Q-95\% & 2.380 & 3.845 & 3.232 & 1.707 & 2.262 & 2.342 & 3.657 & 5.257 & 2.909 \\
 Max & 22.741 & 23.357 & 33.698 & 12.229 & 15.680 & 13.033 & 13.905 & 23.443 & 20.427 \\ \\
\\[-0.2cm]
  &  \multicolumn{5}{l}{\textit{ Realized volatilities (in annual units) }} &  &  &  &   \\
\\[-0.1cm]
 Mean & 0.220 & 0.345 & 0.291 & 0.167 & 0.209 & 0.220 & 0.301 & 0.451 & 0.271 \\
 Std. & 0.146 & 0.240 & 0.271 & 0.103 & 0.119 & 0.137 & 0.186 & 0.238 & 0.171 \\
 Skewness & 5.411 & 4.677 & 12.466 & 4.371 & 3.810 & 4.094 & 3.480 & 3.038 & 2.648 \\
 Kurtosis & 56.606 & 44.835 & 335.461 & 35.263 & 28.941 & 31.864 & 26.516 & 19.475 & 14.031 \\
\\[-0.1cm]
 Min & 0.072 & 0.095 & 0.073 & 0.051 & 0.063 & 0.064 & 0.054 & 0.115 & 0.064 \\
 Q-05\% & 0.107 & 0.151 & 0.131 & 0.085 & 0.107 & 0.109 & 0.125 & 0.232 & 0.112 \\
 Q-25\% & 0.142 & 0.214 & 0.177 & 0.112 & 0.141 & 0.145 & 0.184 & 0.308 & 0.165 \\
 Q-50\% & 0.184 & 0.284 & 0.231 & 0.139 & 0.177 & 0.182 & 0.258 & 0.389 & 0.223 \\
 Q-75\% & 0.251 & 0.389 & 0.313 & 0.188 & 0.237 & 0.247 & 0.359 & 0.511 & 0.311 \\
 Q-95\% & 0.434 & 0.727 & 0.619 & 0.336 & 0.414 & 0.456 & 0.619 & 0.889 & 0.610 \\
 Max & 2.807 & 4.608 & 9.674 & 1.402 & 1.830 & 2.145 & 2.548 & 3.511 & 1.981 \\ \\
\\[-0.5cm]
\midrule
\bottomrule
\end{tabularx}
\par\end{centering}
\caption{\footnotesize  Summary statistics for daily returns and annualized
realized volatilities, where the latter are based on the realized
kernel estimator. The sample period runs from January 2nd, 2002 to
December 31st, 2020 (4,744 trading days).\label{tab:SumStats}}
\end{table}

Summary statistics for the realized correlations are presented in
Table \ref{tab:SumStatCorr}, where the shaded regions illustrate
the block structure we use in a sector-based factor model. The numbers
below the main diagonal are the average realized correlations (the
off-diagonal elements of $\bar{Y}=\tfrac{1}{T}\sum_{t=1}^{T}Y_{t}$)
and the numbers above the diagonal are the corresponding averages
for the transformed quantities (the off-diagonal elements of $\tfrac{1}{T}\sum_{t=1}^{T}\log Y_{t}$).
Note that the realized correlations within each of the blocks have
similar averages. The three assets from the energy sector are highly
correlated, with correlations of about 0.55 on average. The average
within-sector correlations for the Health Care sector and Information
Technology sector stocks are about 0.39 and 0.31, respectively. The
between-sector correlations tend to be smaller, and these range between
0.17 and 0.28. Similar patterns are seen for the logarithmically transformed
correlation variables.
\begin{table}[H]
\begin{centering}

\scriptsize
\begin{tabularx}{\textwidth}{ >{\hsize=0.3\hsize}X >{\hsize=0.8\hsize}Y  >{\hsize=1.1\hsize}Y  >{\hsize=1.1\hsize}Y  >{\hsize=1.1\hsize}Y  >{\hsize=1.1\hsize}Y  >{\hsize=1.1\hsize}Y  >{\hsize=1.1\hsize}Y  >{\hsize=1.1\hsize}Y  >{\hsize=1.1\hsize}Y  >{\hsize=1.1\hsize}Y  }
\\%[-1.0cm]
\toprule
\midrule
  & &  \multicolumn{3}{c}{Energy} & \multicolumn{3}{c}{Health Care} & \multicolumn{3}{c}{Information Tech.} \\
\cmidrule(l){3-5}
\cmidrule(l){6-8}
\cmidrule(l){9-11}
  &  & CVX & MRO & OXY & JNJ & LLY & MRK & AAPL & MU & ORCL \\
\midrule
\\[0.0cm]
 \parbox[c]{1cm}{\multirow{8}{*}{\rotatebox[origin=c]{90}{\footnotesize Energy}}} & CVX &  & 0.491 & 0.513 & \cellcolor{red!15}0.154 & \cellcolor{red!15}0.109 & \cellcolor{red!15}0.128 & 0.157 & 0.108 & 0.154 \\
  &  &  & \textit{(0.162)} & \textit{(0.171)} & \cellcolor{red!15}\textit{(0.123)} & \cellcolor{red!15}\textit{(0.123)} & \cellcolor{red!15}\textit{(0.123)} & \textit{(0.122)} & \textit{(0.110)} & \textit{(0.123)} \\
  &  &  &  &  & \cellcolor{red!15} & \cellcolor{red!15} & \cellcolor{red!15} &  &  & 
\\[0.0cm]
  & MRO & 0.553 &  & 0.494 & \cellcolor{red!15}0.060 & \cellcolor{red!15}0.066 & \cellcolor{red!15}0.074 & 0.119 & 0.125 & 0.111 \\
  &  & \textit{(0.143)} &  & \textit{(0.174)} & \cellcolor{red!15}\textit{(0.119)} & \cellcolor{red!15}\textit{(0.115)} & \cellcolor{red!15}\textit{(0.115)} & \textit{(0.114)} & \textit{(0.120)} & \textit{(0.119)} \\
  &  &  &  &  & \cellcolor{red!15} & \cellcolor{red!15} & \cellcolor{red!15} &  &  & 
\\[0.0cm]
  & OXY & 0.566 & 0.549 &  & \cellcolor{red!15}0.080 & \cellcolor{red!15}0.070 & \cellcolor{red!15}0.082 & 0.116 & 0.095 & 0.112 \\
  &  & \textit{(0.145)} & \textit{(0.147)} &  & \cellcolor{red!15}\textit{(0.116)} & \cellcolor{red!15}\textit{(0.114)} & \cellcolor{red!15}\textit{(0.114)} & \textit{(0.117)} & \textit{(0.114)} & \textit{(0.118)} \\
\\[0.0cm]
 \parbox[c]{1cm}{\multirow{8}{*}{\rotatebox[origin=c]{90}{\footnotesize Health Care}}} & JNJ & \cellcolor{red!15}0.255 & \cellcolor{red!15}0.178 & \cellcolor{red!15}0.193 &  & 0.306 & 0.329 & \cellcolor{red!15}0.135 & \cellcolor{red!15}0.080 & \cellcolor{red!15}0.164 \\
  &  & \cellcolor{red!15}\textit{(0.174)} & \cellcolor{red!15}\textit{(0.172)} & \cellcolor{red!15}\textit{(0.170)} &  & \textit{(0.149)} & \textit{(0.168)} & \cellcolor{red!15}\textit{(0.120)} & \cellcolor{red!15}\textit{(0.115)} & \cellcolor{red!15}\textit{(0.123)} \\
  &  & \cellcolor{red!15} & \cellcolor{red!15} & \cellcolor{red!15} &  &  &  & \cellcolor{red!15} & \cellcolor{red!15} & \cellcolor{red!15}
\\[0.0cm]
  & LLY & \cellcolor{red!15}0.222 & \cellcolor{red!15}0.170 & \cellcolor{red!15}0.177 & 0.382 &  & 0.352 & \cellcolor{red!15}0.120 & \cellcolor{red!15}0.084 & \cellcolor{red!15}0.143 \\
  &  & \cellcolor{red!15}\textit{(0.178)} & \cellcolor{red!15}\textit{(0.171)} & \cellcolor{red!15}\textit{(0.170)} & \textit{(0.160)} &  & \textit{(0.164)} & \cellcolor{red!15}\textit{(0.110)} & \cellcolor{red!15}\textit{(0.108)} & \cellcolor{red!15}\textit{(0.117)} \\
  &  & \cellcolor{red!15} & \cellcolor{red!15} & \cellcolor{red!15} &  &  &  & \cellcolor{red!15} & \cellcolor{red!15} & \cellcolor{red!15}
\\[0.0cm]
  & MRK & \cellcolor{red!15}0.241 & \cellcolor{red!15}0.183 & \cellcolor{red!15}0.192 & 0.397 & 0.409 &  & \cellcolor{red!15}0.128 & \cellcolor{red!15}0.086 & \cellcolor{red!15}0.149 \\
  &  & \cellcolor{red!15}\textit{(0.177)} & \cellcolor{red!15}\textit{(0.169)} & \cellcolor{red!15}\textit{(0.169)} & \textit{(0.172)} & \textit{(0.166)} &  & \cellcolor{red!15}\textit{(0.118)} & \cellcolor{red!15}\textit{(0.109)} & \cellcolor{red!15}\textit{(0.122)} \\
\\[0.0cm]
 \parbox[c]{1cm}{\multirow{8}{*}{\rotatebox[origin=c]{90}{\footnotesize Information Tech.}}} & AAPL & 0.276 & 0.237 & 0.236 & \cellcolor{red!15}0.240 & \cellcolor{red!15}0.226 & \cellcolor{red!15}0.234 &  & 0.235 & 0.284 \\
  &  & \textit{(0.173)} & \textit{(0.163)} & \textit{(0.167)} & \cellcolor{red!15}\textit{(0.160)} & \cellcolor{red!15}\textit{(0.152)} & \cellcolor{red!15}\textit{(0.160)} &  & \textit{(0.144)} & \textit{(0.146)} \\
  &  &  &  &  & \cellcolor{red!15} & \cellcolor{red!15} & \cellcolor{red!15} &  &  & 
\\[0.0cm]
  & MU & 0.221 & 0.216 & 0.198 & \cellcolor{red!15}0.173 & \cellcolor{red!15}0.171 & \cellcolor{red!15}0.176 & 0.301 &  & 0.193 \\
  &  & \textit{(0.153)} & \textit{(0.152)} & \textit{(0.153)} & \cellcolor{red!15}\textit{(0.148)} & \cellcolor{red!15}\textit{(0.142)} & \cellcolor{red!15}\textit{(0.147)} & \textit{(0.148)} &  & \textit{(0.127)} \\
  &  &  &  &  & \cellcolor{red!15} & \cellcolor{red!15} & \cellcolor{red!15} &  &  & 
\\[0.0cm]
  & ORCL & 0.274 & 0.229 & 0.231 & \cellcolor{red!15}0.266 & \cellcolor{red!15}0.248 & \cellcolor{red!15}0.256 & 0.355 & 0.272 &  \\
  &  & \textit{(0.178)} & \textit{(0.171)} & \textit{(0.173)} & \cellcolor{red!15}\textit{(0.164)} & \cellcolor{red!15}\textit{(0.160)} & \cellcolor{red!15}\textit{(0.165)} & \textit{(0.164)} & \textit{(0.150)} &  \\
\\[0.0cm]
\\[-0.5cm]
\midrule
\bottomrule
\end{tabularx}
\par\end{centering}
\caption{\footnotesize  Summary statistics for the realized correlations where
the sector-based block structure is illustrated with the shaded regions.
The average realized correlations and corresponding standard deviations
(in parentheses) are shown below the diagonal. The corresponding statistics
for the transformed elements, $\gamma_{t}$, are shown above the diagonal.
Statistics are based on the sample period from January 2nd, 2002,
to December 31st, 2020 (4,744 trading days).\label{tab:SumStatCorr}}
\end{table}

\begin{figure}[H]
\begin{centering}
\includegraphics[scale=0.63]{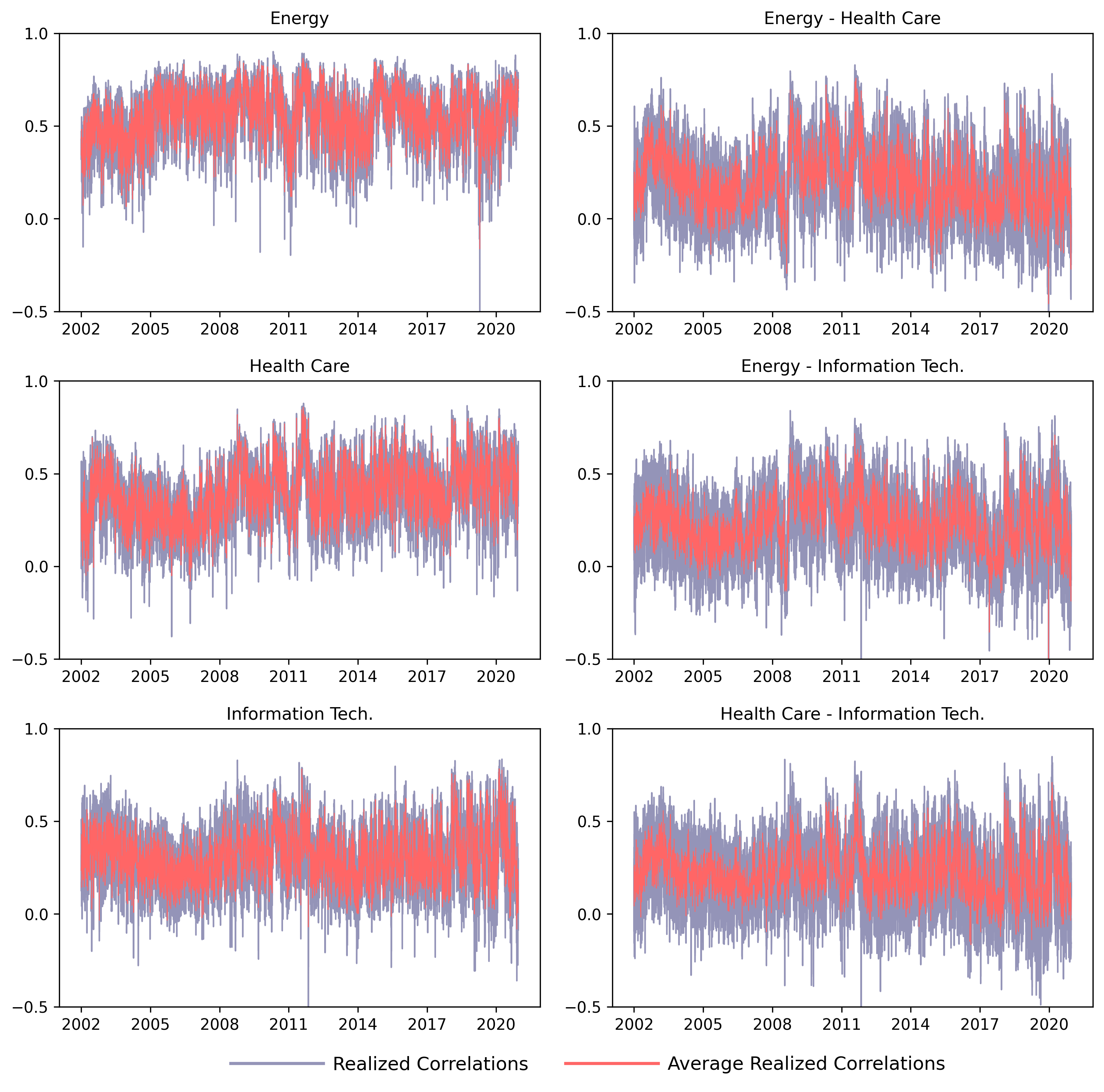}
\par\end{centering}
\caption{\footnotesize  Daily realized correlations for the nine return series
over the full sample period. Left subplots present intra-sector correlations
(gray lines) and their average (red line) for each sector. Right subplots
present inter-sector correlations (gray lines) and their average (red
line) for each pair of sectors.\label{fig:CorrSeriesSector}}
\end{figure}

The time series of realized correlations are shown in Figure \ref{fig:CorrSeriesSector}.
The left subplots present the within-sector correlations (gray lines)
and their average daily correlation (red line) for each of the three
sectors. The right subplots present the between-sector correlations
(gray lines) and their daily average (red line) for the three sector
pairs. The 36 correlation time series are computed from the $9\times9$
multivariate Realized Kernel estimator. Correlations within each of
the six categories tend to move together; however, there are notable
differences across the six types of within-sector and between-sector
correlations, both in terms of their average level and their variation
over time. For instance, the average between-sectors correlation for
Health Care returns and Information Technology returns does not have
a sharp decline in late 2008, as can be seen for the two other between-sectors
correlations involving Energy sector returns. Figure \ref{fig:CorrSeriesSector}
provides additional motivation for exploring a block structure for
the correlation matrix.

\subsection{Empirical Analysis of the Multivariate Realized GARCH model \label{subsec:estimation_results}}

The empirical results are based on the two-stage estimation procedure,
described in Section \ref{subsec:Two-Stage-Estimation}, and we use
three different specifications for the correlation matrix: equicorrelation
(Equi), block correlation (Block), and unrestricted (Full). The simplest
model is the equicorrelation model that assumes a single common dynamic
correlation coefficient that is common for all correlations in $C_{t}$.
The most general specification is the fully dynamic correlation matrix
with 36 dynamic correlation coefficients. The equicorrelation model
has a single latent variable to model the time variation in the correlation
matrix, such that $\zeta_{t}$ is univariate ($r=1)$ in this case.
The block correlation model employs six latent variables ($r=6)$,
whereas the most flexible specification, Full, has $d=36$ latent
variables. 

\begin{table}[H]
\begin{centering}

\scriptsize
\begin{tabularx}{\textwidth}{ >{\hsize=0.55\hsize}Y  >{\hsize=1.05\hsize}Y  >{\hsize=1.05\hsize}Y  >{\hsize=1.05\hsize}Y  >{\hsize=1.05\hsize}Y  >{\hsize=1.05\hsize}Y  >{\hsize=1.05\hsize}Y  >{\hsize=1.05\hsize}Y  >{\hsize=1.05\hsize}Y  >{\hsize=1.05\hsize}Y  }
\toprule
\midrule
  &  \multicolumn{3}{c}{Energy} & \multicolumn{3}{c}{Health Care} & \multicolumn{3}{c}{Information Tech.} \\
\cmidrule(l){2-4}
\cmidrule(l){5-7}
\cmidrule(l){8-10}
   & CVX & MRO & OXY & JNJ & LLY & MRK & AAPL & MU & ORCL \\
\midrule
\\[-0.1cm]
 $\mu$ & 0.036 & 0.075 & 0.050 & 0.037 & 0.034 & 0.027 & 0.151 & 0.065 & 0.033 \\ & {\tiny (0.017)} & {\tiny (0.027)} & {\tiny (0.022)} & {\tiny (0.012)} & {\tiny (0.017)} & {\tiny (0.019)} & {\tiny (0.023)} & {\tiny (0.038)} & {\tiny (0.018)}
\\[0.1cm]
 $\omega$ & 0.087 & 0.121 & 0.097 & 0.001 & 0.081 & 0.208 & 0.182 & 0.156 & 0.053 \\ & {\tiny (0.012)} & {\tiny (0.020)} & {\tiny (0.015)} & {\tiny (0.012)} & {\tiny (0.016)} & {\tiny (0.074)} & {\tiny (0.026)} & {\tiny (0.035)} & {\tiny (0.022)}
\\[0.1cm]
 $\beta$ & 0.618 & 0.638 & 0.644 & 0.669 & 0.651 & 0.563 & 0.441 & 0.554 & 0.376 \\ & {\tiny (0.021)} & {\tiny (0.022)} & {\tiny (0.020)} & {\tiny (0.022)} & {\tiny (0.032)} & {\tiny (0.033)} & {\tiny (0.031)} & {\tiny (0.045)} & {\tiny (0.036)}
\\[0.1cm]
 $\alpha$ & 0.336 & 0.343 & 0.352 & 0.327 & 0.347 & 0.297 & 0.499 & 0.418 & 0.607 \\ & {\tiny (0.021)} & {\tiny (0.024)} & {\tiny (0.024)} & {\tiny (0.026)} & {\tiny (0.034)} & {\tiny (0.048)} & {\tiny (0.030)} & {\tiny (0.045)} & {\tiny (0.037)}
\\[0.1cm]
 $\tau_1$ & -0.050 & -0.043 & -0.046 & -0.049 & -0.033 & -0.027 & -0.052 & -0.036 & -0.031 \\ & {\tiny (0.005)} & {\tiny (0.006)} & {\tiny (0.006)} & {\tiny (0.007)} & {\tiny (0.008)} & {\tiny (0.009)} & {\tiny (0.009)} & {\tiny (0.007)} & {\tiny (0.009)}
\\[0.1cm]
 $\tau_2$ & 0.020 & 0.026 & 0.026 & 0.012 & 0.004 & -0.002 & 0.036 & 0.005 & 0.003 \\ & {\tiny (0.003)} & {\tiny (0.005)} & {\tiny (0.005)} & {\tiny (0.003)} & {\tiny (0.004)} & {\tiny (0.001)} & {\tiny (0.006)} & {\tiny (0.004)} & {\tiny (0.005)}
\\[0.1cm]
\\[0.0cm]
 $\xi$ & -0.202 & -0.261 & -0.211 & -0.013 & -0.171 & -0.569 & -0.185 & -0.120 & 0.010 \\ & {\tiny (0.032)} & {\tiny (0.056)} & {\tiny (0.038)} & {\tiny (0.031)} & {\tiny (0.034)} & {\tiny (0.297)} & {\tiny (0.042)} & {\tiny (0.075)} & {\tiny (0.032)}
\\[0.1cm]
 $\varphi$ & 1.045 & 0.996 & 0.956 & 0.921 & 0.899 & 1.296 & 0.974 & 0.948 & 0.915 \\ & {\tiny (0.030)} & {\tiny (0.040)} & {\tiny (0.033)} & {\tiny (0.048)} & {\tiny (0.041)} & {\tiny (0.195)} & {\tiny (0.029)} & {\tiny (0.035)} & {\tiny (0.024)}
\\[0.1cm]
 $\delta_1$ & -0.082 & -0.052 & -0.052 & -0.042 & -0.058 & -0.038 & -0.110 & -0.039 & -0.060 \\ & {\tiny (0.007)} & {\tiny (0.007)} & {\tiny (0.007)} & {\tiny (0.009)} & {\tiny (0.010)} & {\tiny (0.017)} & {\tiny (0.009)} & {\tiny (0.008)} & {\tiny (0.009)}
\\[0.1cm]
 $\delta_2$ & 0.075 & 0.063 & 0.075 & 0.078 & 0.099 & 0.015 & 0.093 & 0.084 & 0.081 \\ & {\tiny (0.005)} & {\tiny (0.005)} & {\tiny (0.005)} & {\tiny (0.008)} & {\tiny (0.007)} & {\tiny (0.005)} & {\tiny (0.005)} & {\tiny (0.005)} & {\tiny (0.006)}
\\[0.1cm]
\\[0.0cm]
 $\pi$ & 0.969 & 0.979 & 0.980 & 0.970 & 0.962 & 0.949 & 0.927 & 0.950 & 0.931 \\[0.1cm]
 $\sigma^2_v$ & 0.158 & 0.156 & 0.157 & 0.219 & 0.235 & 0.293 & 0.284 & 0.229 & 0.279
\\[0.1cm]
 {\tiny $-2T^{-1}\ell_{r,i}$} & 3.4314 & 4.3803 & 3.9976 & 2.7274 & 3.4129 & 3.6057 & 4.0683 & 4.9674 & 3.6855
\\[0.1cm]
\\[-0.2cm]
\midrule
\bottomrule
\end{tabularx}
\par\end{centering}
\caption{\footnotesize  Parameter estimates with standard errors (in parentheses)
for the nine Realized GARCH models (first-stage estimation). The average
partial log-likelihood for each of the return series are reported
in the last row. \label{tab:Parameter-estimates-var}}
\end{table}

Parameter estimates from the first-stage estimation (univariate Realized
GARCH models) are reported in Table \ref{tab:Parameter-estimates-var}.
The estimates are based on the full sample period from January 3,
2002, to December 31, 2020. Each column in Table \ref{tab:Parameter-estimates-var}
corresponds to one of nine assets in our analysis. We report parameter
estimates along with their corresponding standard errors, shown below
in parentheses.\footnote{We obtain standard errors numerically by calculating the Hessian matrix
and the information matrix via the outer gradient product.} The model implies an AR(1) model for $h_{i,t}$ with $\pi=\beta+\alpha\varphi$
as the autoregressive coefficient. This can be seen by substituting
the measurement equation into the corresponding GARCH equation. In
the bottom of Table \ref{tab:Parameter-estimates-var}, we report
this persistence parameter for each of the conditional variances.
These estimates are all close to unity, as expected, since volatility
is known to be persistent. Overall, the parameter estimates are in
line with results in the existing literature on GARCH models and Realized
GARCH models. In the last row of Table \ref{tab:Parameter-estimates-var},
we report the average partial log-likelihoods, $-2T^{-1}\ell_{r_{i}}=\log2\pi+\frac{1}{T}\sum_{t=1}^{T}\left\{ \log h_{i,t}+z_{i,t}^{2}\right\} $,
$i=1,\ldots,9$, that summarize how well the estimated univariate
Realized GARCH models describe the conditional (marginal) distribution
of returns, for each of the nine asset.
\begin{table}[H]
\begin{centering}

\scriptsize
\begin{tabularx}{\textwidth}{ >{\hsize=0.55\hsize}Y  >{\hsize=1\hsize}Y  >{\hsize=1\hsize}Y  >{\hsize=1\hsize}Y  >{\hsize=1\hsize}Y  >{\hsize=1\hsize}Y  >{\hsize=1\hsize}Y  >{\hsize=1\hsize}Y  >{\hsize=1.45\hsize}Y  }
\toprule
\midrule
  &  \multicolumn{1}{c}{} & \multicolumn{6}{c}{Block model} & \multicolumn{1}{c}{} \\
\cmidrule(l){3-8}
   & Equi & En & En-HC & En-IT & HC & HC-IT & IT & Full model \\
\midrule
\\[-0.1cm]
 $\tilde{\omega}$ & -0.005 & 0.040 & -0.003 & 0.000 & 0.024 & -0.001 & 0.005 & [-0.010, 0.026] \\ & {\tiny (0.008)} & {\tiny (0.012)} & {\tiny (0.002)} & {\tiny (0.003)} & {\tiny (0.013)} & {\tiny (0.004)} & {\tiny (0.008)} & 
\\[0.1cm]
 $\tilde{\beta}$ & 0.539 & 0.720 & 0.790 & 0.762 & 0.765 & 0.737 & 0.712 & [0.828, 0.950] \\ & {\tiny (0.025)} & {\tiny (0.020)} & {\tiny (0.011)} & {\tiny (0.012)} & {\tiny (0.015)} & {\tiny (0.015)} & {\tiny (0.024)} & 
\\[0.1cm]
 $\tilde{\alpha}$ & 0.535 & 0.277 & 0.300 & 0.288 & 0.195 & 0.257 & 0.322 & [0.016, 0.217] \\ & {\tiny (0.042)} & {\tiny (0.031)} & {\tiny (0.025)} & {\tiny (0.030)} & {\tiny (0.035)} & {\tiny (0.039)} & {\tiny (0.045)} & 
\\[0.1cm]
\\[0.0cm]
 $\tilde{\xi}$ & 0.030 & -0.073 & 0.017 & 0.010 & -0.082 & 0.020 & 0.031 & [-0.218, 0.110] \\ & {\tiny (0.012)} & {\tiny (0.046)} & {\tiny (0.007)} & {\tiny (0.010)} & {\tiny (0.074)} & {\tiny (0.013)} & {\tiny (0.023)} & 
\\[0.1cm]
 $\tilde{\varphi}$ & 0.753 & 0.899 & 0.641 & 0.753 & 1.094 & 0.865 & 0.725 & [0.440, 2.731] \\ & {\tiny (0.057)} & {\tiny (0.074)} & {\tiny (0.043)} & {\tiny (0.068)} & {\tiny (0.179)} & {\tiny (0.114)} & {\tiny (0.077)} & 
\\[0.1cm]
\\[0.0cm]
 $\tilde{\pi}$ & 0.942 & 0.969 & 0.982 & 0.978 & 0.979 & 0.960 & 0.946 & [0.956, 0.993] \\[0.1cm]
 $\sigma^2_{\tilde{v}}$ & 0.001 & 0.009 & 0.002 & 0.002 & 0.007 & 0.002 & 0.006 & [0.010, 0.020]
\\[0.1cm]
 {\tiny $-2T^{-1}\ell_{r}$} & 23.3645 &  &  &  &  &  & 22.3006 & 22.2351
\\[0.1cm]
 {\tiny $T^{-1} BIC$ } & 23.3734 &  &  &  &  &  & 22.3542 & 22.5563
\\[0.1cm]
\\[-0.2cm]
\midrule
\bottomrule
\end{tabularx}
\par\end{centering}
\caption{\footnotesize  Estimates with standard errors (in parentheses) for
the parameters related to the dynamics of conditional correlations
(second-stage estimation) for each factor structure, Equi, Block,
and Full. The most general specification, Full, has 36 latent factors,
and we report the ranges of the 36 estimates for each type of parameter.
The (multivariate) return log-likelihood is reported in the last row
for each of the three specifications. \label{tab:Parameter-estimates-corr}}
\end{table}

Table \ref{tab:Parameter-estimates-corr} presents the parameter estimates
from the second-stage estimation of the Multivariate Realized GARCH
model, where the model for the dynamic correlation matrix is estimated.
We report parameter estimates for the three specifications. The first
column is for the equicorrelation model (Equi), the next six columns
are for the six dynamic correlations in the 3$\times$3 block-correlation
structure (Block), and the last column is for the unrestricted correlation
structure (Full), where we present the range of the estimates from
the 36 models for $\gamma_{1,t},\ldots,\gamma_{36,t}$. For instance,
the estimated intercepts in the GARCH equation, $\tilde{\omega}_{j}$,
ranged between $-0.01$ and $0.026$. We also report the persistence
parameter, $\tilde{\pi}=\tilde{\beta}+\tilde{\alpha}\cdot\tilde{\varphi}$,
for the conditional correlations, and these are quite similar to those
of the conditional variances. Thus, both conditional variances and
conditional correlations are found to be persistent.

It is not meaningful to compare the total log-likelihood for models
with different measurement equations. It is, however, meaningful to
compare their log-likelihood for returns, $\ell_{r}$, see (\ref{eq:logLr}),
which reflects how well the model describes the conditional distribution
of returns. We can decompose $\ell_{r}$ as
\[
-2\ell_{r}=\underset{-2\sum_{i=1}^{n}\ell_{r_{i}}}{\underbrace{c_{n}+\sum_{i=1}^{n}\sum_{t=1}^{T}\log h_{i,t}+z_{i,t}^{2}}}+\sum_{t=1}^{T}\log\det C_{t}+z_{t}^{\prime}(C_{t}^{-1}-I)z_{t},
\]
where the first term is the log-likelihoods for the marginal distributions
of returns (which is common for all model-specifications in our analysis),
and the last term captures the effect of different correlation models.
We therefore report $-2T^{-1}\ell_{r}$ in Table \ref{tab:Parameter-estimates-corr}
along with the corresponding BIC that includes a penalty for model
complexity. Each latent variables, $\zeta_{j,t}$, adds five parameters,
$(\tilde{\omega}_{j},\tilde{\alpha}_{j},\tilde{\beta}_{j},\tilde{\xi}_{j},\tilde{\varphi}_{j})$,
to the complexity of the model.
\begin{figure}[H]
\begin{centering}
\includegraphics[scale=0.65]{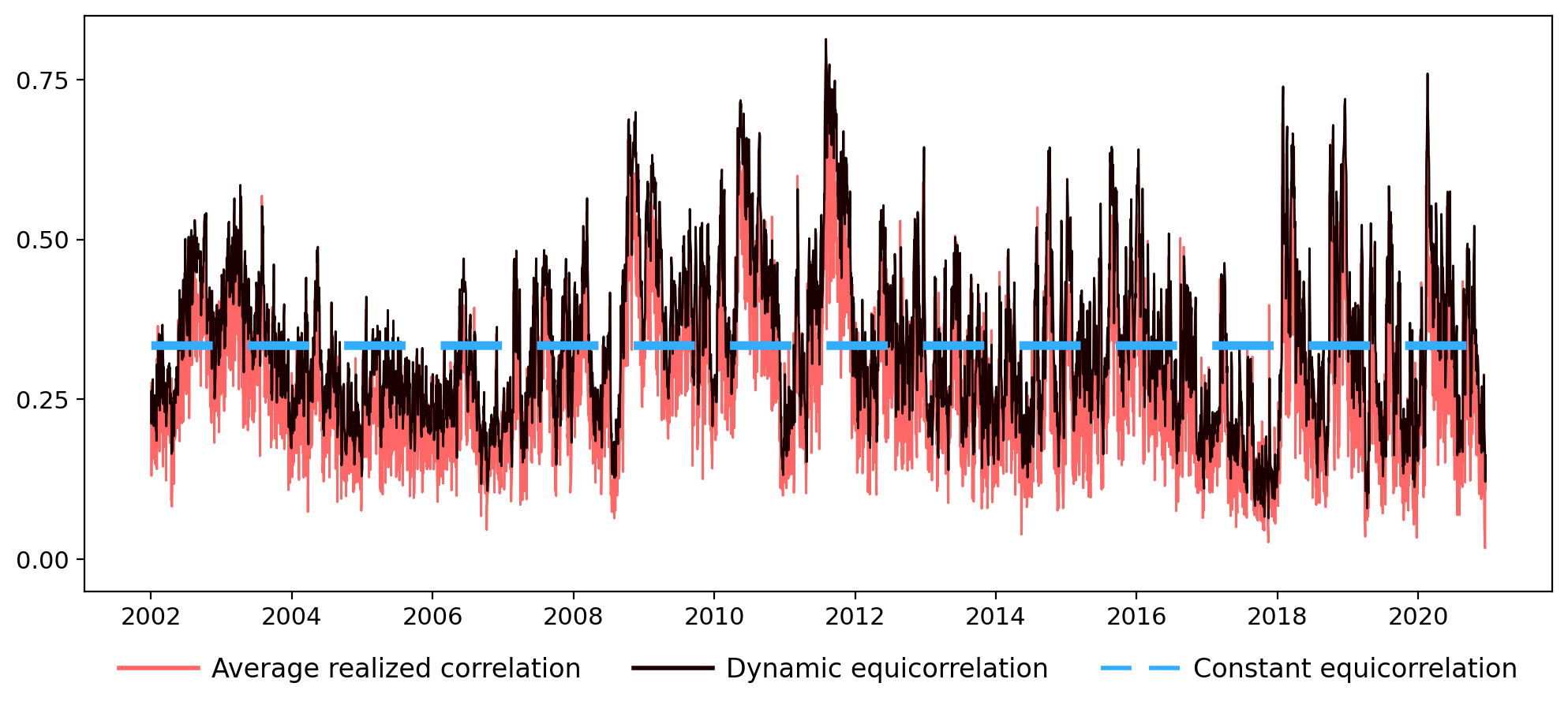}
\par\end{centering}
\caption{\footnotesize  The model-implied correlation coefficient from the
estimated equicorrelation specification (black line) and the daily
average of the 36 realized correlations (red line). The blue dashed
line is the estimate of the equicorrelation coefficient when assumed
to be constant over the sample period.\label{fig:DynamicsCorrEqui}}
\end{figure}

The equicorrelation structure has an average log-likelihood for returns
that is much smaller than those of the more flexible specification.
The average daily difference is about one unit, which add up to a
substantial difference with $T=4,744$ days in the sample. The difference
between the block specification and full specification is more modest
at 0.0655 units per day, which adds up to about 310 units of $2\ell_{r}$
over the full sample. This improvement is achieved with 150 additional
parameters, and the Bayesian information criterion (BIC) prefers the
block specification in this case. In the next section, we evaluate
these specifications in out-of-sample comparisons, and compare them
to DCC-type benchmark models.

The estimated dynamic equicorrelation time series is presented in
Figure \ref{fig:DynamicsCorrEqui}, along with the daily average realized
correlation (red line). The horizontal blue dashed line is the estimated
equicorrelation under the assumption that the correlation is constant
over the sample period.
\begin{figure}[H]
\begin{centering}
\includegraphics[scale=0.63]{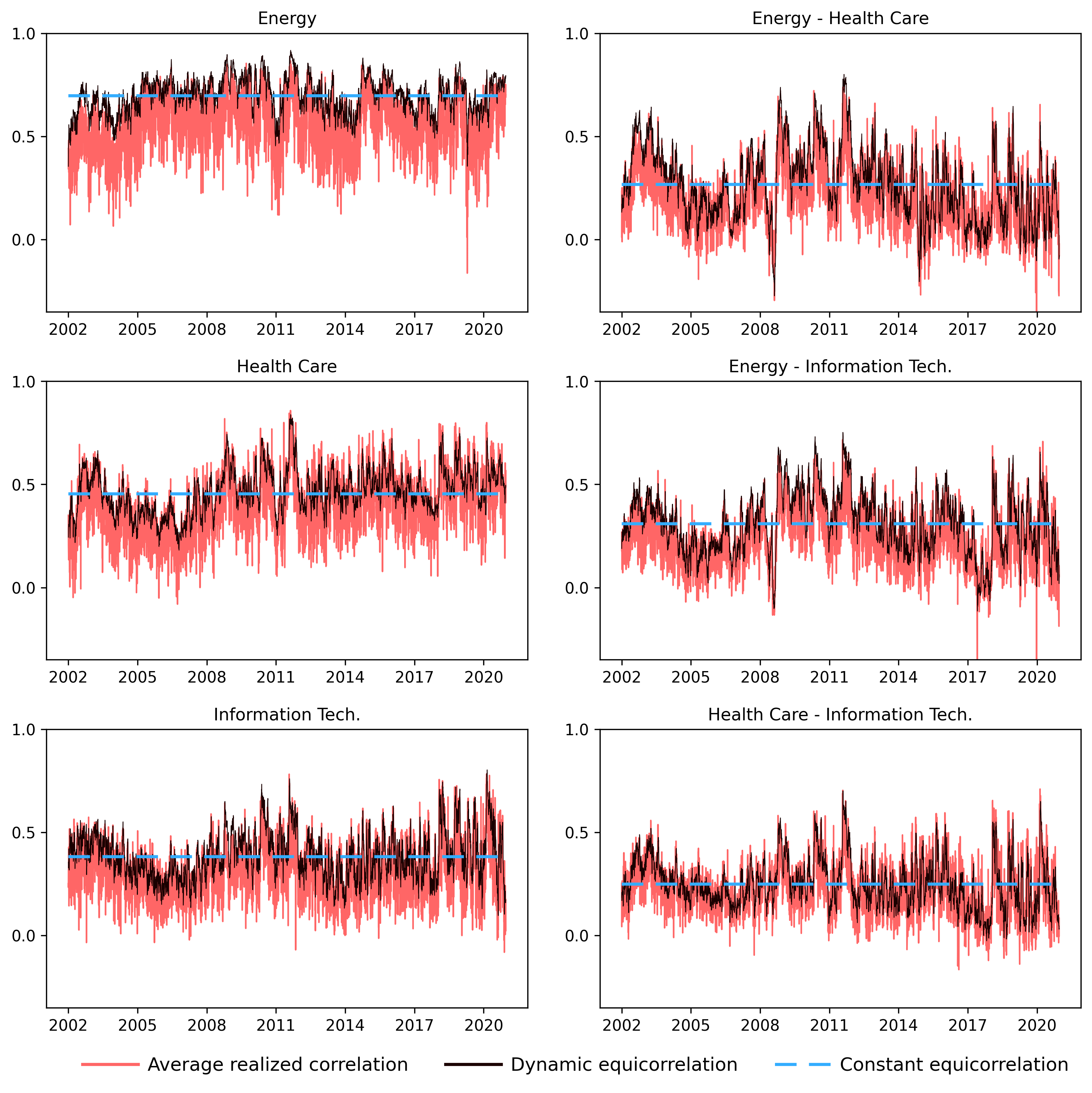}
\par\end{centering}
\caption{\footnotesize  Within-sector and between-sector correlations. The
left subplots present the model-based correlations within each of
the three considered sectors (black lines) and the corresponding daily
averages of realized correlations (red lines). The right panels present
the correlations between sectors. The estimates from the constant
correlation model are indicated by blue dashed lines. \label{fig:DynamicsCorrBlock}}
\end{figure}
 It is comforting that the estimated correlation is in line with with
the average realized correlation, and the gradual variation in the
series strongly suggests that the conditional correlation is time-varying. 

In Figure \ref{fig:DynamicsCorrBlock} we present the corresponding
results for the estimated block specification. In the left panels,
we present the estimated within-sector correlation (black lines) and
the corresponding daily average of the realized correlations (red
lines), and in the right panels we present the results for between-sectors
correlations. The horizontal dashed line in each of the plots is the
estimated block correlation, under the assumption that it is constant
over the sample period.

\subsection{Transformed Realized Correlations are Approximately Gaussian\label{subsec:TransRealCorrGaussian}}

\citet{ABDE:2001} and \citet{ABDL:2001} found that the logarithmic
realized variances for stock returns and exchange rate data are approximately
normally distributed. \citet{ABDE:2001} also found realized correlations
to be approximately normally distributed. Here, we find that the elements
of $y_{t}=\gamma(Y_{t})$, which are the transformed realized correlations,
are also approximately normally distributed. 

\begin{figure}[H]
\begin{centering}
\subfloat[Transformed realized correlations]{\begin{centering}
\includegraphics[scale=0.6]{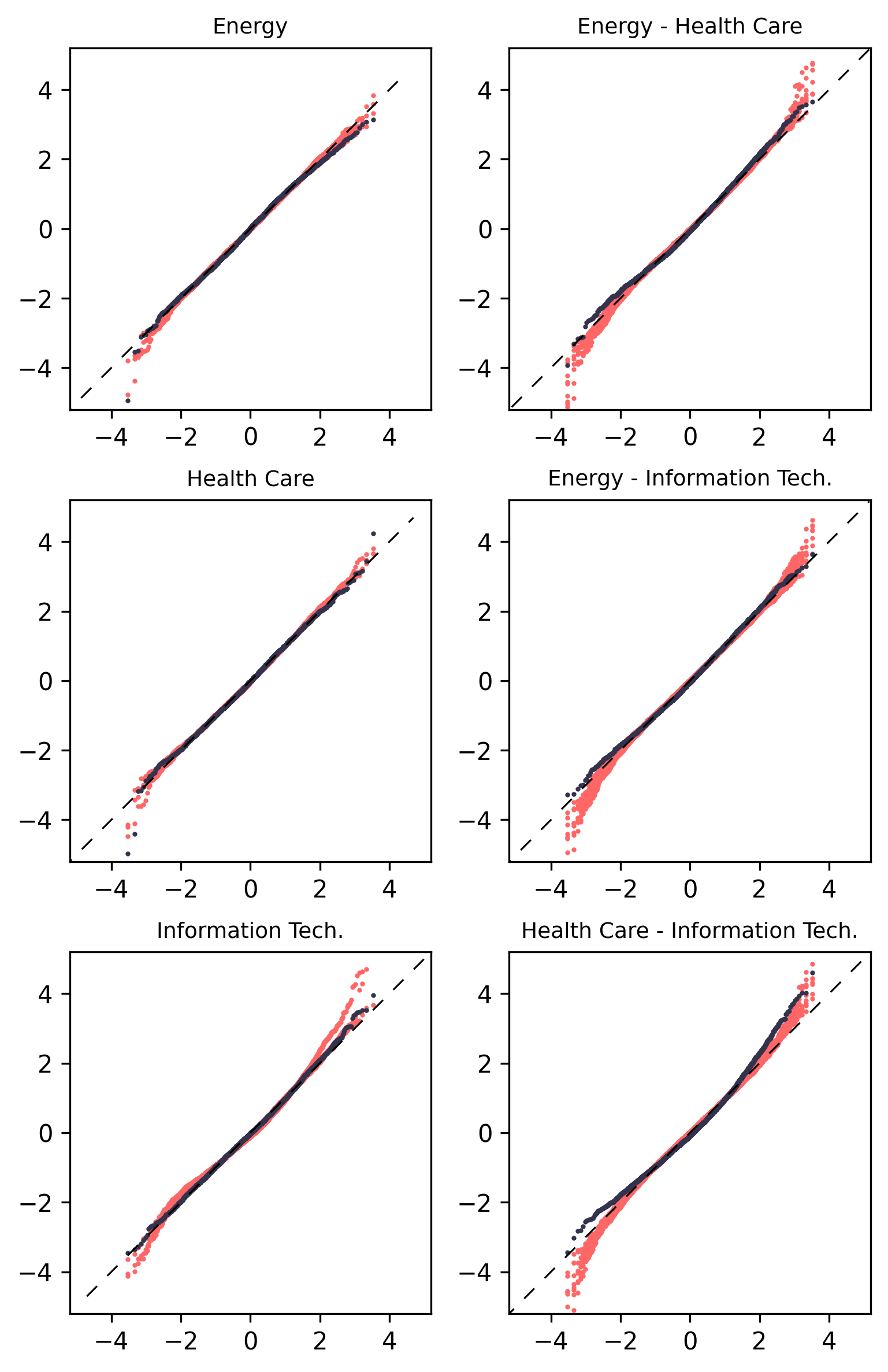}
\par\end{centering}
}\subfloat[Residuals for transformed realized correlations]{\begin{centering}
\includegraphics[scale=0.6]{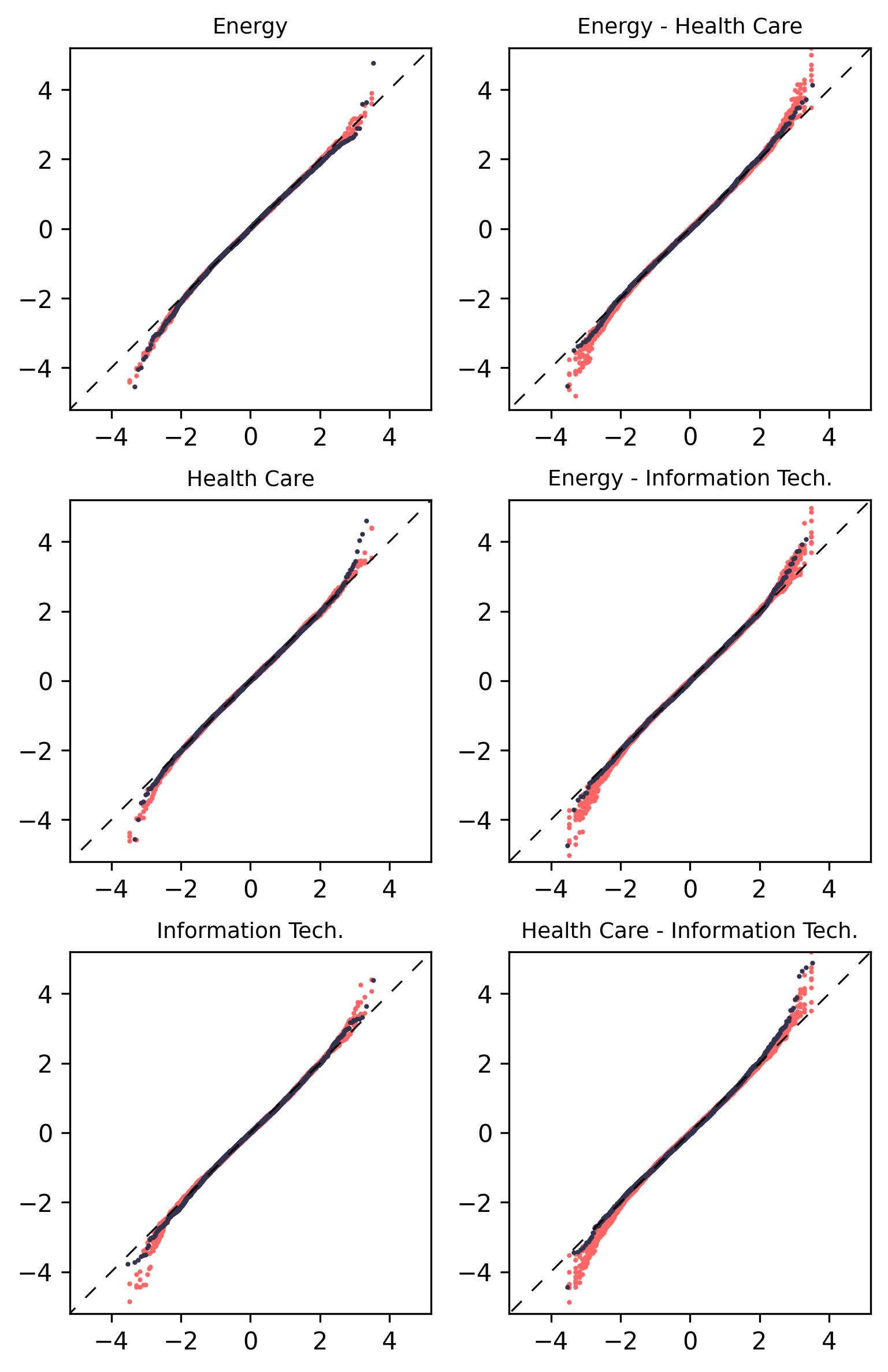}
\par\end{centering}
}
\par\end{centering}
\caption{\label{fig:qq_plots}\footnotesize  Panels (a) and (b) present Q-Q
plots for transformed realized correlations and the corresponding
measurement equation residuals, respectively. Results are shown for
each of six blocks of the correlation matrix, with three within-sector
blocks and three between-sectors blocks. Red dots represent variables
in the Full specification, $y_{t}$ and $\tilde{v}_{t}$, and black
dots represent variables in the Block specification, $\check{y}_{t}$
and $\check{v}_{t}$. The sample period is from January 2, 2002, to
December 31, 2020 (4,744 day).}
\end{figure}

Panel (a) in Figure \ref{fig:qq_plots} presents Q-Q plots for the
empirical distribution of the transformed realized correlations against
the normal distribution. The results are based on the sample period
from January 2, 2002 to December 31, 2020, which has 4,744 daily observations.
The quantiles of their empirical distributions are plotted against
the corresponding quantiles of the normal distribution. The left column
of panel (a) has Q-Q plots for the three unique series within each
of the three diagonal blocks of $Y_{t}$, and the Q-Q plots in the
right column of panel (a) are for the nine series in each of the three
off-diagonal blocks of $Y_{t}$. The black dots are based on elements
of the transformed realized correlations, $y_{t}$, that are used
in the Full specification, and the red dots represent within-block
variables, $\check{y}_{t}$, that are the variables used in the Block
specification. The plots in panel (b) present the corresponding results
for the residuals in the measurement equations, where red dots represent
$\tilde{v}_{t}$ and black dots represent $\check{v}_{t}$.

The Q-Q plots show that both transformed realized correlations and
the corresponding model residuals have empirical distributions that
are reasonably well approximated by Gaussian distributions, albeit
there are some deviations in the tail regions. The discrepancies are
most pronounced for the between-sectors blocks, as can be seen in
the right columns in both panels of Figure \ref{fig:qq_plots}. The
Q-Q plots for the variables in the Block specification appear to approximate
Gaussian distributions more closely than those in the Full specification.
This is not entirely unexpected because the elements of $\check{y}_{t}$
and $\check{v}_{t}$ are effectively defined as averages over elements
of $y_{t}$ and $\tilde{v}_{t}$. The results for $\tilde{v}_{t}$
and $\check{v}_{t}$ provide some justification for adopting a Gaussian
specification in the measurement equation.

\begin{figure}[H]
\begin{centering}
\includegraphics[width=0.9\textwidth]{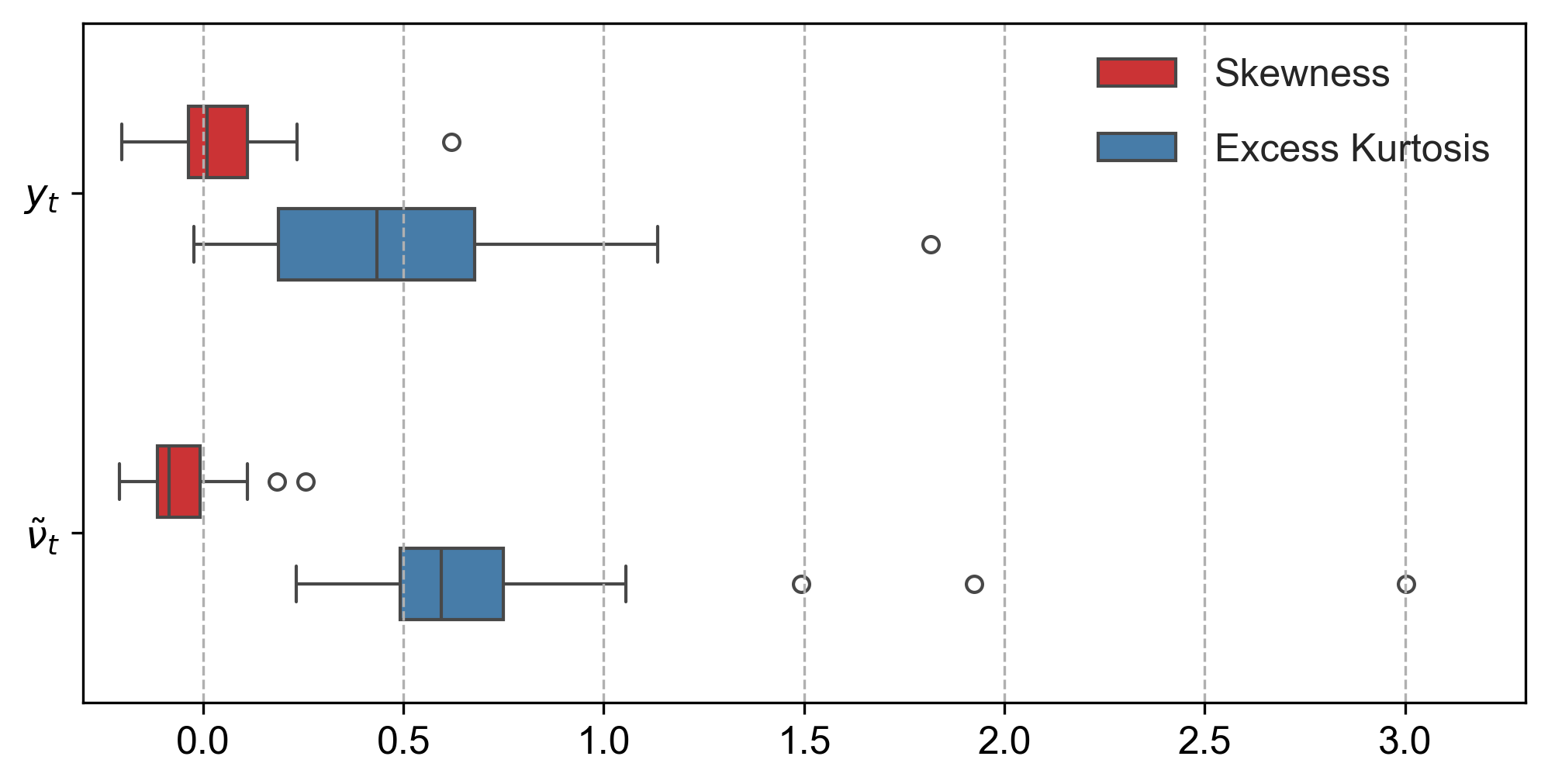}
\par\end{centering}
\caption{\footnotesize  Box-and-whiskers plots for skewness and excess kurtosis
for the transformed realized correlations, $y_{t}$, and the corresponding
``measurement'' errors $\tilde{v}_{t}$ for the full sample period.
Most of the 36 variables have skewness near zero and excess kurtosis
less than one. \label{fig:SkewnessKurtosis}}
\end{figure}

In addition to Q-Q plots we compute the skewness and excess kurtosis
for the variables in the Full specification, $y_{j,t}$ and $\tilde{v}_{j,t}$,
$j=1,\ldots,36$, and present these with box-and-whiskers plots in
Figure \ref{fig:SkewnessKurtosis}. The boxes cover the inter-quartile
range, whiskers the observations that are no more than 3/2 times the
interquartile range away from the edge of a box, and circles represent
observations outside the whiskers (outliers). The variables have,
with one exception, a level of skewness that is fairly close to zero,
while the excess kurtosis is about 0.5 for most variables and a handful
of variables have excess kurtosis larger than one. These are closer
to Gaussian moments than the skewness and kurtosis statistics reported
in \citet{ABDE:2001} for (log) realized variances.

\section{Out-of-Sample Model Performance\label{sec:Out-of-sample}}

In this section, we compare the three specifications of the Multivariate
Realized GARCH model with several benchmark models based on their
out-of-sample performance. Our objective is to compare the different
specifications for the correlation matrix out-of-sample and compare
the Multivariate Realized GARCH model with natural and suitable benchmark
models.

In order not to confound the comparisons with features that relate
to other parts of the model, the first-stage estimation will be identical
for all model in the comparisons. Specifically, we estimate the univariate
Realized GARCH models, (\ref{eq:return})-(\ref{eq:Measurement}),
for each return series, such that $h_{i,t}$ and $z_{i,t}$, $i=1,\ldots,n$,
$t=1,\ldots,T$, are common for all model-specifications.

\subsection{Benchmark Models for Empirical Evaluation }

We adopt the Constant Conditional Correlation (CCC) model by \citet{Bollerslev:1990}
and the Dynamic Conditional Correlation (DCC) model by \citet{Engle2002}
as benchmark models for $C_{t}$. The former has (as its name suggests)
a constant conditional correlation matrix and the latter uses GARCH-type
dynamic for updating $C_{t}$. We label these models as CCC$^{+}$
and DCC$^{+}$, respectively, because they are enhanced CCC and DCC
models that utilize realized measures of volatility for modeling the
univariate conditional variances. The key features of the three types
of models are summarized in Table \ref{tab:models}.

We consider three specifications for the correlation matrix: Equi,
Block, and Full, for each model type: MRG, CCC, and DCC. The DCC$^{+}$
with equicorrelation is similar to the DECO model by \citet{EngleKelly:2012},
with the key difference being that DCC$^{+}$ employs Realized GARCH
models for each of the nine return series. Similarly, DCC$^{+}$ with
block correlation can be viewed as an enhanced version of Block-DECO
by \citet{EngleKelly:2012}.\setlength\extrarowheight{8pt}
\begin{table}[H]
\begin{centering}
\begin{tabular}{c|>{\centering}p{3.5cm}|>{\centering}p{3.5cm}|>{\centering}p{3.5cm}|}
\multicolumn{1}{c}{} & \multicolumn{1}{>{\centering}p{3.5cm}}{{\footnotesize Dynamic variances \newline(Realized GARCH)}} & \multicolumn{1}{>{\centering}p{3.5cm}}{{\footnotesize Dynamic conditional correlations}} & \multicolumn{1}{>{\centering}p{3.5cm}}{{\footnotesize Realized measures of correlations}}\tabularnewline
\cline{2-4}
{\small CCC$^{+}$} & \textcolor{blue}{\cmark} & \textcolor{red}{\xmark} & \textcolor{red}{\xmark}\tabularnewline
\cline{2-4}
{\small DCC$^{+}$} & \textcolor{blue}{\cmark} & \textcolor{blue}{\cmark} & \textcolor{red}{\xmark}\tabularnewline
\cline{2-4}
{\small MRG} & \textcolor{blue}{\cmark} & \textcolor{blue}{\cmark} & \textcolor{blue}{\cmark}\tabularnewline
\cline{2-4}
\end{tabular}
\par\end{centering}
\caption{\footnotesize  The three types of models used in the empirical evaluation
and comparison. Identical univariate Realized GARCH models are used
to model the individual return series, while the models differ in
how the conditional correlation matrices, $C_{t}$, are modeled. Each
model is estimated with three different specifications for $C_{t}$:
Equi, Block, and Full. \label{tab:models}}
\end{table}
\setlength\extrarowheight{0pt}

\subsection{Estimation of Benchmark Models}

Estimation of the CCC$^{+}$ with constant correlations simply amounts
to maximum likelihood estimation of the correlation matrix for $z_{t}$.
For the Full specification this is simply the sample correlation matrix,
and the estimation of Equi and Block specifications is simple using
the results in \citet{ArchakovHansen:CanonicalBlockMatrix}. The DCC
model with equicorrelation and block-correlation matrices were studied
in \citet{EngleKelly:2012}, and we estimate Equi and Block variants
of the DCC$^{+}$ models using the method described in \citet[section 2.1]{EngleKelly:2012}.
The dynamic correlation part of the DCC$^{+}$-Full model is estimated
as a standard DCC model, see \citet{Aielli:2013}.

The estimated CCC$^{+}$, DCC$^{+}$, and MRG models are evaluated
and compared out-of-sample using log-likelihood criteria and in terms
of performance for portfolio construction with a minimum-variance
objective.

\subsection{Out-of-Sample Estimation Scheme}

Of the 19 years of daily data, we use the last nine years (2,226 trading
days), from January 3rd, 2012 to December 31st, 2020, for out-of-sample
evaluation (testing sample). We estimate the models using (in-sample)
data from January 2nd, 2002, to December 30th, 2011 (2,518 trading
days), and evaluate and compare the estimated models with data from
2012. By the end of each calendar year, we update the model estimates,
such that model evaluation is based on the most recent ten calendar
years of data. For instance, out-of-sample comparisons during 2013
are based on models that were estimated with data from the calendar
years, 2003 to 2012, and out-of-sample comparisons during 2020 are
based on model-specifications that were estimated with data from the
ten years from 2010 to 2019.
\begin{figure}[H]
\begin{centering}
\includegraphics[scale=0.75]{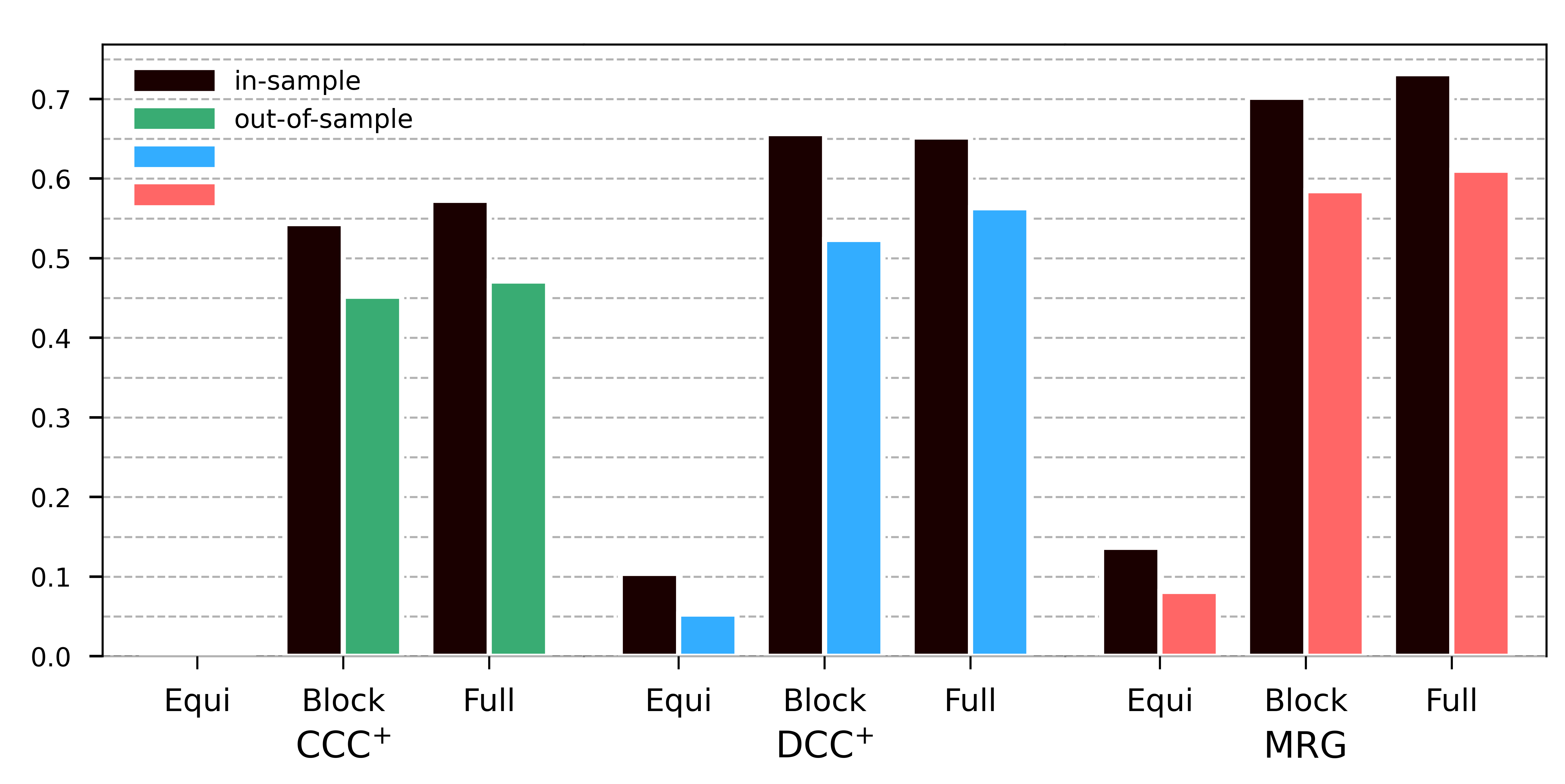}
\par\end{centering}
\caption{\footnotesize  Daily averages of in-sample (black bars) and out-of-sample
(colored bars) log-likelihoods for returns relative to CCC$^{+}$-Equi.
The in-sample period is from January 2, 2002 to December 30, 2011
(2,496 trading days) and the out-of-sample period is from January
3, 2012 to December 31, 2020 (2,248 trading days).\label{fig:Partial-log-lik-barplot}}
\end{figure}

\subsection{Log-Likelihood Analysis for Returns}

Multivariate GARCH models aim to describe the conditional distribution
of the vector of returns, which can be quantified with the log-likelihood
function for the vectors of returns. So, one natural way to compare
the different models and specifications is in terms of their log-likelihoods
for returns, $\ell_{r}$. In this subsection, we evaluate and compare
the specifications in terms of their in-sample and out-of-sample log-likelihoods
for returns, $\ell_{r,t}(\hat{\theta})$. Here $\hat{\theta}$ denotes
the parameter estimates obtained from in-sample data. This comparison
amounts to one-day-ahead density forecasting of the return vector
with the predictive log-likelihood as a gain function, see \citet{Amisano2007},
\citet{GewekeAmisano:2010}, and references therein.

We compute the average in-sample and average out-of-sample return
log-likelihoods, for each of the nine model-specifications. These
are presented in Figure \ref{fig:Partial-log-lik-barplot} with a
bar chart that reports the average log-likelihoods relative to the
simplest model-specification, CCC$^{+}$-Equi, which has the smallest
log-likelihood, both in-sample and out-of-sample. Good in-sample fit
is no guarantee of good out-of-sample fit, but in this application
we find that the relative rankings of model-specifications is largely
preserved. The best model-specification in-sample, MRG-Full, is also
the best model-specification out-of-sample. The three model-specifications
with the worst in-sample fit all use equicorrelation structures, and
these are also the three model-specifications with the worst out-of-sample
fit. This is evidence that equicorrelation structures are too restrictive
in this application. The block specifications are substantially better,
but the full specifications have the best out-of-sample performance
for all three model types. The dynamic correlation models, DCC$^{+}$
and MRG, clearly dominate the static CCC$^{+}$ model, with the MRG
model having the best performance across all specifications for $C_{t}$.
Once again, these results are found to be true in-sample as well as
out-of-sample.
\begin{table}[H]
\begin{centering}
\medskip{}
\par\end{centering}
\begin{centering}

\scriptsize
\begin{tabularx}{\textwidth}{>{\hsize=1.54\hsize}X>{\hsize=0.94\hsize}Y>{\hsize=0.94\hsize}Y>{\hsize=0.94\hsize}Y>{\hsize=0.94\hsize}Y>{\hsize=0.94\hsize}Y>{\hsize=0.94\hsize}Y>{\hsize=0.94\hsize}Y>{\hsize=0.94\hsize}Y>{\hsize=0.94\hsize}Y}
\\[-1.0cm]
\toprule
\midrule
  & \multicolumn{3}{c}{CCC$^+$} & \multicolumn{3}{c}{DCC$^+$} & \multicolumn{3}{c}{MRG}  \\
\cmidrule(l){2-4}
\cmidrule(l){5-7}
\cmidrule(l){8-10}
 \multicolumn{1}{c}{Period} & \multicolumn{1}{c}{Equi} & \multicolumn{1}{c}{Block} & \multicolumn{1}{c}{Free} & \multicolumn{1}{c}{Equi} & \multicolumn{1}{c}{Block} & \multicolumn{1}{c}{Free} & \multicolumn{1}{c}{Equi} & \multicolumn{1}{c}{Block} & \multicolumn{1}{c}{Free}  \\
\midrule
\\[-0.1cm]
\\[-0.1cm]
 In-sample & 0  & 0.544  & 0.573  & 0.104  & 0.657  & 0.652  & 0.137  & 0.702  & 0.732 \\[0.0cm]
 \tiny [2002-2011] &  &  &  &  &  &  &  &  &  \\[0.1cm]
 Out-of-sample & 0  & 0.452  & 0.471  & 0.053  & 0.524  & 0.563  & 0.082  & \textbf{0.585}   & \textbf{0.611}  \\[0.0cm]
  \tiny [2012-2020] & {\tiny (0.000)} & {\tiny (0.000)} & {\tiny (0.000)} & {\tiny (0.000)} & {\tiny (0.000)} & {\tiny (0.024)} & {\tiny (0.000)} & {\tiny (0.135)} & {\tiny (1.000)} \\[0.1cm]
\\[0.0cm]
 2012 & 0  & \textbf{0.413}   & \textbf{0.419}   & 0.021  & \textbf{0.425}   & \textbf{0.460}   & 0.014  & \textbf{0.427}   & \textbf{0.470}  \\[0.0cm]
  & {\tiny (0.000)} & {\tiny (0.246)} & {\tiny (0.246)} & {\tiny (0.000)} & {\tiny (0.246)} & {\tiny (0.760)} & {\tiny (0.000)} & {\tiny (0.246)} & {\tiny (1.000)} \\[0.1cm]
 2013 & 0  & 0.131  & 0.172  & -0.015  & 0.226  & 0.265  & 0.028  & \textbf{0.276}   & \textbf{0.309}  \\[0.0cm]
  & {\tiny (0.001)} & {\tiny (0.001)} & {\tiny (0.001)} & {\tiny (0.001)} & {\tiny (0.011)} & {\tiny (0.011)} & {\tiny (0.001)} & {\tiny (0.292)} & {\tiny (1.000)} \\[0.1cm]
 2014 & 0  & 0.360  & \textbf{0.384}   & 0.029  & \textbf{0.398}   & \textbf{0.391}   & 0.065  & \textbf{0.449}   & \textbf{0.402}  \\[0.0cm]
  & {\tiny (0.010)} & {\tiny (0.049)} & {\tiny (0.180)} & {\tiny (0.010)} & {\tiny (0.180)} & {\tiny (0.180)} & {\tiny (0.027)} & {\tiny (1.000)} & {\tiny (0.180)} \\[0.1cm]
 2015 & 0  & 0.574  & 0.593  & -0.006  & 0.570  & \textbf{0.717}   & 0.024  & 0.556  & \textbf{0.678}  \\[0.0cm]
  & {\tiny (0.000)} & {\tiny (0.004)} & {\tiny (0.004)} & {\tiny (0.000)} & {\tiny (0.001)} & {\tiny (1.000)} & {\tiny (0.000)} & {\tiny (0.004)} & {\tiny (0.374)} \\[0.1cm]
 2016 & 0  & \textbf{0.405}   & \textbf{0.452}   & 0.133  & \textbf{0.406}   & \textbf{0.464}   & 0.179  & \textbf{0.571}   & \textbf{0.574}  \\[0.0cm]
  & {\tiny (0.002)} & {\tiny (0.110)} & {\tiny (0.445)} & {\tiny (0.003)} & {\tiny (0.445)} & {\tiny (0.445)} & {\tiny (0.003)} & {\tiny (0.969)} & {\tiny (1.000)} \\[0.1cm]
 2017 & 0  & 0.440  & 0.398  & 0.267  & \textbf{0.693}   & \textbf{0.628}   & 0.310  & \textbf{0.758}   & \textbf{0.753}  \\[0.0cm]
  & {\tiny (0.000)} & {\tiny (0.021)} & {\tiny (0.000)} & {\tiny (0.000)} & {\tiny (0.336)} & {\tiny (0.110)} & {\tiny (0.000)} & {\tiny (1.000)} & {\tiny (0.847)} \\[0.1cm]
 2018 & 0  & \textbf{0.647}   & \textbf{0.633}   & 0.096  & \textbf{0.723}   & \textbf{0.691}   & 0.084  & \textbf{0.768}   & \textbf{0.739}  \\[0.0cm]
  & {\tiny (0.000)} & {\tiny (0.126)} & {\tiny (0.114)} & {\tiny (0.000)} & {\tiny (0.432)} & {\tiny (0.252)} & {\tiny (0.000)} & {\tiny (1.000)} & {\tiny (0.432)} \\[0.1cm]
 2019 & 0  & 0.298  & 0.300  & 0.082  & \textbf{0.442}   & \textbf{0.418}   & \textbf{0.170}   & \textbf{0.514}   & \textbf{0.516}  \\[0.0cm]
  & {\tiny (0.013)} & {\tiny (0.013)} & {\tiny (0.013)} & {\tiny (0.013)} & {\tiny (0.333)} & {\tiny (0.302)} & {\tiny (0.224)} & {\tiny (0.983)} & {\tiny (1.000)} \\[0.1cm]
 2020 & 0  & 0.798  & 0.886  & -0.128  & 0.827  & \textbf{1.031}   & -0.136  & \textbf{0.939}   & \textbf{1.048}  \\[0.0cm]
  & {\tiny (0.000)} & {\tiny (0.000)} & {\tiny (0.004)} & {\tiny (0.000)} & {\tiny (0.004)} & {\tiny (0.696)} & {\tiny (0.000)} & {\tiny (0.529)} & {\tiny (1.000)} \\[0.1cm] \\
\\[-0.5cm]
\midrule
\bottomrule
\end{tabularx}
\par\end{centering}
\caption{\label{tab:Partial-log-likelihood}\footnotesize  Average in-sample
and out-of-sample partial log-likelihoods for each of the nine model-specifications,
measured relative to CCC$^{+}$-Equi. MCS $p$-values are in parentheses
and bold font identifies the model-specifications in the 95\% MCS.
In-sample statistics are based on the period from January 2, 2002,
to December 30, 2011 (2,496 trading days), and the out-of-sample period
is from January 3, 2012 to December 31, 2020 (2,248 trading days).}
\end{table}

Table \ref{tab:Partial-log-likelihood} has additional details about
the model comparisons and the statistical significance of these. The
first two rows of Table \ref{tab:Partial-log-likelihood} report the
numerical values for the bar plot in Figure \ref{fig:Partial-log-lik-barplot},
and the subsequent rows have the analogous out-of-sample values for
each out-of-sample calendar year. Below each of the out-of-sample
statistics, we report the corresponding model confidence set (MCS)
$p$-values, see \citet{HansenLundeNasonMCS}. A small MCS $p$-value
is evidence that the model-specification is inferior to other model-specifications
in the comparison, and we use a bold font to identify the model-specifications
that are included in the 95\% MCS.

The best out-of-sample performance is achieved with MRG-Full, and
all alternative model-specifications, except for MRG-Block, are significantly
worse over the entire out-of-sample period. The MCSs for the calendar
year tend to include a larger number of model-specifications, which
is to expected because the shorter samples offers less information
to discriminate between the competitors. MRG-Full is the only model-specification
that is included in every MCS. We note that DCC$^{+}$-Block performs
particularly well in 2015, but this could be a statistical anomaly,
since this model-specification is inferior to MRG-Full in all other
years.

The last calendar year, 2020, was a special year with the turbulence
surrounding the outbreak of COVID-19 and the subsequent stock market
rally. This is also the year where heterogeneous correlations were
most beneficial, as evident by the large gap between equicorrelation
specifications and more flexible specifications. This likely reflects
the heterogeneous impact that the pandemic had on different sectors.

\subsection{Global Minimum-Variance Portfolio}

Next, we compare the model-specifications in terms of their ability
to produce a low-variance portfolio out-of-sample. At each point in
time, we compute the one-period-ahead optimal minimum-variance portfolio
weights, as defined by the predicted covariance structure for each
model-specification. 

Let $H_{(m),t}$ denote the conditional covariance matrix for $r_{t}$,
as predicted by the $m$-th model-specification at time $t-1$. We
deduce the corresponding global minimum-variance (GMV) portfolio by
solving
\[
\begin{array}{ccc}
\min_{\omega_{t}\in\mathbb{R}^{n}}\omega_{t}^{\prime}H_{(m),t}\omega_{t}, &  & \text{s.t. }\omega_{t}'\iota=1,\end{array}
\]
where $\iota=(1,\ldots,1)^{\prime}$ is a $n$-dimensional vector
of ones. In the absence of leverage constraints, such as no-shortening
constraints, the well-known solution is: 
\[
\omega_{(m),t}^{*}=\frac{H_{(m),t}^{-1}\iota}{\iota^{\prime}H_{(m),t}^{-1}\iota},
\]
and the resulting portfolio returns are given by
\[
R_{(m),t}^{\mathrm{mv}}=\omega_{(m),t}^{*\prime}r_{t}.
\]
Different model-specifications yield different $H_{(m),t}$, $m=1,\ldots,9$,
and the resulting portfolio weights, returns, and returns-variances,
will therefore be different. To this comparison, we also add the simple
equal-weighted portfolio as another benchmark portfolio. The returns
of the equal-weighted portfolio are given by 
\[
R_{t}^{\mathrm{ew}}=\frac{1}{n}\iota^{\prime}r_{t},
\]
such that each asset is weighted by $1/n$, where $n=9$ in this application.
\begin{table}[H]
\begin{centering}

\scriptsize
\begin{tabularx}{\textwidth}{>{\hsize=1.8\hsize}X>{\hsize=0.92\hsize}Y>{\hsize=0.92\hsize}Y>{\hsize=0.92\hsize}Y>{\hsize=0.92\hsize}Y>{\hsize=0.92\hsize}Y>{\hsize=0.92\hsize}Y>{\hsize=0.92\hsize}Y>{\hsize=0.92\hsize}Y>{\hsize=0.92\hsize}Y>{\hsize=0.92\hsize}Y}
\\
\toprule
\midrule
  & \multicolumn{1}{c}{Equal} & \multicolumn{3}{c}{CCC$^+$} & \multicolumn{3}{c}{DCC$^+$} & \multicolumn{3}{c}{MRG}  \\
\cmidrule(l){3-5}
\cmidrule(l){6-8}
\cmidrule(l){9-11}
 \multicolumn{1}{c}{Period} & \multicolumn{1}{c}{weights} & \multicolumn{1}{c}{Equi} & \multicolumn{1}{c}{Block} & \multicolumn{1}{c}{Free} & \multicolumn{1}{c}{Equi} & \multicolumn{1}{c}{Block} & \multicolumn{1}{c}{Free} & \multicolumn{1}{c}{Equi} & \multicolumn{1}{c}{Block} & \multicolumn{1}{c}{Free}  \\
\midrule
\\[-0.1cm]
\\[-0.1cm]
 In-sample  & 0.212  & 0.158  & 0.160  & 0.160  & 0.165  & 0.166  & 0.158  & 0.163  & 0.155  & 0.157 \\[0.0cm]
 \tiny [2002-2011] \\[0.1cm]
 Out-of-sample  & 0.247  & 0.189  & 0.188  & 0.192  & 0.184  & 0.181  & 0.189  & 0.183  & \textbf{0.176}   & 0.186 \\[0.0cm]
  \tiny [2012-2020] & {\tiny (0.000)} & {\tiny (0.002)} & {\tiny (0.002)} & {\tiny (0.001)} & {\tiny (0.002)} & {\tiny (0.024)} & {\tiny (0.023)} & {\tiny (0.024)} & {\tiny (1.000)} & {\tiny (0.024)} \\[0.1cm]
\\[0.0cm]
 2012  & 0.157  & \textbf{0.105}   & \textbf{0.103}   & \textbf{0.107}   & \textbf{0.105}   & \textbf{0.103}   & \textbf{0.107}   & \textbf{0.103}   & \textbf{0.102}   & \textbf{0.107}  \\[0.0cm]
  & {\tiny (0.000)} & {\tiny (0.297)} & {\tiny (0.414)} & {\tiny (0.297)} & {\tiny (0.256)} & {\tiny (0.297)} & {\tiny (0.077)} & {\tiny (0.667)} & {\tiny (1.000)} & {\tiny (0.077)} \\[0.1cm]
 2013  & 0.125  & \textbf{0.111}   & \textbf{0.109}   & \textbf{0.109}   & \textbf{0.110}   & \textbf{0.109}   & \textbf{0.108}   & \textbf{0.109}   & \textbf{0.109}   & \textbf{0.109}  \\[0.0cm]
  & {\tiny (0.013)} & {\tiny (0.843)} & {\tiny (0.981)} & {\tiny (0.905)} & {\tiny (0.843)} & {\tiny (0.981)} & {\tiny (1.000)} & {\tiny (0.981)} & {\tiny (0.981)} & {\tiny (0.981)} \\[0.1cm]
 2014  & 0.137  & \textbf{0.119}   & \textbf{0.120}   & \textbf{0.119}   & \textbf{0.118}   & \textbf{0.119}   & \textbf{0.119}   & \textbf{0.117}   & \textbf{0.118}   & \textbf{0.119}  \\[0.0cm]
  & {\tiny (0.012)} & {\tiny (0.452)} & {\tiny (0.452)} & {\tiny (0.559)} & {\tiny (0.467)} & {\tiny (0.559)} & {\tiny (0.559)} & {\tiny (0.559)} & {\tiny (1.000)} & {\tiny (0.467)} \\[0.1cm]
 2015  & 0.203  & \textbf{0.164}   & \textbf{0.160}   & \textbf{0.160}   & 0.167  & \textbf{0.164}   & \textbf{0.164}   & \textbf{0.165}   & \textbf{0.160}   & \textbf{0.163}  \\[0.0cm]
  & {\tiny (0.002)} & {\tiny (0.054)} & {\tiny (0.470)} & {\tiny (0.470)} & {\tiny (0.040)} & {\tiny (0.054)} & {\tiny (0.050)} & {\tiny (0.470)} & {\tiny (1.000)} & {\tiny (0.050)} \\[0.1cm]
 2016  & 0.197  & \textbf{0.124}   & \textbf{0.136}   & \textbf{0.131}   & \textbf{0.129}   & \textbf{0.134}   & \textbf{0.129}   & \textbf{0.130}   & \textbf{0.134}   & \textbf{0.123}  \\[0.0cm]
  & {\tiny (0.000)} & {\tiny (0.521)} & {\tiny (0.107)} & {\tiny (0.114)} & {\tiny (0.114)} & {\tiny (0.114)} & {\tiny (0.114)} & {\tiny (0.114)} & {\tiny (0.114)} & {\tiny (1.000)} \\[0.1cm]
 2017  & 0.094  & 0.080  & 0.081  & 0.079  & 0.073  & \textbf{0.073}   & \textbf{0.074}   & \textbf{0.073}   & \textbf{0.070}   & \textbf{0.072}  \\[0.0cm]
  & {\tiny (0.000)} & {\tiny (0.000)} & {\tiny (0.000)} & {\tiny (0.000)} & {\tiny (0.045)} & {\tiny (0.144)} & {\tiny (0.144)} & {\tiny (0.124)} & {\tiny (1.000)} & {\tiny (0.144)} \\[0.1cm]
 2018  & 0.198  & 0.165  & \textbf{0.158}   & \textbf{0.163}   & 0.166  & \textbf{0.157}   & \textbf{0.163}   & 0.166  & \textbf{0.155}   & 0.168 \\[0.0cm]
  & {\tiny (0.000)} & {\tiny (0.002)} & {\tiny (0.143)} & {\tiny (0.136)} & {\tiny (0.002)} & {\tiny (0.486)} & {\tiny (0.079)} & {\tiny (0.002)} & {\tiny (1.000)} & {\tiny (0.002)} \\[0.1cm]
 2019  & 0.164  & \textbf{0.118}   & 0.123  & 0.130  & 0.119  & \textbf{0.120}   & 0.125  & \textbf{0.114}   & \textbf{0.116}   & 0.125 \\[0.0cm]
  & {\tiny (0.000)} & {\tiny (0.328)} & {\tiny (0.011)} & {\tiny (0.011)} & {\tiny (0.011)} & {\tiny (0.252)} & {\tiny (0.011)} & {\tiny (1.000)} & {\tiny (0.328)} & {\tiny (0.011)} \\[0.1cm]
 2020  & 0.435  & \textbf{0.308}   & \textbf{0.317}   & \textbf{0.316}   & \textbf{0.341}   & \textbf{0.348}   & \textbf{0.308}   & \textbf{0.336}   & \textbf{0.307}   & \textbf{0.306}  \\[0.0cm]
  & {\tiny (0.000)} & {\tiny (0.833)} & {\tiny (0.833)} & {\tiny (0.416)} & {\tiny (0.353)} & {\tiny (0.458)} & {\tiny (0.833)} & {\tiny (0.416)} & {\tiny (1.000)} & {\tiny (0.833)} \\[0.1cm] \\
\\[-0.7cm]
\midrule
\bottomrule
\end{tabularx}
\par\end{centering}
\caption{\footnotesize  The annualized volatilities for the equal-weighted
portfolio (left column) and for the GMV portfolios implied by the
predictions of the nine competing model-specifications. Numbers in
parentheses are MCS $p$-values based on absolute portfolio returns,
$|R_{t}^{\mathrm{ew}}|$ and $|R_{(j),t}^{\mathrm{mv}}|$, for $j=1,...,9$.
Bold font identifies those included in the 95\% MCS. The in-sample
period is January 2, 2002, to December 30, 2011 (2,496 trading days)
and the out-of-sample period is January 3, 2012 to December 31, 2020
(2,248 trading days). We also include the corresponding results for
each calendar year in the out-of-sample period. \label{tab:gmvp}}
\end{table}

We compare the empirical variances of the ten portfolios, both in-sample
and out-of-sample, and we report the results in units of annualized
volatilities in Table \ref{tab:gmvp}. We also report the analogous
results for each of the nine years in the out-of-sample period. 

A result that stands out from Table \ref{tab:gmvp}, is that the equal-weighted
portfolio is inferior to all nine model-specifications. This is not
too surprising, because the equal-weighted portfolio does not utilize
information about the covariance structure. The model-based portfolios
benefit from modeling $H_{t}$, which reduce the portfolio variance
by as much as 50\%, which translates to a reduction in annualized
volatility by a factor of about $\sqrt{2}$. 

The MRG-based specifications have the smallest variances, with MRG-Block
being the best model-specification in this application -- both in-sample
and out-of-sample. MRG-Block reduces the annualized volatility by
50 to 150 basis points, relative to other model-based portfolios.\footnote{In practice, one would also want to account for portfolio turnover,
because transaction costs may offset the gain from the reduction in
variance.} The parsimonious structure of MRG-Block helps reduce the portfolio
variance and MRG-Block is the only model-specification that is included
in all MCSs. The MCSs for the individual calendar years are less informative,
because there is not enough information to discriminate between competitors,
with the exception of the equal-weighted portfolio, which is not included
in any MCS.

\begin{figure}[H]
\begin{centering}
\includegraphics[scale=0.65]{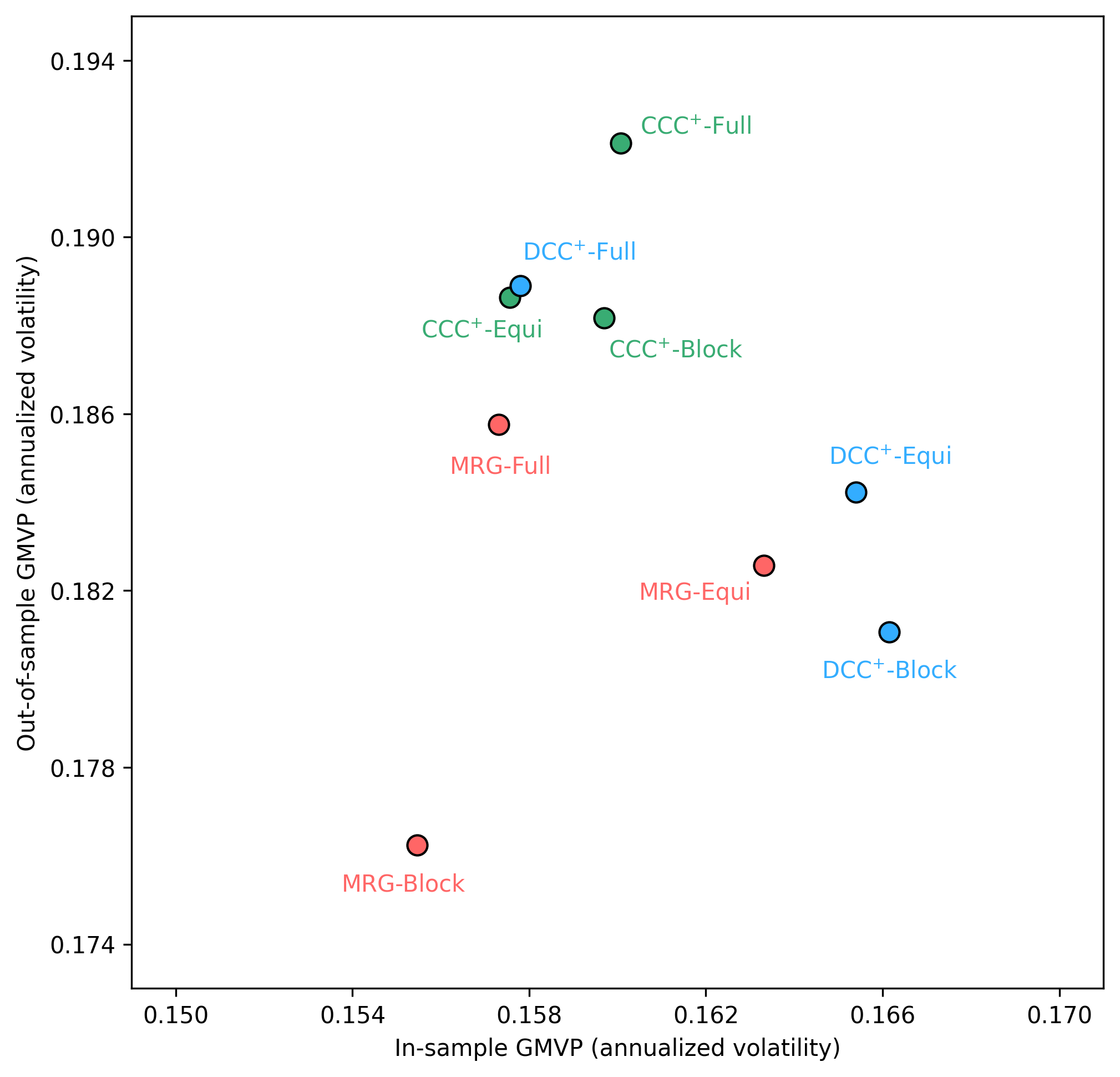}
\par\end{centering}
\caption{\footnotesize  Annualized volatility of GMV portfolios, where the
out-of-sample results (vertical axis) are plotted against the in-sample
results (horizontal axis), for each of the nine model-specifications.
The in-sample period is January 2, 2002, to December 30, 2011 (2,496
trading days) and the out-of-sample period is January 3, 2012 to December
31, 2020 (2,248 trading days).\label{fig:gmvp_vol}}
\end{figure}

Figure \ref{fig:gmvp_vol} is a graphical illustration of the in-sample
and out-of-sample performance of the nine model-specifications. The
equal-weighted portfolio cannot be seen because its very inferior
performance places it outside the range shown in Figure \ref{fig:gmvp_vol}.
It is easy to see that the CCC$^{+}$ specifications (green dots)
have the worst performance among the model-based portfolios. Interestingly,
the most flexible specification, Full, have the worst out-of-sample
portfolio performance for all types of models, which is evidence that
the most flexible correlation specification, Full, suffers from overfitting
in this application.

The different outcomes in the two out-of-sample comparisons highlight
at important feature of model diagnostics. The ``best'' model-specification
depends on the intended use of the model, and the empirical objective
should be taken into account when applying model diagnostic tests,
see \citet{HansenDumitrescu:2022}. In the first comparison, we saw
that the Full specification has better in-sample and out-of-sample
performance in terms of the predicted log-likelihood. Whereas in the
second comparison, the Block specification was significantly better
for constructing minimum-variance portfolios.

\section{\label{sec:Conclusion-and-Outline}Conclusion and Discussion}

In this paper, we have introduced a novel framework for multivariate
GARCH modeling. Our key methodological contribution is the dynamic
modeling of the conditional correlation matrix $C_{t}\in\mathbb{R}^{n\times n}$
using a vector parametrization, $\gamma_{t}=\gamma(C_{t})$. Importantly,
this approach facilitates simple linear factor models of the correlation
structure, while utilizing realized measures of correlations. In its
most general form, the model does not impose any restrictions on $C_{t}$,
while it is guaranteed to produce a valid positive definite correlation
matrix. In many situations, it will be desirable to impose additional
structure, especially in high-dimensional system, and the factor model
for $\gamma_{t}$ serves this purpose by reducing the number of free
parameters and latent variables. 

Imposing block correlation structures is one way to reduce the number
of free parameters. We have shown that a block structure is equivalent
to a simple linear factor model, $\gamma_{t}=A\zeta_{t}\in\mathbb{R}^{d}$,
where $\zeta_{t}\in\mathbb{R}^{r}$ with $r<d$ and $A$ is a matrix
defined by the block structure in $C_{t}$. While the block structure
is a useful example, the linear factor structure is not specific to
block correlation specifications. Other choices for $A$ can be entertained,
and it will be interesting to explore data-dependent choices for $A$
and non-linear factor models, $\gamma_{t}=\varrho(\zeta_{t})$. These
are topics we leave for future research.

The MRG model can be seen as the natural generalization of \citet{HansenLundeVoev:2014}
to higher dimensions, because $\gamma(C)$ is identical to the Fisher
transformed correlation when $n=2$, and the Fisher transformation
was the transformation used in the bivariate structure proposed in
\citet{HansenLundeVoev:2014}.

We have applied the MRG model to nine assets from three sectors, and
we used three different specifications for the correlation matrix.
We compared the MRG model to CCC and DCC style models and found the
MRG model to be superior in terms of the predicted likelihood of returns
and in terms of portfolio construction with a minimum variance objective.
For all three types of models, MRG, CCC, and DCC, we also compared
different structures on the conditional correlation matrix, Equi,
Block, and Full. The equicorrelation structure was found to be too
restrictive, with a performance that was uniformly inferior to more
flexible structures. For the portfolio construction problem, the sector-based
block correlation structure had the best out-of-sample performance,
and a close second to the Full specification in terms of predictive
log-likelihood. Conveniently, the Block specification, with a fixed
number of blocks, is easy to scale to high dimensional applications. 

The structure we have developed is also very beneficial for applications
of the Block-DECO model by \citet{EngleKelly:2012}. Specifically,
the simplified expressions for the inverse block correlation matrix
and its determinant by \citet{ArchakovHansen:CanonicalBlockMatrix}
make it straightforward to evaluate the likelihood function, thus
eliminating the need to resort to alternative estimation methods when
$K>2$.

We have established a wide range of attractive computational and empirical
features of the MRG framework, but several theoretical properties
remain unresolved. It would, for instance, be desirable to establish
conditions that ensure stationarity, ergodicity, and the existence
of moments for time series whose data generating process is MRG. Similarly,
it would also be desirable to establish sufficient conditions that
ensure the likelihood-based estimators are asymptotically normally
distributed. Some results are available for univariate Realized GARCH
model, see \citet{HansenHuangShek:2012}. Moreover, \citet{BougerolPicard:1992},
\citet{CarrascoChen:2002}, \citet{JensenRahbek:2004}, \citet{StraumannMikosch:06},
\citet{MeitzSaikkonen:2008}, \citet{Kristensen:2009}, and \citet[section 10]{FrancqZakoian2019},
offer ideas for establishing these results for the MRG model in future
research.

\appendix

\section{Analytical Derivatives\label{sec:app_estimation}}

\setcounter{equation}{0}\renewcommand{\theequation}{A.\arabic{equation}}

The model estimation outlined in Section 3 is based on maximization
of the log-likelihood function, and this can be done using various
numerical optimization algorithms. Gradient-based algorithms (e.g.,
Newton type) are significantly faster if analytical expressions for
the gradient vector and the corresponding Hessian matrix are available.
In this appendix, we provide analytical expressions for the derivatives
of the partial log-likelihood component maximized in the second stage
of the two-stage estimation procedure (see \eqref{eq:logLcorrelationmatrix}
in Section \ref{subsec:Two-Stage-Estimation}). The first stage estimation
is computationally simple, because it only involves univariate Realized
GARCH models. 

\subsection{Full Model Estimation (Second Stage, $\vartheta_{2}$)\label{subsec:Full-Model-Estimation}}

\subsubsection{Notation }

Let $K_{n}$ be the commutation matrix, characterized by $K_{n}\,\text{vec}M=\text{vec}M^{\prime}$
for any $M\in\mathbb{R}^{n\times n}$, and $E_{l}$, $E_{u}$, and
$E_{d}$ are elimination matrices defined by $\text{vecl}M=E_{l}\,\text{vec}M$,
$\text{vecl}M^{\prime}=E_{u}\,\text{vec}M$, and $\mathrm{diag}M=E_{d}\,\text{vec}M$,
respectively. The identity matrix of size $m\times m$ is denoted
by $I_{m}$ and we use $\otimes$ to represent the Kronecker product.

\subsubsection{Preliminaries\label{subsec:PrelimA}}

Any correlation matrix, $C\in\mathbb{R}^{n\times n}$, can be expressed
as, $C=Q\Lambda Q^{\prime}$, where $Q^{\prime}Q=I_{n}$ and $\Lambda=\text{diag}(\lambda_{1},\lambda_{2},...,\lambda_{n})\in\mathbb{R}^{n\times n}$
is a diagonal matrix with the eigenvalues of $C$. The matrix logarithm
of $C$ is well-defined if $C$ is positive definite, in which case
we write $G=\log C$, where $\log C=Q\tilde{\Lambda}Q^{\prime}$ with
$\tilde{\Lambda}=\mathrm{diag}(\log\lambda_{1},\ldots,\lambda_{n})$.
Note that $G=G(\gamma)$ and $\gamma=\text{vecl}\log C=\text{vecl}G(\gamma)$. 

The Jacobian matrix of the vector of correlations with respect to
$\gamma$ is 
\begin{equation}
J=\frac{\partial\text{vecl}C}{\partial\gamma^{\prime}}=E_{l}\Bigl(I_{d}-J_{G}E_{d}^{\prime}\Bigl(E_{d}J_{G}E_{d}\Bigl)^{\prime}E_{d}\Bigl)J_{G}(E_{l}+E_{u})^{\prime},\label{eq:dCdgamma}
\end{equation}
where $J_{G}=\frac{\partial\text{vec}C}{(\partial\text{vec}G)^{\prime}}$,
see \citet[proposition 3]{ArchakovHansen:Correlation}, and from \citet{LintonMcCrorie:1995}
we have the expression, $J_{G}=(Q\otimes Q)\Xi\bigl(Q\otimes Q\bigl)^{\prime}$,
where $\Xi$ is an $n^{2}\times n^{2}$ diagonal matrix with diagonal
elements
\[
\Xi_{(i-1)n+j,(i-1)n+j}=\begin{cases}
\lambda_{i}, & \text{if}\qquad\lambda_{i}=\lambda_{j},\\
\tfrac{\lambda_{i}-\lambda_{j}}{\log\lambda_{i}-\log\lambda_{j}}, & \text{if}\qquad\lambda_{i}\neq\lambda_{j},
\end{cases}
\]
for $i=1,\ldots,n$ and $j=1,\ldots,n$. 

\subsubsection{Gradient Vector\label{subsec:GradientA}}

In the second stage estimation of the unrestricted MRG model, we maximize
\begin{equation}
\ell=-\frac{1}{2}\sum_{t=1}^{T}\underbrace{\Bigl\{\log\det C_{t}+z_{t}^{\prime}C_{t}^{-1}z_{t}\Bigl\}}_{-2\ell_{c,t}}-\frac{1}{2}\sum_{t=1}^{T}\underbrace{\Bigl\{\log\det\tilde{\Omega}+\tilde{v}_{t}^{\prime}\tilde{\Omega}^{-1}\tilde{v}_{t}\Bigl\}}_{-2\ell_{y,t}},\label{eq:log-lik-full}
\end{equation}
with respect to the vector of parameters, $\vartheta_{2}=(\tilde{\omega}^{\prime},\tilde{\beta}^{\prime},\tilde{\alpha}^{\prime},\tilde{\xi}^{\prime},\tilde{\varphi}^{\prime})^{\prime}$.
It is convenient to write $\vartheta_{2}=(\eta^{\prime},\phi^{\prime})^{\prime}$,
where $\eta=(\tilde{\omega}^{\prime},\tilde{\beta}^{\prime},\tilde{\alpha}^{\prime})^{\prime}$
includes the parameters from the GARCH equations and $\phi=(\tilde{\xi}^{\prime},\tilde{\varphi}^{\prime})^{\prime}$
has the parameters from the measurement equations. (The estimate of
the covariance matrix, $\tilde{\Omega}=\mathrm{var}(\tilde{v}_{t})$,
is computed straightforwardly from the residuals of $\tilde{v}_{t}$.)

The gradients of $\ell$ with respect to these two parameter vectors
are
\[
\frac{\partial\ell}{\partial\eta^{\prime}}=\sum_{t=1}^{T}\Biggl(\frac{\partial\ell_{c,t}}{\partial(\text{vec}C_{t})^{\prime}}\cdot\frac{\partial\text{vec}C_{t}}{\partial\gamma_{t}^{\prime}}+\frac{\partial\ell_{y,t}}{\partial\tilde{v}_{t}^{\prime}}\cdot\frac{\partial\tilde{v}_{t}}{\partial\gamma_{t}^{\prime}}\Biggl)\cdot\frac{\partial\gamma_{t}}{\partial\eta^{\prime}},\qquad\text{and}\qquad\frac{\partial\ell}{\partial\phi^{\prime}}=\sum_{t=1}^{T}\frac{\partial\ell_{y,t}}{\partial\tilde{v}_{t}^{\prime}}\cdot\frac{\partial\tilde{v}_{t}}{\partial\phi^{\prime}},
\]
respectively, and we have
\[
-2\mathrm{d}\ell_{c,t}=\text{tr}\bigl(C_{t}^{-1}\mathrm{d}C_{t}\bigl)+\text{tr}\bigl(\mathrm{d}C_{t}^{-1}z_{t}z_{t}^{\prime}\bigl)=\bigl(\text{vec}C_{t}^{-1}\bigl)^{\prime}\text{vec}\mathrm{d}C_{t}-\text{vec}\Bigl(C_{t}^{-1}z_{t}z_{t}^{\prime}C_{t}^{-1}\Bigl)^{\prime}\text{vec}(\mathrm{d}C_{t}),
\]
and $-2\mathrm{d}\ell_{y,t}=\mathrm{d}(\tilde{v}_{t}^{\prime}\tilde{\Omega}^{-1}\tilde{v}_{t})=2\tilde{v}_{t}^{\prime}\tilde{\Omega}^{-1}\mathrm{d}\tilde{v}_{t}$,
such that $-2\frac{\partial\ell_{c,t}}{\partial(\text{vec}C_{t})^{\prime}}=g_{c,t}^{\prime}$,
where $g_{c,t}=\text{vec}\Bigl(C_{t}^{-1}-C_{t}^{-1}z_{t}z_{t}^{\prime}C_{t}^{-1}\Bigl)$,
and $-2\frac{\partial\ell_{y,t}}{\partial\tilde{v}_{t}^{\prime}}=2\tilde{v}_{t}^{\prime}\tilde{\Omega}^{-1}$.

We proceed to obtain the derivatives with respect to $\gamma_{t}$.
We have $\frac{\partial\text{vec}C_{t}}{\partial\gamma_{t}^{\prime}}=\frac{\partial\text{vec}C_{t}}{\partial(\text{vecl}C_{t})^{\prime}}\frac{\partial\text{vecl}C_{t}}{\partial\gamma_{t}^{\prime}}=(E_{l}+E_{u})^{\prime}J_{t}$,
where $J_{t}$ is defined in (\ref{eq:dCdgamma}) and $\frac{\partial\tilde{v}_{t}}{\partial\gamma_{t}^{\prime}}=-\Lambda_{\tilde{\varphi}}$
with $\Lambda_{\tilde{\varphi}}=\text{diag}(\tilde{\varphi}_{1},...,\tilde{\varphi}_{d})$.

Next, we derive the derivatives with respect to $\eta$ and $\phi$.
From the GARCH equations for $\gamma_{t}$, we have $\gamma_{t}=\tilde{F}_{t}\eta$,
where $\tilde{F}_{t}=\Bigl[I_{d},\:\Lambda_{\gamma_{t-1}},\:\Lambda_{y_{t-1}}\Bigl]$
is $d\times3d$ matrix, with $\Lambda_{\gamma_{t-1}}=\text{diag}(\gamma_{t-1})$
and $\Lambda_{y_{t-1}}=\text{diag}(y_{t-1})$. So, we have the recursions,
\[
\dot{\gamma}_{t}\equiv\frac{\partial\gamma_{t}}{\partial\eta^{\prime}}=\tilde{F}_{t}+\Lambda_{\tilde{\beta}}\dot{\gamma}_{t-1}\qquad\text{with }\dot{\gamma}_{0}=0,
\]
where $\Lambda_{\tilde{\beta}}=\text{diag}(\tilde{\beta}_{1},...,\tilde{\beta}_{d})$,
and it follows that $\dot{\gamma}_{t}=\sum_{j=0}^{t-1}\Lambda_{\tilde{\beta}}^{j}\tilde{F}_{t-j}$. 

From the measurement equations for $\gamma_{t}$ we have $\tilde{v}_{t}=y_{t}-\tilde{M}_{t}\phi$,
where $\tilde{M}_{t}=\Bigl[I_{d},\:\Lambda_{\gamma_{t}}\Bigl]$ and
$\Lambda_{\gamma_{t}}=\text{diag}(\gamma_{t})$. Such that $\dot{\tilde{v}}_{t}=-\tilde{M}_{t}$,
and $\frac{\partial\ell}{\partial\vartheta_{2}}=\Bigl(\frac{\partial\ell}{\partial\eta^{\prime}},\frac{\partial\ell}{\partial\phi^{\prime}}\Bigl)^{\prime}$,
where
\[
\frac{\partial\ell}{\partial\eta^{\prime}}=-\frac{1}{2}\sum_{t=1}^{T}\Biggl(g_{c,t}^{\prime}(E_{l}+E_{u})^{\prime}J_{t}+2\tilde{v}_{t}^{\prime}\tilde{\Omega}^{-1}\Lambda_{\tilde{\varphi}}\Biggl)\dot{\gamma}_{t},\qquad\text{and}\quad\frac{\partial\ell}{\partial\phi^{\prime}}=\sum_{t=1}^{T}\tilde{v}_{t}^{\prime}\tilde{\Omega}^{-1}\tilde{M}_{t}.
\]

\subsubsection{Approximation of Hessian Matrix\label{subsec:Approximation-of-Hessian_1}}

In the optimization algorithm, we approximate the Hessian matrix by
the Fisher information matrix,

\[
\mathbb{E}\left(\frac{\partial\ell}{\partial\vartheta_{2}}\frac{\partial\ell}{\partial\vartheta_{2}^{\prime}}\right)=-\left(\begin{array}{cc}
\mathbb{E}\frac{\partial\ell}{\partial\eta}\frac{\partial\ell}{\partial\eta^{\prime}} & \mathbb{E}\frac{\partial\ell}{\partial\eta}\frac{\partial\ell}{\partial\phi^{\prime}}\\
\mathbb{E}\frac{\partial\ell}{\partial\phi}\frac{\partial\ell}{\partial\eta^{\prime}} & \mathbb{E}\frac{\partial\ell}{\partial\phi}\frac{\partial\ell}{\partial\phi^{\prime}}
\end{array}\right).
\]
The two are not equal if the model is misspecified, but the approximating
Fisher information matrix can still be very beneficial for the optimization
algorithm.\footnote{In our Matlab implementation, we found it useful to run the first
first few iterations with a numerical Hessian (we used 25), before
using the analytical approximation for the Hessian.}

For the upper-left element of the Fisher Information matrix, $\mathbb{E}\frac{\partial\ell}{\partial\eta}\frac{\partial\ell}{\partial\eta^{\prime}}$
, we have
\[
\begin{aligned}g_{c,t}= & \text{vec}C_{t}^{-1}-\text{vec}[C_{t}^{-1}z_{t}z_{t}^{\prime}C_{t}^{-1}]=\text{vec}C_{t}^{-1}-\Bigl(C_{t}^{-1}\otimes I_{n}\Bigl)\text{vec}\bigl(C_{t}^{-1}z_{t}z_{t}^{\prime}\bigl),\end{aligned}
\]
such that
\[
\begin{aligned}\mathbb{E}(g_{c,t}g_{c,t}^{\prime})= & \text{vec}C_{t}^{-1}\Bigl(\text{vec}C_{t}^{-1}\Bigl)^{\text{\ensuremath{\prime}}}-2\mathbb{E}\Bigl(C_{t}^{-1}\otimes I_{n}\Bigl)\text{vec}\bigl(C_{t}^{-1}z_{t}z_{t}^{\prime}\bigl)\Bigl(\text{vec}C_{t}^{-1}\Bigl)^{\text{\ensuremath{\prime}}}\\
 & +\mathbb{E}\Bigl(C_{t}^{-1}\otimes I_{n}\Bigl)\text{vec}\bigl(C_{t}^{-1}z_{t}z_{t}^{\prime}\bigl)\text{vec}\bigl(C_{t}^{-1}z_{t}z_{t}^{\prime}\bigl)^{\prime}\Bigl(C_{t}^{-1}\otimes I_{n}\Bigl)\\
= & -\text{vec}C_{t}^{-1}\Bigl(\text{vec}C_{t}^{-1}\Bigl)^{\text{\ensuremath{\prime}}}+\Bigl(C_{t}^{-1}\otimes I_{n}\Bigl)\mathbb{E}[\text{vec}\bigl(\varepsilon_{t}\varepsilon_{t}^{\prime}\bigl)\text{vec}\bigl(\varepsilon_{t}\varepsilon_{t}^{\prime}\bigl)^{\prime}]\Bigl(C_{t}^{-1}\otimes I_{n}\Bigl)\\
= & -\text{vec}C_{t}^{-1}\Bigl(\text{vec}C_{t}^{-1}\Bigl)^{\text{\ensuremath{\prime}}}+\Bigl(C_{t}^{-1}\otimes I_{n}\Bigl)\Bigl(I_{n^{2}}+K_{n}-\text{vec}\bigl(I_{n}\bigl)\text{vec}\bigl(I_{n}\bigl)^{\prime}\Bigl)\Bigl(C_{t}^{-1}\otimes I_{n}\Bigl),
\end{aligned}
\]
where $\varepsilon_{t}\sim N_{n}(0,I_{n})$ and we have used the identity,
$\mathbb{E}[\text{vec}\bigl(\varepsilon_{t}\varepsilon_{t}^{\prime}\bigl)\text{vec}\bigl(\varepsilon_{t}\varepsilon_{t}^{\prime}\bigl)^{\prime}]=I_{n^{2}}+K_{n}-\text{vec}\bigl(I_{n}\bigl)\text{vec}\bigl(I_{n}\bigl)^{\prime}$,
which is a property of the commutation matrix, $K_{n}$, see \citet{Neudecker:1968}
and \citet{MagnusNeudecker:1979}. This proves that
\[
\begin{aligned}\mathbb{E}\frac{\partial\ell}{\partial\eta}\frac{\partial\ell}{\partial\eta^{\prime}}= & \sum_{t=1}^{T}\dot{\gamma}_{t}^{\prime}\Bigl[\frac{1}{4}J_{t}^{\prime}(E_{l}+E_{u})\mathbb{E}(g_{c,t}g_{c,t}^{\prime})(E_{l}+E_{u})^{\prime}J_{t}+\Lambda_{\tilde{\varphi}}^{\prime}\tilde{\Omega}^{-1}\Lambda_{\tilde{\varphi}}\Bigl]\dot{\gamma}_{t}\end{aligned}
.
\]
The two remaining terms of the Fisher information matrix are given
by
\[
\begin{aligned}\mathbb{E}\frac{\partial\ell}{\partial\phi}\frac{\partial\ell}{\partial\eta^{\prime}}= & \sum_{t=1}^{T}\tilde{M}_{t}^{\prime}\tilde{\Omega}^{-1}\mathbb{E}(\tilde{v}_{t}\tilde{v}_{t}^{\prime})\tilde{\Omega}^{-1}\Lambda_{\tilde{\varphi}}\dot{\gamma}_{t}=\sum_{t=1}^{T}\tilde{M}_{t}^{\prime}\tilde{\Omega}^{-1}\Lambda_{\tilde{\varphi}}\dot{\gamma}_{t}\end{aligned}
,
\]
and
\[
\begin{aligned}\mathbb{E}\frac{\partial\ell}{\partial\phi}\frac{\partial\ell}{\partial\phi^{\prime}}= & \sum_{t=1}^{T}\tilde{M}_{t}^{\prime}\tilde{\Omega}^{-1}\mathbb{E}(\tilde{v}_{t}\tilde{v}_{t}^{\prime})\tilde{\Omega}^{-1}\tilde{M}_{t}=\sum_{t=1}^{T}\tilde{M}_{t}^{\prime}\tilde{\Omega}^{-1}\tilde{M}_{t}\end{aligned}
.
\]

\subsection{Block Model Estimation (Second Stage, $\vartheta_{2}$)\label{subsec:Block-Model-Estimation}}

Suppose $M\in\mathbb{R}^{K\times K}$ is some real matrix of size
$K\times K$. We introduce elimination matrices $E_{h}$ and $E_{q}$
such that $\text{vech}M=E_{h}\,\text{vec}M$ and $\text{vech}M+\text{vech}M^{\prime}-\text{vech}(\Lambda_{\mathrm{diag}(M)})=E_{q}\,\text{vec}M$,
respectively, where by $\Lambda_{\mathrm{diag}(M)}=\mathrm{diag}(M_{11},\ldots,M_{KK})$.

Recall that a $K\times K$ block correlation matrix is given by,{\small
\[
C=\left(\begin{array}{cccc}
C_{[1,1]} & C_{[1,2]} & \cdots & C_{[1,K]}\\
C_{[2,1]} & C_{[2,2]}\\
\vdots &  & \ddots\\
C_{[K,1]} &  &  & C_{[K,K]}
\end{array}\right)\in\mathbb{R}^{n\times n},
\]
}where the blocks given by,{\small
\begin{equation}
C_{[i,i]}=\left(\begin{array}{cccc}
1 & \rho_{i,i} & \cdots & \rho_{i,i}\\
\rho_{i,i} & 1 & \ddots & \vdots\\
\vdots & \ddots & \ddots & \rho_{i,i}\\
\rho_{i,i} & \cdots & \rho_{i,i} & 1
\end{array}\right)\in\mathbb{R}^{n_{i}\times n_{i}}\quad\text{and}\quad C_{[i,j]}=\left(\begin{array}{ccc}
\rho_{i,j} & \cdots & \rho_{i,j}\\
\vdots & \ddots & \vdots\\
\rho_{i,j} & \cdots & \rho_{i,j}
\end{array}\right)\in\mathbb{R}^{n_{i}\times n_{j}},\quad i\neq j.\label{eq:BlockG}
\end{equation}
}The diagonal blocks, $C_{[i,i]}\in\mathbb{R}^{n_{i}\times n_{i}}$,
have ones along the diagonal and $\rho_{i,i}\in(-1,1)$ in all off-diagonal
elements (i.e., equicorrelation structure), and the off-diagonal blocks,
$C_{[i,j]}\in\mathbb{R}^{n_{i}\times n_{j}}$, for $i\neq j$, have
all elements equal to $\rho_{i,j}\in(-1,1)$. Symmetry is guaranteed
with $\rho_{i,j}=\rho_{j,i}$. The correlations $\rho_{i,j}$, $i,j=1,\ldots,K$,
must also be such that $C$ is positive definite. 

The analysis of the Block model is based on the canonical representation
of the block matrices provided in \citet{ArchakovHansen:CanonicalBlockMatrix}.
According to the canonical representation, the conditional correlation
matrix with the block structure can be represented as $C=QDQ^{\prime}$,
where $Q$ is an orthonormal matrix such that $Q^{\prime}Q=I$ and
$D$ is a block-diagonal matrix with the following structure{\small
\[
D=\left(\begin{array}{cccc}
S &  &  & 0\\
 & I_{n_{1-1}}\cdot\lambda_{1}\\
 &  & \ddots\\
0 &  &  & I_{n_{K-1}}\cdot\lambda_{K}
\end{array}\right).
\]
}The $S$-matrix in the upper-left corner is a $K\times K$ matrix
with entries 
\[
S_{i,j}=\begin{cases}
1+(n_{i}-1)\rho_{i,i} & \text{for }i=j,\\
\sqrt{n_{i}n_{j}}\rho_{i,j} & \text{for }i\neq j,
\end{cases}
\]
and $\lambda_{i}=1-\rho_{i,i}$ are eigenvalues of $C$ with the multiplicity
$n_{i}-1$, for $i=1,...,K$. The remaining $K$ eigenvalues of $C$
are also eigenvalues of matrix $S$. Note that the low dimensional
matrix, $S$, contains sufficient information about the correlation
coefficients in $C$.

A remarkable property of this block structure is that it is preserved
under a range of matrix transformations, including the matrix logarithm.
For $G=\log C$ we have,{\small
\[
G=\left(\begin{array}{cccc}
G_{[1,1]} & G_{[1,2]} & \cdots & G_{[1,K]}\\
G_{[2,1]} & G_{[2,2]}\\
\vdots &  & \ddots\\
G_{[K,1]} &  &  & G_{[K,K]}
\end{array}\right)\in\mathbb{R}^{n\times n},
\]
}where{\small
\begin{equation}
G_{[i,i]}=\left(\begin{array}{cccc}
x_{i} & \gamma_{i,i} & \cdots & \gamma_{i,i}\\
\gamma_{i,i} & x_{i} & \ddots & \vdots\\
\vdots & \ddots & \ddots & \gamma_{i,i}\\
\gamma_{i,i} & \cdots & \gamma_{i,i} & x_{i}
\end{array}\right)\in\mathbb{R}^{n_{i}\times n_{i}}\qquad G_{[i,j]}=\left(\begin{array}{ccc}
\gamma_{i,j} & \cdots & \gamma_{i,j}\\
\vdots & \ddots & \vdots\\
\gamma_{i,j} & \cdots & \gamma_{i,j}
\end{array}\right)\in\mathbb{R}^{n_{i}\times n_{j}},\quad i\neq j.\label{eq:BlockG-1}
\end{equation}
}Therefore, a block structure for $C$ implies the same block structure
for $G$, and the number of distinct parameters in $\text{vecl}G$
is identical to the number of distinct correlation parameters in $C$.
Let $\Gamma$ denote a $K\times K$ symmetric matrix, with elements,
$\Gamma_{ij}=\gamma_{ij}$, such that $\Gamma$ contains all the distinct
off-diagonal elements of $G$.

From the canonical representation, we have that $G=\log C=Q(\log D)Q^{\prime}$,
where is the following block-diagonal matrix{\small
\[
\log D=\left(\begin{array}{cccc}
\log S &  &  & 0\\
 & I_{n_{1-1}}\cdot\log\lambda_{1}\\
 &  & \ddots\\
0 &  &  & I_{n_{K-1}}\cdot\log\lambda_{K}
\end{array}\right).
\]
}Furthermore, we have the following relationship between $\log S$
and $S$,
\[
\log S=\Lambda_{n}\Gamma\Lambda_{n}+\log\Lambda_{\lambda},
\]
where $\Lambda_{n}=\text{diag}(n_{1},n_{2},...,n_{K})$ and $\Lambda_{\lambda}=\text{diag}(\lambda_{1},\lambda_{2},...,\lambda_{K})$
are $K\times K$ diagonal matrices. We then can express $\text{vec}\Gamma$
as follows, 
\begin{equation}
\text{vec}\Gamma=\bigl(\Lambda_{n}^{-1}\otimes\Lambda_{n}^{-1}\bigl)\text{vec}\bigl(\log S-\log\Lambda_{\lambda}\bigl).\label{eq:vec_Gamma}
\end{equation}
Then, $\zeta=\text{vech}\Gamma=E_{h}\text{vec}\Gamma$ is a vector
with transformed conditional correlations which is lower dimensional
than $\gamma=\text{vecl}G$.\footnote{Note that we can express $\gamma=A\zeta$, where $A$ is a duplication
matrix with compatible dimensions (see Section 2.4 in the paper).} In the Block MRG model, we model dynamics of vector $\zeta$ instead
of $\gamma$, and this effectively reduces the number of GARCH and
measurement equations from the dimension of $\gamma$ to the dimension
of $\zeta$.

\subsubsection{Gradient Vector}

In the Block MRG model, in the second stage of estimation, we maximize
\begin{equation}
\ell=-\frac{1}{2}\sum_{t=1}^{T}\underbrace{\Bigl\{\log\det D_{t}+\check{z}_{t}^{\prime}D_{t}^{-1}\check{z}_{t}\Bigl\}}_{-2\ell_{c,t}}-\frac{1}{2}\sum_{t=1}^{T}\underbrace{\Bigl\{\log\det\check{\Omega}+\check{v}_{t}^{\prime}\check{\Omega}^{-1}\check{v}_{t}\Bigl\}}_{-2\ell_{y,t}},\label{eq:log-lik-block}
\end{equation}
where term $\ell_{c,t}$ can be further simplified due to the block-diagonal
structure of $D_{t}$,
\[
\begin{aligned}\ell_{c,t}= & -\frac{1}{2}\Bigl(\log\det S_{t}+\check{z}_{0,t}^{\prime}S_{t}^{-1}\check{z}_{0,t}\Bigl)-\frac{1}{2}\sum_{k=1}^{K}\Bigl((n_{k}-1)\log\lambda_{k,t}+\check{z}_{k,t}^{\prime}\check{z}_{k,t}\lambda_{k,t}^{-1}\Bigl)\end{aligned}
\]
(\citet[section 4]{ArchakovHansen:CanonicalBlockMatrix}). The rotated
vector of residuals $z_{t}$, obtained in the first stage, we denote
by $\check{z}_{t}=Q^{\prime}z_{t}=(\check{z}_{0,t}^{\prime},\check{z}_{1,t}^{\prime},...,\check{z}_{K,t}^{\prime})^{\prime}$,
where $\check{z}_{0,t}\in\mathbb{R}^{K}$ and $\check{z}_{k,t}\in\mathbb{R}^{n_{k}-1}$,
$k=1,...,K$. The minimization is with respect to the vector of model
parameters $\vartheta_{2}=(\check{\omega}^{\prime},\check{\beta}^{\prime},\check{\alpha}^{\prime},\check{\xi}^{\prime},\check{\varphi}^{\prime})^{\prime}$,
conditionally on $\check{z}_{t}$, obtained in the first stage, and
on $\check{y}_{t}$ (see Section 3.2). Again, we split the parameter
vector into two components: parameters related to the GARCH equations,
$\eta=(\check{\omega}^{\prime},\check{\beta}^{\prime},\check{\alpha}^{\prime})^{\prime}$,
and parameters related to the measurement equations, $\phi=(\check{\xi}^{\prime},\check{\varphi}^{\prime})^{\prime}$.
Then, for the parameter vector we have, $\vartheta_{2}=(\eta^{\prime},\phi^{\prime})^{\prime}$.

We can restore $C_{t}$ from the sub-matrix $S_{t}$, and it is convenient
to obtain the gradients using,
\[
\frac{\partial\ell}{\partial\eta^{\prime}}=\sum_{t=1}^{T}\Biggl(\frac{\partial\ell_{c,t}}{\partial(\text{vec}S_{t})^{\prime}}\cdot\frac{\partial\text{vec}S_{t}}{\partial\zeta_{t}^{\prime}}+\frac{\partial\ell_{y,t}}{\partial\check{v}_{t}^{\prime}}\cdot\frac{\partial\check{v}_{t}}{\partial\zeta_{t}^{\prime}}\Biggl)\cdot\frac{\partial\zeta_{t}}{\partial\eta^{\prime}},\qquad\text{and}\qquad\frac{\partial\ell}{\partial\phi^{\prime}}=\sum_{t=1}^{T}\frac{\partial\ell_{y,t}}{\partial\check{v}_{t}^{\prime}}\cdot\frac{\partial\check{v}_{t}}{\partial\phi^{\prime}}.
\]

We begin with term $\frac{\partial\ell_{c,t}}{\partial(\text{vec}S_{t})^{\prime}}$.
Using Proposition 1 in \citet{ArchakovHansen:CanonicalBlockMatrix},
we obtain
\[
\frac{\partial}{\partial(\text{vec}S_{t})^{\prime}}\Bigl(\log\det S_{t}+\check{z}_{0,t}^{\prime}S_{t}^{-1}\check{z}_{0,t}\Bigl)=\text{vec}\Bigl(S_{t}^{-1}-S_{t}^{-1}\check{z}_{0,t}\check{z}_{0,t}^{\prime}S_{t}^{-1}\Bigl)^{\prime},
\]
and
\[
\frac{\partial}{\partial(\text{vec}S_{t})^{\prime}}\sum_{k=1}^{K}\Bigl((n_{k}-1)\log\lambda_{k,t}+\check{z}_{k,t}^{\prime}\check{z}_{k,t}\lambda_{k,t}^{-1}\Bigl)=(\text{vec}W_{t})^{\prime},
\]
where $W_{t}$ is a $K\times K$ diagonal matrix with its $k$-th
diagonal element given by $W_{kk,t}=\frac{1}{\lambda_{k,t}}\Bigl(\frac{1}{(n_{k}-1)\lambda_{k,t}}\check{z}_{k,t}^{\prime}\check{z}_{k,t}-1\Bigl)$,
for $k=1,...,K$, and the latter result is due to $\lambda_{k,t}=1-\rho_{kk,t}=\frac{S_{kk,t}-1}{n_{k}-1}$.
So, we have
\[
-2\frac{\partial\ell_{c,t}}{\partial(\text{vec}S_{t})^{\prime}}=(g_{s,t}+w_{s,t})^{\prime},\qquad\text{and}\quad-2\frac{\partial\ell_{y,t}}{\partial\check{v}_{t}^{\prime}}=2\check{v}_{t}^{\prime}\check{\Omega}^{-1},
\]
where $g_{s,t}=\text{vec}\Bigl(S_{t}^{-1}-S_{t}^{-1}\check{z}_{0,t}\check{z}_{0,t}^{\prime}S_{t}^{-1}\Bigl)$,
$w_{s,t}=\text{vec}W_{t}$, and $\frac{\partial\ell_{y,t}}{\partial\check{v}_{t}^{\prime}}$
is obtained analogously to $\frac{\partial\ell_{y,t}}{\partial\tilde{v}_{t}^{\prime}}$
in \ref{subsec:GradientA}.

The term, $\frac{\partial\text{vec}S_{t}}{\partial\zeta_{t}^{\prime}}=\frac{\partial\text{vec}S_{t}}{\partial(\text{vec}\Gamma_{t})^{\prime}}\cdot\frac{\partial\text{vec}\Gamma_{t}}{\partial\zeta_{t}^{\prime}}$,
can be obtained with (\eqref{eq:vec_Gamma}) and the Implicit Function
Theorem,
\[
\begin{aligned}\frac{\partial\text{vec}S_{t}}{\partial(\text{vec}\Gamma_{t})^{\prime}}= & \Biggl(\frac{\partial\text{vec}\log S_{t}}{\partial(\text{vec}S_{t})^{\prime}}-\frac{\partial\text{vec}\log\Lambda_{1-\rho}}{\partial(\text{vec}S_{t})^{\prime}}\Biggl)^{-1}\bigl(\Lambda_{n}\otimes\Lambda_{n}\bigl)\\
= & \Biggl(J_{s,t}^{-1}-E_{d}^{\prime}H_{t}E_{d}\Biggl)^{-1}\bigl(\Lambda_{n}\otimes\Lambda_{n}\bigl),
\end{aligned}
\]
where $J_{s,t}=\frac{\partial\text{vec}S_{t}}{\partial(\text{vec}\log S_{t})^{\prime}}$
is a $K^{2}\times K^{2}$ Jacobian matrix defined as in Section \ref{subsec:PrelimA},
but for matrix $S_{t}$ instead of $C_{t}$, and $H_{t}$ is a diagonal
$K\times K$ matrix with diagonal elements given by $H_{kk,t}=\frac{1}{(n_{k}-1)\lambda_{k,t}}$
for $k=1,...,K$. The first term involves the inversion of the $K^{2}\times K^{2}$
matrix, and this can be simplified by using the Woodbury matrix identity,
\[
\begin{aligned}\bigl(J_{s,t}^{-1}-E_{d}^{\prime}H_{t}E_{d}\bigl)^{-1}= & J_{s,t}\bigl(I_{K^{2}}-J_{s,t}^{-1}E_{d}^{\prime}H_{t}E_{d}\bigl)^{-1}\\
= & J_{s,t}\bigl(I_{K^{2}}+J_{s,t}^{-1}E_{d}^{\prime}\bigl(I_{K}-H_{t}E_{d}J_{s,t}^{-1}E_{d}^{\prime}\bigl)^{-1}H_{t}E_{d}\bigl)\\
= & J_{s,t}+E_{d}^{\prime}\bigl(I_{K}-H_{t}E_{d}J_{s,t}^{-1}E_{d}^{\prime}\bigl)^{-1}H_{t}E_{d}.
\end{aligned}
\]
So, the matrix to be inverted has dimension $K\times K$. Given that
$\frac{\partial\text{vec}\Gamma_{t}}{\partial\zeta_{t}^{\prime}}=E_{q}^{\prime}$,
we finally obtain
\[
\frac{\partial\text{vec}S_{t}}{\partial\zeta_{t}^{\prime}}=\bigl(J_{s,t}+E_{d}^{\prime}\bigl(I_{K}-H_{t}E_{d}J_{s,t}^{-1}E_{d}^{\prime}\bigl)^{-1}H_{t}E_{d}\bigl)\bigl(\Lambda_{n}\otimes\Lambda_{n}\bigl)E_{q}^{\prime}.
\]
Also, we have $\frac{\partial\check{v}_{t}}{\partial\zeta_{t}^{\prime}}=-\Lambda_{\check{\varphi}}$,
where $\Lambda_{\check{\varphi}}=\text{diag}(\check{\varphi}_{1},...,\check{\varphi}_{r})$.

The derivatives with respect to the model parameter vectors, $\eta$
and $\phi$, are obtained analogously to the derivatives in Section
\ref{subsec:GradientA}. Thus, $\frac{\partial\zeta_{t}}{\partial\eta^{\prime}}=\dot{\zeta}_{t}=\sum_{j=0}^{t-1}\Lambda_{\check{\beta}}^{j}\check{F}_{t-j}$,
where $\Lambda_{\check{\beta}}=\text{diag}(\check{\beta}_{1},...,\check{\beta}_{r})$
and $\check{F}_{t}=\Bigl[I_{r},\:\Lambda_{\zeta_{t-1}},\:\Lambda_{\check{y}_{t-1}}\Bigl]$
is $r\times3r$ matrix, $\Lambda_{\zeta_{t-1}}=\text{diag}(\zeta_{t-1})$
and $\Lambda_{\check{y}_{t-1}}=\text{diag}(\check{y}_{t-1})$. Also,
$\frac{\partial\check{v}_{t}}{\partial\phi^{\prime}}=-\check{M}_{t}$,
where $\check{M}_{t}=\Bigl[I_{r},\:\Lambda_{\zeta_{t}}\Bigl]$ and
$\Lambda_{\zeta_{t}}=\text{diag}(\zeta_{t})$.

Then, we can rewrite the log-likelihood functions as
\[
\frac{\partial\ell}{\partial\eta^{\prime}}=-\frac{1}{2}\sum_{t=1}^{T}\Biggl((g_{s,t}+w_{s,t})^{\prime}\Psi_{t}+2\check{v}_{t}^{\prime}\check{\Omega}^{-1}\Lambda_{\check{\varphi}}\Biggl)\dot{\zeta}_{t},\qquad\text{and}\quad\frac{\partial\ell}{\partial\phi^{\prime}}=\sum_{t=1}^{T}\check{v}_{t}^{\prime}\check{\Omega}^{-1}\check{M}_{t},
\]
where we denote $\Psi_{t}=\frac{\partial\text{vec}S_{t}}{\partial\zeta_{t}^{\prime}}$.
Together, these results provide the gradient for the entire parameter
vector, $\frac{\partial\ell}{\partial\vartheta_{2}}=\Bigl(\frac{\partial\ell}{\partial\eta^{\prime}},\frac{\partial\ell}{\partial\phi^{\prime}}\Bigl)^{\prime}$.

\subsubsection{Approximation of Hessian Matrix}

Analogously to the Full model, we provide an approximation to the
Hessian matrix in the Block specification by means of the Fisher Information
matrix,
\[
\mathbb{E}\biggl(\frac{\partial\ell}{\partial\vartheta_{2}}\frac{\partial\ell}{\partial\vartheta_{2}^{\prime}}\biggl)=\left(\begin{array}{cc}
\mathbb{E}\frac{\partial\ell}{\partial\eta}\frac{\partial\ell}{\partial\eta^{\prime}} & \mathbb{E}\frac{\partial\ell}{\partial\eta}\frac{\partial\ell}{\partial\phi^{\prime}}\\
\mathbb{E}\frac{\partial\ell}{\partial\phi}\frac{\partial\ell}{\partial\eta^{\prime}} & \mathbb{E}\frac{\partial\ell}{\partial\phi}\frac{\partial\ell}{\partial\phi^{\prime}}
\end{array}\right).
\]
For the upper left corner, we note that $\mathbb{E}(g_{s,t}w_{s,t}^{\prime})=\mathbb{E}(g_{s,t}w_{s,t}^{\prime})=0$,
because $\mathbb{E}(w_{s,t})=0$ and the elements in $g_{s,t}$ and
$w_{s,t}$ are uncorrelated. Moreover,
\[
\mathbb{E}(g_{s,t}g_{s,t}^{\prime})=\text{vec}S_{t}^{-1}\Bigl(\text{vec}S_{t}^{-1}\Bigl)^{\text{\ensuremath{\prime}}}+\Bigl(S_{t}^{-1}\otimes I_{K}\Bigl)\Bigl(I_{K^{2}}+K_{K}-\text{vec}\bigl(I_{K}\bigl)\text{vec}\bigl(I_{K}\bigl)^{\prime}\Bigl)\Bigl(S_{t}^{-1}\otimes I_{K}\Bigl),
\]
is obtained analogously to $\mathbb{E}(g_{s,t}g_{s,t}^{\prime})$
in Section \ref{subsec:Approximation-of-Hessian_1}. Let $V_{t}$
be the $K\times K$ diagonal matrix with diagonal elements, $V_{kk,t}=\frac{2}{(n_{k}-1)\lambda_{k,t}^{2}}$,
for $k=1,...,K$. Then, $\mathbb{E}(w_{s,t}w_{s,t}^{\prime})$ is
a $K^{2}\times K^{2}$ diagonal matrix with $\text{vec}V_{t}$ on
the main diagonal. Finally, we have
\[
\begin{aligned}\mathbb{E}\frac{\partial\ell}{\partial\eta}\frac{\partial\ell}{\partial\eta^{\prime}}= & \sum_{t=1}^{T}\dot{\gamma}_{t}^{\prime}\Bigl[\frac{1}{4}\Psi_{t}^{\prime}\Bigl(\mathbb{E}(g_{s,t}g_{s,t}^{\prime})+\mathbb{E}(w_{s,t}w_{s,t}^{\prime})\Bigl)\Psi_{t}+\Lambda_{\check{\varphi}}^{\prime}\check{\Omega}^{-1}\Lambda_{\check{\varphi}}\Bigl]\dot{\gamma}_{t}\end{aligned}
.
\]
The remaining terms of the Fisher Information matrix are obtained
similarly to the analogous terms derived in Section \ref{subsec:Approximation-of-Hessian_1},
\[
\mathbb{E}\frac{\partial\ell}{\partial\phi}\frac{\partial\ell}{\partial\eta^{\prime}}=\sum_{t=1}^{T}\check{M}_{t}^{\prime}\check{\Omega}^{-1}\Lambda_{\check{\varphi}}\dot{\zeta}_{t},\qquad\text{and}\quad\mathbb{E}\frac{\partial\ell}{\partial\phi}\frac{\partial\ell}{\partial\phi^{\prime}}=\sum_{t=1}^{T}\check{M}_{t}^{\prime}\check{\Omega}^{-1}\check{M}_{t}.
\]

\section{Estimation Procedure}

In this section, we provide a description of the estimation procedure
for the Multivariate Realized GARCH model that was briefly outlined
in Section 3.2. We omit the details on the first stage of the model
estimation (parameter vector $\vartheta_{1}$) that involves estimation
of $n$ univariate realized GARCH models. This part closely follows
the estimation routines outlined in \citet{HansenLundeVoev:2014}
and \citet{HansenHuang:2016}.

After the first stage estimation, we have the standardized return
vectors, $z_{t}=(z_{1,t},...,z_{n,t})^{\prime}$, for $t=1,...,T$.
These will be used in the second stage in conjunction with the vector
of transformed correlations, $y_{t}=\text{vecl}\log Y_{t}$, where
$Y_{t}$ is the realized correlation matrix, constructed from high
frequency data on day $t$, for $t=1,...,T$. 

\subsection{Full (unrestricted) Model Estimation ($\vartheta_{2}$)}

In the second stage of the two-stage estimation procedure, we maximize
the log-likelihood component in (\ref{eq:log-lik-full}) with respect
to $\vartheta_{2}$. 
\begin{enumerate}
\item Initialize: The starting values for the transformed conditional correlations,
$\gamma_{1}$, can be treated as a model parameter, in which case
we initialize the static model parameters, $\vartheta_{2}=(\tilde{\omega}^{\prime},\tilde{\beta}^{\prime},\tilde{\alpha}^{\prime},\tilde{\xi}^{\prime},\tilde{\varphi}^{\prime},\gamma_{1}^{\prime})^{\prime}$.
Alternatively, we assign $\gamma_{1}$ a suitable value, such as the
average value of the transformed realized correlations, $y_{t}$,
over the first several months of a sample period, and initialize $\vartheta_{2}=(\tilde{\omega}^{\prime},\tilde{\beta}^{\prime},\tilde{\alpha}^{\prime},\tilde{\xi}^{\prime},\tilde{\varphi}^{\prime})^{\prime}$.
\item Given a value for $\vartheta_{2}$. 
\begin{enumerate}
\item Compute $\gamma_{t}(\vartheta_{2})$ using the GARCH equations, from
which we obtain $C_{t}(\vartheta_{2})=C\bigl(\gamma_{t}(\vartheta_{2})\bigl)$,
using the algorithm provided in \citet[corollary 1]{ArchakovHansen:Correlation},
for $t=1,...,T$. 
\item Compute the vectors of measurement errors, $\tilde{v}_{t}(\vartheta_{2})$,
for $t=1,...,T$, using the measurement equations. From these we compute
$\tilde{\Omega}=T^{-1}\sum_{t=1}^{T}\tilde{v}_{t}(\vartheta_{2})\tilde{v}_{t}(\vartheta_{2})^{\prime}$.
\end{enumerate}
\item Given $z_{t}$, $C_{t}(\vartheta_{2})$, and $\tilde{v}_{t}(\vartheta_{2})$,
we evaluate the partial log-likelihood function given in (\ref{eq:log-lik-full}). 
\end{enumerate}
We maximize the log-likelihood in (\ref{eq:log-lik-full}) with respect
to $\vartheta_{2}$, where steps 2 and 3 are repeated, every time
$\vartheta_{2}$ is updated. If a gradient-based algorithm is used
for the numeric optimization of the partial log-likelihood function,
then it can be highly beneficial to supply it with the analytical
derivatives we obtained in Section \ref{subsec:Full-Model-Estimation}.

\subsection{Block (restricted) Model Estimation ($\vartheta_{2}$)}

Maximizing the log-likelihood in (\ref{eq:log-lik-full}) can be computationally
challenging if $d=\frac{1}{2}n(n-1)$ is large. In this case one can
impose the factor structure, $\gamma_{t}=A\zeta_{t}$, such as the
one induced by block correlation matrices, because a block structure
in $C$ implies the same block structure in $\log C$, see Section
\ref{subsec:Block-Model-Estimation}. Only one unique $\gamma$-element
is needed per block, such that the vector of transformed correlations
can be represented as $\gamma_{t}=A\zeta_{t}$ for some selected elimination
matrix $A$ of size $d\times r$, and all sufficient information about
conditional correlation matrix is incorporated in a smaller $r$-dimensional
vector $\zeta_{t}$. Thus, in the second stage of the two-stage estimation
procedure with a block correlation structure we maximize the log-likelihood
component given in (\ref{eq:log-lik-block}) with respect to $\vartheta_{2}$.
\begin{enumerate}
\item Apply the vector parametrization to the realized correlation matrices,
so we obtain $y_{t}=\text{vecl}\log Y_{t}$, for $t=1,...,T$. Then,
construct a smaller dimensional realized measures of transformed block
correlations, $\check{y}_{t}=(A^{\prime}A)^{-1}A^{\prime}y_{t}$,
for $t=1,...,T$, which are used as realized measures for $\zeta_{t}$.
\item Initialize the static model parameters in $\vartheta_{2}=(\check{\omega}^{\prime},\check{\beta}^{\prime},\check{\alpha}^{\prime},\check{\xi}^{\prime},\check{\varphi}^{\prime})^{\prime}$.
Also, initialize starting values for the transformed conditional correlations,
$\zeta_{1}$. Similarly, $\zeta_{1}$ can be treated as a model parameter
being included in $\theta$, or can be pre-initialized with observed
realized correlations.
\item Compute $\zeta_{t}(\vartheta_{2})$ using the GARCH equations (7)
and then map $\zeta_{t}$ to $C_{t}=C(\zeta_{t})$, for $t=1,...,T$.
The mapping from $\zeta_{t}$ to $C_{t}$  can be done with the algorithm
provided in \citet[theorem 5]{ArchakovHansenLuo-RandomCorr:2024}.
\item For given $\vartheta_{2}$, using the measurement equations (8), compute
the vectors of measurement errors, $\check{v}_{t}(\vartheta_{2})$,
for $t=1,...,T$. This allows to compute $\check{\Omega}=T^{-1}\sum_{t=1}^{T}\check{v}_{t}(\vartheta_{2})\check{v}_{t}(\vartheta_{2})^{\prime}$
and concentrate it out from the log-likelihood function.
\item Using $z_{t}$, $C_{t}(\vartheta_{2})$, $\check{v}_{t}(\vartheta_{2})$,
evaluate the partial log-likelihood function given in (\ref{eq:log-lik-block}). 
\end{enumerate}
We maximize the log-likelihood in (\ref{eq:log-lik-block}) with respect
to $\vartheta_{2}$, by repeating steps 2-5 every time $\vartheta_{2}$
is updated. If gradient based algorithms are used for numeric optimization,
we can accelerate the estimation process by supplying the analytical
derivatives obtained in Section \ref{subsec:Block-Model-Estimation}.

\newpage{}

\bibliographystyle{apalike}
\bibliography{prh}

\begin{thebibliography}{}

\bibitem[Aielli, 2013]{Aielli:2013}
Aielli, G.~P. (2013).
\newblock Dynamic conditional correlation: on properties and estimation.
\newblock {\em Journal of Business and Economic Statistics}, 31:282--299.

\bibitem[Amisano and Giacomini, 2007]{Amisano2007}
Amisano, G. and Giacomini, R. (2007).
\newblock {Comparing Density Forecasts via Weighted Likelihood Ratio Tests}.
\newblock {\em Journal of Business and Economic Statistics}, 25:177--190.

\bibitem[Andersen and Bollerslev, 1998]{AndersenBollerslev:1998a}
Andersen, T.~G. and Bollerslev, T. (1998).
\newblock Answering the skeptics: Yes, standard volatility models do provide
  accurate forecasts.
\newblock {\em International Economic Review}, 39:885--905.

\bibitem[Andersen et~al., 2001a]{ABDE:2001}
Andersen, T.~G., Bollerslev, T., Diebold, F.~X., and Ebens, H. (2001a).
\newblock The distribution of realized stock return volatility.
\newblock {\em Journal of Financial Economics}, 61:43--76.

\bibitem[Andersen et~al., 2001b]{ABDL:2001}
Andersen, T.~G., Bollerslev, T., Diebold, F.~X., and Labys, P. (2001b).
\newblock The distribution of realized exchange rate volatility.
\newblock {\em Journal of the American Statistical Association}, 96:42--55.

\bibitem[Andersen et~al., 2003]{ABDL:2003}
Andersen, T.~G., Bollerslev, T., Diebold, F.~X., and Labys, P. (2003).
\newblock Modeling and forecasting realized volatility.
\newblock {\em Econometrica}, 71:579--625.

\bibitem[Archakov and Hansen, 2021]{ArchakovHansen:Correlation}
Archakov, I. and Hansen, P.~R. (2021).
\newblock A new parametrization of correlation matrices.
\newblock {\em Econometrica}, 89:1699--1715.

\bibitem[Archakov and Hansen, 2024]{ArchakovHansen:CanonicalBlockMatrix}
Archakov, I. and Hansen, P.~R. (2024).
\newblock A canonical representation of block matrices with applications to
  covariance and correlation matrices.
\newblock {\em Review of Economics and Statistics}, 106:1--15.

\bibitem[Archakov et~al., 2024]{ArchakovHansenLuo-RandomCorr:2024}
Archakov, I., Hansen, P.~R., and Luo, Y. (2024).
\newblock A new method for generating random correlation matrices.
\newblock {\em Econometrics Journal}, forthcoming.

\bibitem[Asai and So, 2015]{AsaiSo:2015}
Asai, M. and So, M. (2015).
\newblock Long memory and asymmetry for matrix-exponential dynamic correlation
  processes.
\newblock {\em Journal of Time Series Econometrics}, 7:69--74.

\bibitem[Barndorff-Nielsen et~al., 2009]{BNHLS-RKpractice:2009}
Barndorff-Nielsen, O.~E., Hansen, P.~R., Lunde, A., and Shephard, N. (2009).
\newblock Realized kernels in practice: trades and quotes.
\newblock {\em Econometrics Journal}, 12:C1--C32.

\bibitem[Barndorff-Nielsen et~al., 2011]{BNHLS-MRK:2011}
Barndorff-Nielsen, O.~E., Hansen, P.~R., Lunde, A., and Shephard, N. (2011).
\newblock Multivariate realised kernels: consistent positive semi-definite
  estimators of the covariation of equity prices with noise and non-synchronous
  trading.
\newblock {\em Jounal of Econometrics}, 162:149--169.

\bibitem[Barndorff-Nielsen and Shephard, 2002]{BNS:2002}
Barndorff-Nielsen, O.~E. and Shephard, N. (2002).
\newblock Estimating quadratic variation using realized variance.
\newblock {\em Journal of Applied Econometrics}, 17:457--477.

\bibitem[Barndorff-Nielsen and Shephard, 2004]{BNS:2004}
Barndorff-Nielsen, O.~E. and Shephard, N. (2004).
\newblock Econometric analysis of realized covariation: High frequency based
  covariance, regression, and correlation in financial economics.
\newblock {\em Econometrica}, 72:885--925.

\bibitem[Bauer and Vorkink, 2011]{BauerVorkink:2011}
Bauer, G.~H. and Vorkink, K. (2011).
\newblock Forecasting multivariate realized stock market volatility.
\newblock {\em Journal of Econometrics}, 160:93--101.

\bibitem[Bauwens et~al., 2006]{BauwensLaurentRombouts:2006}
Bauwens, L., Laurent, S., and Rombouts, J. V.~K. (2006).
\newblock Multivariate {GARCH} models: A survey.
\newblock {\em Journal of Applied Econometrics}, 21:79--109.

\bibitem[Bauwens et~al., 2012]{BauwensStortiViolante:2012}
Bauwens, L., Storti, G., and Violante, F. (2012).
\newblock Dynamic conditional correlation models for realized covariance
  matrices.
\newblock {\em CORE Discussion Papers:2012060}.

\bibitem[Black, 1976]{Black:1976}
Black, F. (1976).
\newblock Studies of stock price volatility changes.
\newblock In {\em Proceedings of the 1976 Meetings of the American Statistical
  Association, Business and Economic Statistics Section}, Washington, D.C.
  American Statistical Association.

\bibitem[Bollerslev, 1990]{Bollerslev:1990}
Bollerslev, T. (1990).
\newblock Modelling the coherence in short-run nominal exchange rates: A
  multivariate generalized {ARCH} model.
\newblock {\em The Review of Economics and Statistics}, 72:498--505.

\bibitem[Bougerol and Picard, 1992]{BougerolPicard:1992}
Bougerol, P. and Picard, N. (1992).
\newblock Stationarity of {GARCH} processes and of some nonnegative time
  series.
\newblock {\em Journal of Econometrics}, 52:115--127.

\bibitem[Carrasco and Chen, 2002]{CarrascoChen:2002}
Carrasco, M. and Chen, X. (2002).
\newblock Mixing and moment properties of various {GARCH} and stochastic
  volatility models.
\newblock {\em Econometric Theory}, 18:17--39.

\bibitem[Chiriac and Voev, 2011]{ChiriacVoev:2011}
Chiriac, R. and Voev, V. (2011).
\newblock Modelling and forecasting multivariate realized volatility.
\newblock {\em Journal of Applied Econometrics}, 26:922--947.

\bibitem[Chiu et~al., 1996]{ChiuLeonardTsui:1996}
Chiu, T., Leonard, T., and Tsui, K.-W. (1996).
\newblock The matrix-logarithmic covariance model.
\newblock {\em Journal of the American Statistical Association}, 91:198--210.

\bibitem[Christie, 1982]{Christie:1982}
Christie, A.~A. (1982).
\newblock The stochastic behavior of common stock variances : Value, leverage
  and interest rate effects.
\newblock {\em Journal of Financial Economics}, 10:407--432.

\bibitem[Creal et~al., 2013]{CrealKoopmanLucas:2013}
Creal, D., Koopman, S.~J., and Lucas, A. (2013).
\newblock Generalized autoregressive score models with applications.
\newblock {\em Journal of Applied Econometrics}, 28:777--795.

\bibitem[Engle and Kelly, 2012]{EngleKelly:2012}
Engle, R. and Kelly, B. (2012).
\newblock Dynamic equicorrelation.
\newblock {\em Journal of Business and Economic Statistics}, 30:212--228.

\bibitem[Engle, 1982]{Engle:1982}
Engle, R.~F. (1982).
\newblock Autoregressive conditional heteroskedasticity with estimates of the
  variance of {U.K.} inflation.
\newblock {\em Econometrica}, 45:987--1007.

\bibitem[Engle, 2002a]{Engle2002}
Engle, R.~F. (2002a).
\newblock Dynamic conditional correlation: A simple class of multivariate
  generalized autoregressive conditional heteroskedasticity models.
\newblock {\em Journal of Business and Economic Statistics}, 20:339--350.

\bibitem[Engle, 2002b]{Engle2002b}
Engle, R.~F. (2002b).
\newblock New frontiers for {ARCH} models.
\newblock {\em Journal of Applied Econometrics}, 17:425--446.

\bibitem[Engle and Gallo, 2006]{engle-gallo:06}
Engle, R.~F. and Gallo, G. (2006).
\newblock A multiple indicators model for volatility using intra-daily data.
\newblock {\em Journal of Econometrics}, 131:3--27.

\bibitem[Engle et~al., 2019]{EngleLedoitWolf:2019}
Engle, R.~F., Ledoit, O., and Wolf, M. (2019).
\newblock {Large dynamic covariance matrices}.
\newblock {\em Journal of Business and Economic Statistics}, 37:363--375.

\bibitem[Engle and Ng, 1993]{Engle_Ng_1993}
Engle, R.~F. and Ng, V.~K. (1993).
\newblock { Measuring and Testing the Impact of News on Volatility}.
\newblock {\em Journal of Finance}, 48:1749--78.

\bibitem[Engle and Sheppard, 2001]{EngleSheppard:2001}
Engle, R.~F. and Sheppard, K. (2001).
\newblock {Theoretical and Empirical properties of Dynamic Conditional
  Correlation Multivariate GARCH}.
\newblock NBER Working Papers 8554, National Bureau of Economic Research, Inc.

\bibitem[Francq and Zakoian, 2019]{FrancqZakoian2019}
Francq, C. and Zakoian, J.-M. (2019).
\newblock {\em GARCH Models: Structure, Statistical Inference and Financial
  Applications}.
\newblock Wiley, 2nd edition.

\bibitem[Geweke and Amisano, 2010]{GewekeAmisano:2010}
Geweke, J. and Amisano, G. (2010).
\newblock {Comparing and evaluating Bayesian predictive distributions of asset
  returns}.
\newblock {\em International Journal of Forecasting}, 26:216--230.

\bibitem[Golosnoy et~al., 2012]{Golosnoy_Gribisch_Liesenfeld_2012}
Golosnoy, V., Gribisch, B., and Liesenfeld, R. (2012).
\newblock {The conditional autoregressive Wishart model for multivariate stock
  market volatility}.
\newblock {\em Journal of Econometrics}, 167(1):211--223.

\bibitem[Gorgi et~al., 2019]{GorgiHansenJanusKoopman:2019}
Gorgi, P., Hansen, P.~R., Janus, P., and Koopman, S.~J. (2019).
\newblock Realized {W}ishart-{GARCH}: A score-driven multi-asset volatility
  model.
\newblock {\em Journal of Financial Econometrics}, 17:1--32.

\bibitem[Hafner and Wang, 2023]{HafnerWang:2023}
Hafner, C.~M. and Wang, L. (2023).
\newblock A dynamic conditional score model for the log correlation matrix.
\newblock {\em Journal of Econometrics}, 237.

\bibitem[Hansen and Dumitrescu, 2022]{HansenDumitrescu:2022}
Hansen, P.~R. and Dumitrescu, E. (2022).
\newblock How should parameter estimation be tailored to the objective?
\newblock {\em Journal of Econometrics}, 230:535--558.

\bibitem[Hansen and Huang, 2016]{HansenHuang:2016}
Hansen, P.~R. and Huang, Z. (2016).
\newblock Exponential {GARCH} modeling with realized measures of volatility.
\newblock {\em Journal of Business and Economic Statistics}, 34:269--287.

\bibitem[Hansen et~al., 2012]{HansenHuangShek:2012}
Hansen, P.~R., Huang, Z., and Shek, H. (2012).
\newblock Realized {GARCH}: A joint model of returns and realized measures of
  volatility.
\newblock {\em Journal of Applied Econometrics}, 27:877--906.

\bibitem[Hansen and Lunde, 2011]{HansenLundeVolForecastingHandbook}
Hansen, P.~R. and Lunde, A. (2011).
\newblock Forecasting volatility using high frequency data.
\newblock In Clements, M. and Hendry, D., editors, {\em Handbook of Economic
  Forecasting}, chapter~19, pages 525--556. Oxford University Press.

\bibitem[Hansen et~al., 2011]{HansenLundeNasonMCS}
Hansen, P.~R., Lunde, A., and Nason, J.~M. (2011).
\newblock The model confidence set.
\newblock {\em Econometrica}, 79:456--497.

\bibitem[Hansen et~al., 2014]{HansenLundeVoev:2014}
Hansen, P.~R., Lunde, A., and Voev, V. (2014).
\newblock Realized beta {GARCH}: A multivariate {GARCH} model with realized
  measures of volatility.
\newblock {\em Journal of Applied Econometrics}, 29:774--799.

\bibitem[Hetland et~al., 2023]{HetlandPedersenRahbek:2023}
Hetland, S., Pedersen, R.~S., and Rahbek, A. (2023).
\newblock Dynamic conditional eigenvalue {GARCH}.
\newblock {\em Journal of Econometrics}, 237:105175.

\bibitem[Ishihara et~al., 2016]{IshiharaOmoriAsai:2016}
Ishihara, T., Omori, Y., and Asai, M. (2016).
\newblock Matrix exponential stochastic volatility with cross leverage.
\newblock {\em Computational Statistics \& Data Analysis}, 100:331--350.

\bibitem[Jensen and Rahbek, 2004]{JensenRahbek:2004}
Jensen, S.~T. and Rahbek, A.~C. (2004).
\newblock Asymptotic normality of the {QMLE} estimator of {ARCH} in the
  nonstationary case.
\newblock {\em Econometrica}, 72:641--646.

\bibitem[Kawakatsu, 2006]{Kawakatsu:2006}
Kawakatsu, H. (2006).
\newblock Matrix exponential {GARCH}.
\newblock {\em Journal of Econometrics}, 134:95--128.

\bibitem[Kristensen, 2009]{Kristensen:2009}
Kristensen, D. (2009).
\newblock On stationarity and ergodicity of the bilinear model with
  applications to {GARCH} models.
\newblock {\em Journal of Time Series Analysis}, 30:125--144.

\bibitem[Linton and McCrorie, 1995]{LintonMcCrorie:1995}
Linton, O. and McCrorie, J.~R. (1995).
\newblock Differentiation of an exponential matrix function: Solution.
\newblock {\em Econometric Theory}, 11:1182--1185.

\bibitem[Liu, 2009]{Liu_2009}
Liu, Q. (2009).
\newblock On portfolio optimization: {H}ow and when do we benefit from
  high-frequency data?
\newblock {\em Journal of Applied Econometrics}, 24:560--582.

\bibitem[Lunde and Olesen, 2014]{LundeOlesenEnergyRealizedGARCH}
Lunde, A. and Olesen, K.~V. (2014).
\newblock Modeling and forecasting the volatility of energy forward returns.
\newblock {\em Working Paper}.

\bibitem[Magnus and Neudecker, 1979]{MagnusNeudecker:1979}
Magnus, J.~R. and Neudecker, H. (1979).
\newblock The commutation matrix: some properties and applications.
\newblock {\em The Annals of Statistics}, 7:381--394.

\bibitem[Meitz and Saikkonen, 2008]{MeitzSaikkonen:2008}
Meitz, M. and Saikkonen, P. (2008).
\newblock Ergodicity, mixing and existence of moments of a class of {M}arkov
  models with applications to {GARCH} and {ACD} models.
\newblock {\em Econometric Theory}, 24:1291--1320.

\bibitem[Neudecker, 1968]{Neudecker:1968}
Neudecker, H. (1968).
\newblock The kronecker matrix product and some of its applications in
  econometrics.
\newblock {\em Statistica Neerlandica}, 22:69--82.

\bibitem[Noureldin et~al., 2012]{NoureldinShephardSheppard:2012}
Noureldin, D., Shephard, N., and Sheppard, K. (2012).
\newblock Multivariate high-frequency-based volatility ({HEAVY}) models.
\newblock {\em Journal of Applied Econometrics}, 27:907--933.

\bibitem[Nyblom, 1989]{Nyblom89}
Nyblom, J. (1989).
\newblock Testing for the constancy of parameters over time.
\newblock {\em Journal of the American Statistical Association}, 84:223--230.

\bibitem[Pakel et~al., 2021]{PakelShephardSheppardEngle:2021}
Pakel, C., Shephard, N., Sheppard, K., and Engle, R.~F. (2021).
\newblock Fitting vast dimensional time-varying covariance models.
\newblock {\em Journal of Business and Economic Statistics}, 39:652--668.

\bibitem[Shephard and Sheppard, 2010]{ShephardSheppard:2010}
Shephard, N. and Sheppard, K. (2010).
\newblock {Realising the future: forecasting with high-frequency-based
  volatility (HEAVY) models}.
\newblock {\em Journal of Applied Econometrics}, 25:197--231.

\bibitem[Silvennoinen and Terasvirta, 2009]{SilvennoinenTerasvirta:2009}
Silvennoinen, A. and Terasvirta, T. (2009).
\newblock Multivariate {GARCH} models.
\newblock In Andersen, T.~G., Davis, R.~A., Kreiss, J.-P., and Mikosch, T.,
  editors, {\em Handbook of Financial Econometrics}, pages 201--229. Springer
  Berlin Heidelberg.

\bibitem[Straumann and Mikosch, 2006]{StraumannMikosch:06}
Straumann, D. and Mikosch, T. (2006).
\newblock Quasi-maximum-likelihood estimation in conditionally heteroscedastic
  time series: A stochastic recurrence equation approach.
\newblock {\em Annals of Statistics}, 34:2449--2495.

\bibitem[Weigand, 2014]{Weigand:2014}
Weigand, R. (2014).
\newblock Matrix {B}ox-{C}ox models for multivariate realized volatility.
\newblock {\em Working Paper}.

\end{thebibliography}

\end{document}